\documentclass{article}
\usepackage[T1]{fontenc}

\usepackage{microtype}
\usepackage{graphicx}
\usepackage{subcaption}
\usepackage{booktabs} 
\usepackage{amsmath}
\usepackage{amssymb}
\usepackage{statmath}

\usepackage{hyperref}

\DeclareMathOperator{\Var}{Var}
\DeclareMathOperator{\Cov}{Cov}

\usepackage[accepted]{eiml_icml2026}

\usepackage{amsmath}
\usepackage{amssymb}
\usepackage{mathtools}
\usepackage{amsthm}

\usepackage[capitalize,noabbrev]{cleveref}

\theoremstyle{plain}
\newtheorem{theorem}{Theorem}[section]
\newtheorem{proposition}[theorem]{Proposition}
\newtheorem{lemma}[theorem]{Lemma}
\newtheorem{corollary}[theorem]{Corollary}
\theoremstyle{definition}
\newtheorem{definition}[theorem]{Definition}
\newtheorem{assumption}[theorem]{Assumption}
\theoremstyle{remark}
\newtheorem{remark}[theorem]{Remark}

\usepackage[textsize=tiny]{todonotes}

\icmltitlerunning{Semantic Identifiability in Pairwise Preference Learning}

\begin{document}

\twocolumn[
  \icmltitle{What Does Disagreement Mean? A Semantic Identifiability Frontier for Pairwise Preference Learning}


  \icmlsetsymbol{equal}{*}

  \begin{icmlauthorlist}
    \icmlauthor{Manoj Saravanan}{vt}
    \icmlauthor{Rohit Kumar Salla}{vt}
    \icmlauthor{Shrikar Reddy Kota}{vt}
  \end{icmlauthorlist}

  \icmlaffiliation{vt}{Virginia Tech, Blacksburg, VA, USA}

  \icmlcorrespondingauthor{Manoj Saravanan}{manoj663@vt.edu}

  \icmlkeywords{preference learning, identifiability, annotator disagreement, epistemic uncertainty, pairwise comparisons}

  \vskip 0.3in
]

\printAffiliationsAndNotice{}  

\hypersetup{
  pdftitle={What Does Disagreement Mean? A Semantic Identifiability Frontier for Pairwise Preference Learning},
  pdfauthor={Manoj Saravanan, Rohit Kumar Salla, Shrikar Reddy Kota},
  pdfsubject={2nd Workshop on Epistemic Intelligence in Machine Learning (EIML@ICML 2026)},
  pdfkeywords={preference learning, identifiability, annotator disagreement, epistemic uncertainty, pairwise comparisons}
}

\begin{abstract}
Pairwise comparison data are a standard supervision signal in preference learning,
but disagreement in such data can reflect stable latent heterogeneity, ephemeral
stochasticity, or non-binary responses hidden by a forced-choice interface. We
study which of these semantic distinctions are identifiable from the feedback
channel itself. We introduce a nonparametric latent semantic model with three
querywise quantities: aggregate directional preference $\mu_q$, persistent
disagreement $\tau_q$, and non-binary mass $\alpha_q$. Under the anonymous binary
channel, we prove an exact semantic fiber theorem: $\mu_q$ is point-identified,
whereas $(\tau_q,\alpha_q)$ vary over an explicit feasible region compatible with
the same binary law. Same-annotator repeats identify $\tau_q$ through a
covariance identity, abstention-enabled feedback identifies $\alpha_q$ directly,
and together they point-identify the full semantic triple. We also derive
finite-sample estimators, nonasymptotic concentration bounds, and semantics-aware
action-recovery guarantees, and validate the theory on theorem-matched synthetic
data.
\end{abstract}

\section{Introduction}
\label{sec:introduction}

Pairwise comparison data are a central supervision signal in modern preference
learning, from reinforcement learning from human feedback to direct preference
optimization and related alignment pipelines
\citep{christiano2017deep,ouyang2022training,rafailov2023direct}. In practice,
disagreement is often collapsed into a single aggregate label or treated as simple
noise. That simplification is increasingly questionable. Empirical analyses show
that disagreement in preference datasets often reflects task underspecification,
response style, or refusals rather than mere annotation error
\citep{zhang2025diverging}, while recent theory shows that hidden context and
heterogeneous user types can systematically alter the semantics of the induced
preference objective
\citep{siththaranjan2023distributional,shirali2025direct}.

This leaves an epistemic question that is orthogonal to parametric modeling.
Existing work clarifies what pairwise data recover at the aggregate level: in
particular, \citet{pukdee2026what} identify the conditional pairwise preference
distribution as the preference object encoded by pairwise comparison data,
independent of any particular Bradley--Terry parameterization. Complementary work
shows that latent heterogeneity is generally not identified from binary feedback
alone without richer elicitation \citep{chidambaram2025direct}. But a simpler
question remains open: \emph{what does observed disagreement mean, and which
semantic distinctions are identifiable from the feedback channel itself?} In
particular, can anonymous forced-choice data distinguish persistent latent
disagreement from ephemeral stochasticity, or from non-binary responses hidden by
the elicitation interface?

We answer this question with a channel-theoretic analysis of preference learning.
At the latent level, a query may admit a strict preference, a strict preference in
the opposite direction, or a non-binary response such as abstention, tie, refusal,
or incomparability. Standard pairwise pipelines, however, often expose only a
forced binary choice. For each query $q$, we define three nonparametric semantic
quantities: aggregate directional preference $\mu_q$, persistent disagreement
$\tau_q$, and non-binary mass $\alpha_q$. Under anonymous binary feedback, only
$\mu_q$ is identified: the same binary law is compatible with an explicit
two-dimensional feasible region of $(\tau_q,\alpha_q)$. Same-annotator repeats
identify $\tau_q$ via a covariance identity, abstention-enabled feedback
identifies $\alpha_q$ directly, and only their combination point-identifies the
full semantic triple.

We also derive finite-sample estimators, concentration bounds, and semantics-aware
action-recovery guarantees. Our contributions are fourfold: first, we formalize a
nonparametric latent semantic model that separates aggregate preference,
persistent disagreement, and non-binary mass; second, we prove an exact semantic
fiber theorem for anonymous binary feedback; third, we prove a precise
channel-refinement frontier showing that repeats identify $\tau_q$, abstention
identifies $\alpha_q$, and the combined refinement identifies the full semantic
triple; and fourth, we derive nonasymptotic recovery guarantees for simple
semantics-aware decision rules. Synthetic experiments constructed directly from
the proofs confirm both the fiber geometry and the downstream action-recovery
consequences.

The perspective of the paper is deliberately foundational. We do not claim that
every instance of persistent disagreement is normatively meaningful pluralism, nor
that every non-binary response should be interpreted as incomparability. Rather,
we ask what distinctions are available \emph{from the data alone}. The resulting
decision-theoretic lesson is sharp: from anonymous binary data alone, one cannot
in general determine whether disagreement should be denoised away, preserved via
personalization, or treated as a signal to query more or abstain. Richer
elicitation is therefore not merely a modeling choice; in this setting it is an
identification requirement.

\section{Model and Feedback Channels}
\label{sec:model}

We separate latent preference semantics from the observation channel. All
statements are querywise. Let $\mathcal{Q}$ be a finite set of ordered pairwise
queries. For each $q\in\mathcal Q$, a sampled annotator has a stable latent type
$Z\sim\pi$ on a standard Borel space $(\mathcal Z,\mathcal F,\pi)$. Conditional
on $(Z=z,q)$, the latent response is ternary,
\[
\widetilde Y\in\{+1,0,-1\},
\]
where $0$ denotes a protocol-agnostic non-binary response (e.g., abstention, tie,
refusal, or incomparability). We parameterize the ternary kernel by measurable
functions
\begin{align}
a_q &: \mathcal{Z} \to [0,1], \\
m_q &: \mathcal{Z} \to [-1,1], \\
|m_q(z)| &\le 1 - a_q(z),
\end{align}
and set
\[
p_q^0(z):=a_q(z),
\qquad
p_q^{\pm}(z):=\frac{1-a_q(z)\pm m_q(z)}{2}.
\]
Thus $a_q(z)=\mathbb P(\widetilde Y=0\mid Z=z,q)$ is the conditional non-binary
mass, and
\[
m_q(z)=\mathbb E[\widetilde Y\mid Z=z,q]
\]
is the conditional directional preference mean.

The querywise semantic parameters are
\begin{align}
\mu_q &:= \mathbb{E}[m_q(Z)], \\
\nu_q &:= \mathbb{E}[m_q(Z)^2], \\
\tau_q &:= \Var(m_q(Z)) = \nu_q - \mu_q^2, \\
\alpha_q &:= \mathbb{E}[a_q(Z)].
\end{align}
We interpret $\mu_q$ as the aggregate directional preference, $\tau_q$ as
persistent disagreement, and $\alpha_q$ as non-binary mass. The key point is that
$\tau_q$ is \emph{not} the variance of the observed label; it is the between-type
variance of the conditional mean $m_q(Z)$.

We study three primitive feedback channels. Under the \emph{anonymous binary
channel} $\mathsf B$, the latent zero state is forced into a strict choice by an
independent fair completion coin. Under the \emph{same-annotator repeat channel}
$\mathsf R$, the same latent type $Z$ is reused and two conditionally independent
repeated labels are elicited, each with independent completion. Under the
\emph{abstention channel} $\mathsf A$, the ternary response $\widetilde Y$ is
observed directly. Combined channels such as $\mathsf{BR}$ or $\mathsf{BRA}$
provide independent samples from each constituent primitive channel.

The fair-completion rule in channel $\mathsf B$ is deliberate. It makes forced
completion informationally neutral on latent non-binary responses and therefore
gives anonymous binary feedback its strongest possible chance to preserve
semantics. If completion were asymmetric or unknown, the binary channel would
conflate $\mu_q$ and $\alpha_q$ even more severely. The negative results below
should therefore be read as upper bounds on what anonymous forced choice can
identify.

\begin{lemma}[Observable laws]
\label{lem:observable-laws}
For every query $q\in\mathcal Q$, the primitive channels satisfy
\[
\begin{gathered}
\mathbb P_{\mathsf B}(Y=+1\mid q)=\frac{1+\mu_q}{2},
\\
\mathbb E_{\mathsf R}[Y^{(1)}Y^{(2)}\mid q]=\nu_q=\mu_q^2+\tau_q,
\end{gathered}
\]
and
\[
\begin{gathered}
\mathbb P_{\mathsf A}(\widetilde Y=0\mid q)=\alpha_q,
\\
\mathbb P_{\mathsf A}(\widetilde Y=\pm1\mid q)=\frac{1-\alpha_q\pm\mu_q}{2}.
\end{gathered}
\]
\end{lemma}

For a model $M$ and channel $\mathsf C$, let
$\operatorname{Law}^{\mathsf C}_{M,q}$ denote the induced observable law on query
$q$. We write
\[
M\equiv_{\mathsf C} M'
\]
if $\operatorname{Law}^{\mathsf C}_{M',q}=\operatorname{Law}^{\mathsf C}_{M,q}$
for every $q\in\mathcal Q$. Section~\ref{sec:binary-collapse} characterizes the
exact semantic fiber under $\mathsf B$, while
Section~\ref{sec:refinements} studies how $\mathsf R$ and $\mathsf A$ refine the
corresponding equivalence classes. The proof of
Lemma~\ref{lem:observable-laws} and the full formal setup are deferred to
Appendix~\ref{app:formal-setup}.

\section{Semantic Collapse under Anonymous Binary Feedback}
\label{sec:binary-collapse}

Under anonymous forced choice, the observable law depends on the latent model only
through $\mu_q$, consistent with the aggregate-comparison viewpoint of
\citet{pukdee2026what}. The sharper question is: for a fixed binary mean $\mu_q$,
which values of $(\tau_q,\alpha_q)$ remain compatible with the same observable
binary law? The next theorem answers this exactly.

For $\mu\in[-1,1]$, define the feasible set
\begin{equation}
\label{eq:feasible-set-main}
\mathcal F(\mu)
:=
\left\{
(\alpha,\tau)\in\mathbb R_{\ge 0}^2
:
\begin{aligned}
&\alpha \le 1-|\mu|,\\
&\tau \le 1-\alpha-\mu^2
\end{aligned}
\right\}.
\end{equation}

\begin{theorem}[Exact semantic fiber under anonymous binary feedback]
\label{thm:binary-fiber-main}
Fix a query $q\in\mathcal Q$ and a model $M$. Let
$\mathcal S_q^{\mathsf B}(M)$ denote the set of all triples
$(\mu_q(M'),\tau_q(M'),\alpha_q(M'))$ generated by models $M'$ that are
observationally equivalent to $M$ under the anonymous binary channel $\mathsf B$.
Then
\begin{equation}
\label{eq:binary-fiber-main}
\mathcal S_q^{\mathsf B}(M)
=
\left\{
\bigl(\mu_q(M),\tau,\alpha\bigr)
:
(\alpha,\tau)\in \mathcal F\bigl(\mu_q(M)\bigr)
\right\}.
\end{equation}
Equivalently, every model satisfies
\[
\bigl(\alpha_q,\tau_q\bigr)\in \mathcal F(\mu_q),
\]
and for every $\mu\in[-1,1]$ and every $(\alpha,\tau)\in\mathcal F(\mu)$ there
exists a latent semantic preference model with semantic triple $(\mu,\tau,\alpha)$
and binary law
\[
\mathbb P(Y=+1\mid q)=\frac{1+\mu}{2}.
\]
\end{theorem}

\emph{Proof sketch.} Necessity follows from the pointwise constraint
$|m_q(z)|\le 1-a_q(z)$, which implies
\begin{align}
|\mu_q| &= \bigl|\mathbb{E}[m_q(Z)]\bigr| \le 1-\alpha_q, \\
\tau_q+\mu_q^2 &= \mathbb{E}[m_q(Z)^2] \le 1-\alpha_q.
\end{align}
Hence $(\alpha_q,\tau_q)\in\mathcal F(\mu_q)$. For sharpness, fix $(\mu,\alpha)$.
A homogeneous one-type model realizes $\tau=0$, while a three-type extremal model
with $m\in\{+1,0,-1\}$ and masses
\[
\frac{1-\alpha+\mu}{2},\qquad \alpha,\qquad \frac{1-\alpha-\mu}{2}
\]
realizes $\tau=1-\alpha-\mu^2$. Mixing these two constructions interpolates every
intermediate value of $\tau$. The full proof, including embedding into an ambient
multi-query model, is given in Appendix~\ref{app:proof-theorem-3-1}.

Theorem~\ref{thm:binary-fiber-main} says that anonymous binary feedback collapses a
two-dimensional semantic region to the single scalar $\mu_q$. In particular, the
same fair-coin binary law at $\mu_q=0$ is compatible with
\emph{(i)} a homogeneous noisy world $(\tau_q,\alpha_q)=(0,0)$,
\emph{(ii)} a purely persistent split world $(\tau_q,\alpha_q)=(1,0)$, and
\emph{(iii)} a pure non-binary-completion world $(\tau_q,\alpha_q)=(0,1)$.
Binary marginals alone therefore do not reveal what disagreement \emph{means}.
This sharpened failure mode is distinct from ordinary model misspecification: the
observable binary law is fully pinned down by $\mu_q$, but the latent semantic
preimage beneath that law remains two-dimensional. Aggregate recovery of the
pairwise comparison law therefore does not by itself resolve whether disagreement
is persistent, non-binary, or both.

\begin{corollary}[Identification under the anonymous binary channel]
\label{cor:binary-identification-main}
Under $\mathsf B$, the aggregate directional preference $\mu_q$ is
point-identified. If $|\mu_q|<1$, neither persistent disagreement $\tau_q$ nor
non-binary mass $\alpha_q$ is point-identified.
\end{corollary}

\begin{figure*}[t]
\centering
\includegraphics[width=0.94\textwidth]{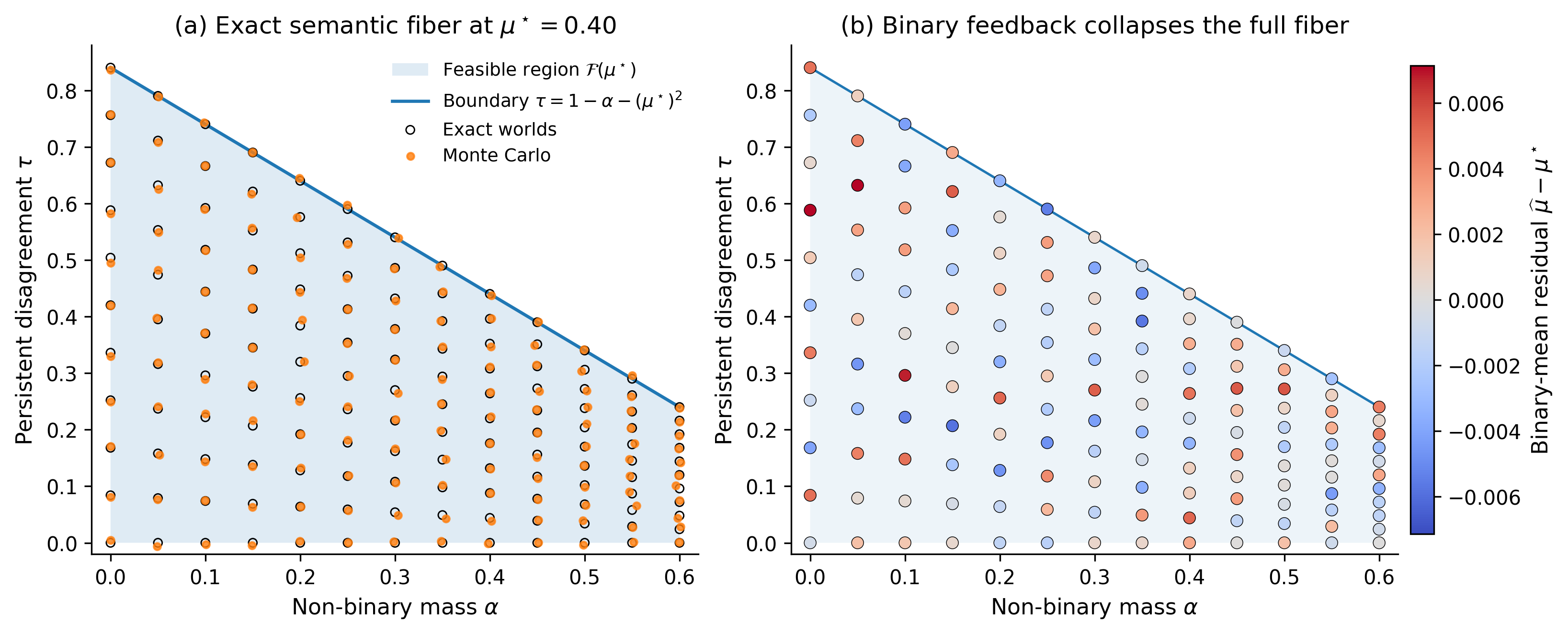}
\caption{\textbf{Exact semantic fiber under anonymous binary feedback.}
For $\mu^\star=0.40$, panel (a) shows the feasible set
$\mathcal F(\mu^\star)$ and the constructive worlds from
Theorem~\ref{thm:binary-fiber-main}; panel (b) colors the same worlds by the
residual $\widehat\mu-\mu^\star$. Across the full feasible region, the induced
anonymous binary law is essentially unchanged.}
\label{fig:semantic-fiber-main}
\end{figure*}

\begin{corollary}[Binary-only testing impossibility]
\label{cor:binary-testing-impossibility-main}
Let $M_0$ and $M_1$ be two models such that
$M_0 \equiv_{\mathsf B} M_1$. Then for any test $T$ based only on anonymous
binary observations,
\[
\max_{j\in\{0,1\}} \mathbb P_{M_j}\!\bigl(T\neq j\bigr)\ge \frac12.
\]
Consequently, if the correct downstream action differs between $M_0$ and $M_1$,
no binary-only procedure can reliably decide between those actions.
\end{corollary}

This is the operational form of semantic collapse: if observationally equivalent
worlds prescribe different responses to disagreement---for example,
\emph{denoise}, \emph{personalize}, or \emph{query/abstain}---anonymous binary
data alone cannot determine which response is warranted.
Section~\ref{sec:refinements} therefore asks for the minimal refinements that
break this ambiguity.

\section{Minimal Channel Refinements}
\label{sec:refinements}

Which minimal refinement recovers the missing dimensions? In our model,
same-annotator repeats identify $\tau_q$, abstention identifies $\alpha_q$, and
the two refinements are complementary
\citep{siththaranjan2023distributional,shirali2025direct,chidambaram2025direct}.
We state the first result for the combined channel $\mathsf{BR}$, matching the
intended design in which anonymous binary labels are the default supervision
source and repeat data are collected on a smaller calibration subset.

\begin{theorem}[Repeat covariance identifies persistent disagreement]
\label{thm:repeat-cov-main}
Under the combined channel $\mathsf{BR}$,
\[
\begin{aligned}
\tau_q &= \Cov\!\bigl(Y^{(1)},Y^{(2)} \mid q\bigr)\\
&= \mathbb E\!\left[Y^{(1)}Y^{(2)}\mid q\right]
-\Bigl(\mathbb E[Y\mid q]\Bigr)^2.
\end{aligned}
\]
Consequently, $\tau_q$ is point-identified under $\mathsf{BR}$.
\end{theorem}

\emph{Proof sketch.} Conditional on the latent type $Z$, repeated responses are
independent and each has conditional mean $m_q(Z)$. Hence
\[
\mathbb E\!\left[Y^{(1)}Y^{(2)}\mid Z,q\right]=m_q(Z)^2.
\]
Taking expectations gives
\[
\mathbb E\!\left[Y^{(1)}Y^{(2)}\mid q\right]=\nu_q,
\qquad
\mathbb E[Y\mid q]=\mu_q,
\]
and subtraction yields $\tau_q=\nu_q-\mu_q^2$. The full proof is given in
Appendix~\ref{app:proofs-section-4}.

\begin{corollary}[Same-annotator versus different-annotator disagreement gap]
\label{cor:disagreement-gap-main}
Let $D_q^{\mathrm{diff}}$ denote the disagreement probability between two
independent anonymous binary labels on query $q$, and let
$D_q^{\mathrm{same}}$ denote the disagreement probability between two repeated
labels from the same annotator. Then
\[
D_q^{\mathrm{diff}}-D_q^{\mathrm{same}}=\frac{\tau_q}{2}.
\]
\end{corollary}

Thus the gap between inter-annotator disagreement and intra-annotator
disagreement is exactly the persistent component.

\begin{theorem}[Abstention identifies non-binary mass]
\label{thm:abstention-main}
Under the abstention channel $\mathsf A$,
\begin{align}
\alpha_q &= \mathbb{P}(\widetilde{Y}=0\mid q), \\
\mu_q &= \mathbb{P}(\widetilde{Y}=+1\mid q) - \mathbb{P}(\widetilde{Y}=-1\mid q).
\end{align}
Consequently, the pair $(\mu_q,\alpha_q)$ is point-identified under
$\mathsf A$.
\end{theorem}

The proof is immediate from Lemma~\ref{lem:observable-laws}: the abstention
channel reveals the full trinomial law, and hence both the aggregate directional
preference and the non-binary mass.

\begin{theorem}[Channel-refinement frontier]
\label{thm:frontier-main}
For any two models $M,M'$,
\begin{align*}
M \equiv_{\mathsf B} M'
&\iff
\mu(M)=\mu(M'),\\
M \equiv_{\mathsf{BR}} M'
&\iff
(\mu,\tau)(M)=(\mu,\tau)(M'),\\
M \equiv_{\mathsf A} M'
&\iff
(\mu,\alpha)(M)=(\mu,\alpha)(M'),\\
M \equiv_{\mathsf{BRA}} M'
&\iff
(\mu,\tau,\alpha)(M)=(\mu,\tau,\alpha)(M').
\end{align*}
Moreover,
\[
\equiv_{\mathsf{BRA}}
\subsetneq
\equiv_{\mathsf{BR}}
\subsetneq
\equiv_{\mathsf B},
\qquad
\equiv_{\mathsf{BRA}}
\subsetneq
\equiv_{\mathsf A}
\subsetneq
\equiv_{\mathsf B}.
\]
\end{theorem}

\emph{Proof sketch.} Under $\mathsf B$, the observable law depends only on $\mu_q$
by Lemma~\ref{lem:observable-laws}. Under $\mathsf{BR}$, the repeat law determines
$(\mu_q,\nu_q)$, hence $(\mu_q,\tau_q)$ since $\tau_q=\nu_q-\mu_q^2$. Under
$\mathsf A$, the trinomial law determines $(\mu_q,\alpha_q)$. Under
$\mathsf{BRA}$, factorization of independent channel outputs combines these
characterizations. Strictness follows from one-query witness models with matched
binary means but different semantics; see Appendix~\ref{app:proofs-section-4}.

Thus repeats recover the between-type dimension $\tau_q$, abstention recovers
$\alpha_q$, and only the combined refinement $\mathsf{BRA}$ point-identifies the
full semantic triple.

\section{Finite-Sample Recovery and Testing}
\label{sec:finite-sample}

Assume the channel-separated product experiment of
Appendix~\ref{app:formal-setup}: for each query $q\in\mathcal Q$, we observe
$n_{B,q}$ anonymous binary labels, $n_{R,q}$ same-annotator repeat pairs, and
$n_{A,q}$ abstention-enabled labels, with the three channel blocks independent.
Define
\begin{align*}
\widehat\mu_q
&:=
\frac{1}{n_{B,q}}\sum_{i=1}^{n_{B,q}} Y^B_{q,i},
\\
\widehat\nu_q
&:=
\frac{1}{n_{R,q}}\sum_{i=1}^{n_{R,q}}
Y^{R,(1)}_{q,i}Y^{R,(2)}_{q,i}, \\
\widehat\alpha_q
&:=
\frac{1}{n_{A,q}}\sum_{i=1}^{n_{A,q}}
\mathbf 1\{\widetilde Y^A_{q,i}=0\},
\\
\widehat\tau_q
&:=
\widehat\nu_q-\widehat\mu_q^2.
\end{align*}
Here $\widehat\mu_q$ estimates the binary mean, $\widehat\nu_q$ the repeat moment,
$\widehat\alpha_q$ the non-binary mass, and $\widehat\tau_q$ the persistent
disagreement.

\begin{proposition}[Exact finite-sample moments]
\label{prop:finite-sample-moments-main}
For every query $q\in\mathcal Q$,
\[
\mathbb E[\widehat\mu_q]=\mu_q,
\qquad
\mathbb E[\widehat\nu_q]=\nu_q,
\qquad
\mathbb E[\widehat\alpha_q]=\alpha_q,
\]
and
\[
\mathbb E[\widehat\tau_q]
=
\tau_q-\frac{1-\mu_q^2}{n_{B,q}}.
\]
Hence the plug-in estimator is downward biased by exactly
$(1-\mu_q^2)/n_{B,q}$. A bias-corrected unbiased version is given in
Appendix~\ref{app:proofs-finite-sample}.
\end{proposition}

Since all channel outputs are bounded, concentration follows directly from
Hoeffding's inequality \citep{hoeffding1963probability}; see
Appendix~\ref{app:proofs-finite-sample}.

\begin{theorem}[Querywise concentration]
\label{thm:finite-sample-main}
For every query $q\in\mathcal Q$ and every $t,\varepsilon>0$,
\begin{align}
\mathbb{P}\!\left(|\widehat{\mu}_q-\mu_q|>t\right)
&\le
2\exp\!\left(-\frac{n_{B,q}t^2}{2}\right), \\
\mathbb{P}\!\left(|\widehat{\alpha}_q-\alpha_q|>t\right)
&\le
2\exp\!\left(-2n_{A,q}t^2\right),
\end{align}
and
\[
\begin{aligned}
\mathbb P\!\left(|\widehat\tau_q-\tau_q|>\varepsilon\right)
&\le 2\exp\!\left(-\frac{n_{R,q}\varepsilon^2}{8}\right)\\
&\quad +2\exp\!\left(-\frac{n_{B,q}\varepsilon^2}{32}\right).
\end{aligned}
\]
Since $\mathcal Q$ is finite, the corresponding uniform-over-$\mathcal Q$ bounds
follow by a union bound.
\end{theorem}

Fix thresholds $t_\tau,t_\alpha$, and define the plug-in action
\[
\widehat a_q=
\begin{cases}
\mathsf{query}, & \widehat\alpha_q>t_\alpha,\\
\mathsf{personalize}, & \widehat\alpha_q\le t_\alpha \text{ and } \widehat\tau_q>t_\tau,\\
\mathsf{denoise}, & \widehat\alpha_q\le t_\alpha \text{ and } \widehat\tau_q\le t_\tau.
\end{cases}
\]

\begin{corollary}[Exact recovery of semantic actions]
\label{cor:finite-sample-actions-main}
Let
\[
\Gamma
:=
\min_{q\in\mathcal Q}
\min\!\left\{
|\tau_q-t_\tau|,
|\alpha_q-t_\alpha|
\right\},
\]
and suppose $\Gamma>0$. Under balanced allocations
$n_{B,q}\equiv n_B$, $n_{R,q}\equiv n_R$, and $n_{A,q}\equiv n_A$, the event
$\widehat a_q=a_q^\star$ for all $q\in\mathcal Q$ holds with probability at least
$1-\delta$ whenever
\begin{align}
n_R &\ge \frac{32}{\Gamma^2}\log\!\left(\frac{6|\mathcal{Q}|}{\delta}\right), \\
n_B &\ge \frac{128}{\Gamma^2}\log\!\left(\frac{6|\mathcal{Q}|}{\delta}\right), \\
n_A &\ge \frac{2}{\Gamma^2}\log\!\left(\frac{6|\mathcal{Q}|}{\delta}\right).
\end{align}
\end{corollary}

Once the channel is minimally refined, the semantic distinction between denoising,
personalization, and abstention becomes not only identified but estimable at the
standard bounded-observation rate
$O(\Gamma^{-2}\log(|\mathcal Q|/\delta))$. The asymmetry between the constants is
itself interpretable: $\alpha_q$ is directly observable under abstention, whereas
$\tau_q$ is recovered through a covariance identity that depends on both repeat
and binary samples.

\section{Synthetic Validation}
\label{sec:synthetic-validation}

Our empirical goal is theorem-matched validation rather than benchmark
optimization. Because the main claims are about channel-level identifiability and
finite-sample recovery, all experiments are synthetic and are generated directly
from the constructive proofs in Appendices~\ref{app:proof-theorem-3-1},
\ref{app:proofs-section-4}, and \ref{app:proofs-finite-sample}.
Figure~\ref{fig:semantic-fiber-main} already validates the exact semantic fiber
and the collapse of anonymous binary feedback. The appendix further validates the
strict channel frontier via explicit witness worlds and the separate finite-sample
recovery of $\tau_q$ and $\alpha_q$. In the main text we emphasize the strongest
downstream consequence: whether the refined channels suffice to recover the
\emph{correct epistemic action}.

\begin{figure*}[!t]
\centering
\includegraphics[width=0.94\textwidth]{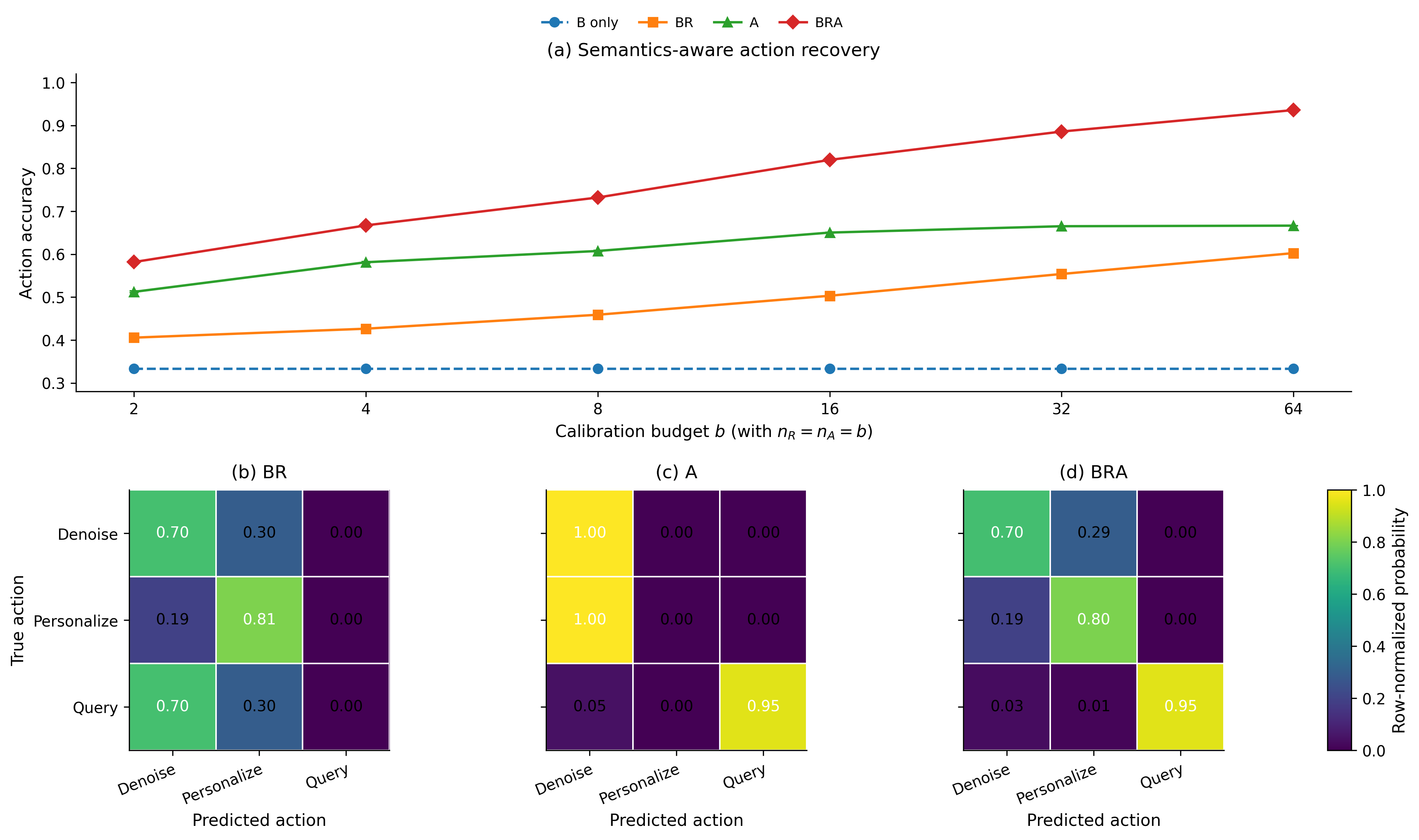}
\caption{\textbf{Semantics-aware action recovery under channel refinement.}
The three regimes use thresholds $t_\tau=t_\alpha=0.15$:
\textsc{Denoise} $(0.02,0.02)$,
\textsc{Personalize} $(0.35,0.02)$, and
\textsc{Query} $(0.02,0.35)$.
All three regimes are constructed to share the same empirical binary-law
distribution by reusing the same set of $\mu$-values across regimes. Panel (a)
plots action accuracy versus calibration budget $b$, with $n_R=n_A=b$ and
$n_B=512$. Panels (b)--(d) show row-normalized confusion matrices at
representative budget $b=16$ for $\mathsf{BR}$, $\mathsf A$, and
$\mathsf{BRA}$. Repeats and abstention recover complementary action dimensions,
while the combined channel recovers the full semantics-aware policy.}
\label{fig:action-recovery-main}
\end{figure*}

For Figure~\ref{fig:action-recovery-main}, we define three semantic regimes
around the thresholds $t_\tau=t_\alpha=0.15$: \textsc{Denoise} with
$(\tau,\alpha)=(0.02,0.02)$, \textsc{Personalize} with $(0.35,0.02)$, and
\textsc{Query} with $(0.02,0.35)$. To make binary underidentification sharp, the
same set of querywise $\mu$-values is reused across all three regimes, so the
anonymous binary laws are matched while the semantics differ. With $n_B=512$ and
calibration budgets $n_R=n_A=b\in\{2,4,8,16,32,64\}$, the resulting curves show
exactly the predicted complementarity: a $\mathsf B$-only rule remains near
chance, $\mathsf{BR}$ identifies the high-$\tau$ regime but confuses
\textsc{Query}, and $\mathsf A$ identifies the high-$\alpha$ regime but collapses
\textsc{Denoise} and \textsc{Personalize}. Only the combined channel
$\mathsf{BRA}$ reliably recovers all three actions, matching both the
channel-refinement frontier of Theorem~\ref{thm:frontier-main} and the
finite-sample regime-selection guarantee of
Corollary~\ref{cor:finite-sample-actions-main}.

These experiments are not intended as benchmark claims. Their role is to
instantiate the exact witness constructions behind the proofs and to show that the
downstream action boundary inherits the same channel-level identifiability
structure as the population theory.

\section{Discussion}
\label{sec:discussion}

Our results separate two questions that are often conflated in pairwise
preference learning: what aggregate comparison law is encoded by the data, and
what semantic source of disagreement generated that law. In the anonymous binary
setting, the data identify only the aggregate directional preference $\mu_q$, but
not whether disagreement arises from persistent latent variation, non-binary
responses hidden by forced choice, or both. Theorems~\ref{thm:binary-fiber-main},
\ref{thm:repeat-cov-main}, \ref{thm:abstention-main}, and
\ref{thm:frontier-main} show that same-annotator repeats and abstention are
minimal complementary refinements: the former identifies $\tau_q$, the latter
identifies $\alpha_q$, and only their combination point-identifies the full
semantic triple. The consequence is operational rather than merely philosophical:
semantics-aware choices such as denoising, personalization, and abstention cannot
be justified from anonymous binary data alone.

Our scope is deliberately channel-theoretic. The parameter $\alpha_q$ is defined
as total non-binary mass, so interpreting it specifically as tie, refusal, or
incomparability requires richer output alphabets or additional structural
assumptions; similarly, $\tau_q$ isolates stable between-type variation but does
not by itself determine whether that variation is normatively meaningful
pluralism or nuisance context. These are limitations of the available
information rather than of a particular estimator. The practical implication is
therefore methodological: if downstream systems are expected to distinguish noise
from persistent disagreement, or to abstain on semantically underspecified
comparisons, then richer elicitation protocols are not optional regularizers but
identification requirements. A natural next step is to study which minimal
refinements separate distinct forms of non-binary mass while preserving the same
nonparametric identification viewpoint.

\bibliography{example_paper}
\bibliographystyle{icml2026}

\newpage
\appendix
\onecolumn

\section{Positioning and Novelty Map}
\label{app:positioning}

This appendix clarifies the precise gap filled by the paper relative to the
closest theory and empirical literature. The central novelty is \emph{not} a new
parametric preference model, a new reward objective, or a new alignment
algorithm. Rather, the paper gives a \emph{channel-theoretic} answer to a prior
question: before choosing an aggregation objective or fitting a preference model,
what semantic distinctions about disagreement are identifiable from the feedback
channel itself? Our main theorems show that anonymous binary data identify only
the aggregate directional preference $\mu_q$, whereas same-annotator repeats and
abstention are minimal complementary refinements that identify the missing
semantic quantities $\tau_q$ and $\alpha_q$.

\begin{table}[t]
\centering
\small
\caption{Novelty map relative to the closest lines of work.}
\label{tab:novelty-map}
\renewcommand{\arraystretch}{1.15}
\begin{tabular}{|p{0.20\linewidth}|p{0.20\linewidth}|p{0.23\linewidth}|p{0.27\linewidth}|}
\hline
\textbf{Line of work} &
\textbf{Representative references} &
\textbf{Primary object of analysis} &
\textbf{How this paper differs} \\
\hline

Aggregate recovery from pairwise data &
\citet{pukdee2026what} &
The conditional pairwise preference distribution (CPRD), i.e., the aggregate
preference object encoded by pairwise data &
We study the \emph{latent semantic preimage} of a fixed binary law. Theorem~\ref{thm:binary-fiber-main}
characterizes the exact fiber over $(\mu_q)$, which is a different question from
recovering or representing the aggregate law itself. \\
\hline

Hidden context in RLHF / distributional preference learning &
\citet{siththaranjan2023distributional} &
How hidden context induces aggregation effects (e.g.\ Borda-type aggregation) and
how utility distributions can account for it &
We abstract away from any particular score-based learning rule and ask what the
feedback channel identifies about disagreement semantics \emph{before} one
chooses an aggregation mechanism. \\
\hline

Heterogeneous alignment with user types &
\citet{shirali2025direct} &
Alignment when user types are heterogeneous and extra annotator information may be
available &
We do not assume a target average-reward objective or a user-type model. Our
results are upstream: they characterize when disagreement semantics are
identifiable from data at all. \\
\hline

Richer feedback for latent heterogeneity &
\citet{chidambaram2025direct} &
Identifiability of latent preference heterogeneity under binary versus ternary or
ranking-based feedback &
We study a more primitive question: whether observed disagreement reflects
persistent latent variation, non-binary mass hidden by forced choice, or both.
Our refinements are repeats and abstention rather than rankings. \\
\hline

Empirical taxonomies of disagreement &
\citet{zhang2025diverging} &
Empirical causes of disagreement in real preference datasets &
These works motivate our latent decomposition, but do not provide exact
nonparametric identifiability regions or channel-refinement theorems. \\
\hline
\end{tabular}
\end{table}

\paragraph{Aggregate preference recovery versus semantic identifiability.}
The closest theoretical paper is \citet{pukdee2026what}, which formalizes the
conditional pairwise preference distribution (CPRD) as the aggregate preference
object encoded by pairwise comparison data. Their contribution is to clarify what
pairwise data recover at the \emph{preference-law} level. Our contribution begins
where that analysis stops. Fix the anonymous binary law for a single query, or
equivalently fix $\mu_q$. We ask: which latent semantic worlds remain compatible
with that law? The answer is Theorem~\ref{thm:binary-fiber-main}, which gives the
exact fiber
\[
\mathcal S_q^{\mathsf B}(M)
=
\left\{
(\mu_q(M),\tau,\alpha):(\alpha,\tau)\in\mathcal F(\mu_q(M))
\right\}.
\]
Thus the object of the paper is not the aggregate law itself, but the exact
semantic multiplicity \emph{beneath} that law. In particular, the paper proves a
result that is orthogonal to CPRD recovery: anonymous binary data may determine
the aggregate directional preference while leaving a full two-dimensional region
of disagreement semantics unresolved.

\paragraph{Hidden context versus channel-level identifiability.}
The hidden-context literature, especially
\citet{siththaranjan2023distributional}, asks how unobserved context alters the
behavior of preference learning procedures and can induce implicit social-choice
aggregation. Our framework is compatible with that viewpoint---indeed, one can
interpret the latent type $Z$ as hidden context---but the question is different.
Those works analyze what a learning rule does \emph{given hidden context}. We ask
what the \emph{observation channel} alone reveals about the semantic source of
disagreement. For this reason, our results do not depend on Bradley--Terry,
Plackett--Luce, utility representability, or any particular scoring model. The
identification frontier in Theorem~\ref{thm:frontier-main} is therefore a
channel-level statement, not an algorithm-level one.

\paragraph{Heterogeneous alignment versus semantic-source identification.}
Work on heterogeneous alignment, such as \citet{shirali2025direct}, starts from a
population of user types and studies how one should align a single policy when
preferences differ across types. That line of work is objective-first: it asks
which social or average objective is learnable under which side-information
assumptions. Our paper is prior to that choice. We do not assume a user-type
space that is finite, discrete, or even normatively meaningful; we only assume a
stable latent source of variation. The quantity
\[
\tau_q=\Var(m_q(Z))
\]
therefore measures persistent latent disagreement without committing to whether it
should ultimately be interpreted as pluralistic value disagreement or nuisance
context. In this sense, our paper identifies a precondition for heterogeneous
alignment: before optimizing a heterogeneous objective, one must first know
whether the observed disagreement is evidence of stable heterogeneity at all.

\paragraph{Richer feedback for latent heterogeneity.}
The closest paper in spirit on richer elicitation is
\citet{chidambaram2025direct}, which shows that binary comparisons are
insufficient to identify latent heterogeneous preferences in their setting, while
richer feedback such as ternary or ranking-based responses can restore
identifiability. That result is complementary to ours, not redundant with it.
Their question is: \emph{within a heterogeneous-preference model, can one identify
the latent structure?} Our question is earlier: \emph{does the data tell us that
the disagreement is persistent latent heterogeneity rather than hidden non-binary
mass or binary stochasticity?} Accordingly, the two papers use different
refinements for different goals. We use same-annotator repeats and abstention to
separate semantic sources of disagreement; they use richer rankings to identify
latent heterogeneity inside a heterogeneous-preference family. One can therefore
view the two results as compositional: our results identify when a persistent
component is present, while theirs can then be used to identify structure within
that component.

\paragraph{Empirical disagreement taxonomies versus theoremic identification.}
Empirical work such as \citet{zhang2025diverging} documents that disagreement in
real preference data often arises from underspecification, response style, or
refusal rather than simple noise. This empirical observation is one of the main
motivations for our semantic decomposition. But the role of our paper is
different. We do not infer a taxonomy from real data; instead, we prove which
distinctions are identifiable from each feedback channel. In particular, the
paper gives exact characterization results---Theorem~\ref{thm:binary-fiber-main}
for anonymous binary feedback and Theorem~\ref{thm:frontier-main} for channel
refinement---that are not provided by empirical disagreement audits alone.

\paragraph{What is new, precisely.}
To the best of our knowledge, the paper is the first to provide all three of the
following within a single nonparametric framework:
\begin{enumerate}
    \item an \emph{exact querywise semantic fiber} over the anonymous binary law,
    not merely a statement that identifiability fails;
    \item an \emph{exact channel-refinement frontier}
    \[
    \equiv_{\mathsf{BRA}}
    \subsetneq
    \equiv_{\mathsf{BR}}
    \subsetneq
    \equiv_{\mathsf B},
    \qquad
    \equiv_{\mathsf{BRA}}
    \subsetneq
    \equiv_{\mathsf A}
    \subsetneq
    \equiv_{\mathsf B},
    \]
    showing which minimal refinements recover which semantic coordinates; and
    \item \emph{finite-sample and decision-theoretic consequences} of that
    frontier, namely nonasymptotic recovery guarantees for
    $(\widehat\tau_q,\widehat\alpha_q)$ and a binary-only impossibility result for
    semantics-aware actions.
\end{enumerate}
This is the sense in which the contribution is foundational rather than
incremental. The paper does not add another model to the existing preference
learning pipeline; it identifies which semantic distinctions that pipeline can and
cannot support.

\paragraph{Non-claims.}
The paper does not claim that every persistent component of disagreement is
normatively meaningful pluralism, nor that every non-binary response is genuine
incomparability. Those stronger semantic interpretations require richer output
alphabets or additional structural assumptions, which is precisely why
Appendix~\ref{app:extensions} studies tie/incomparability separation and repeated
abstention. Our claim is narrower and more robust: under the standard anonymous
binary channel, these distinctions are not identifiable from the data alone, and
therefore cannot be justified merely by improving the estimator.
\section{Full Formal Setup}
\label{app:formal-setup}

This appendix specifies the statistical experiment underlying the main theorems.
Pairwise comparison data are the dominant supervision signal in preference learning,
RLHF, and direct alignment pipelines
\citep{christiano2017deep,ouyang2022training,rafailov2023direct}.
Recent work clarifies that such data identify an aggregate comparison law---formalized
as the conditional pairwise preference distribution (CPRD)---even when common
parametric models are misspecified \citep{pukdee2026what}. Complementary work shows
that hidden context and heterogeneous annotators can induce nontrivial aggregation
effects and identifiability failures
\citep{siththaranjan2023distributional,shirali2025direct,chidambaram2025direct},
while empirical analyses find that disagreement often reflects underspecification,
response style, or refusals rather than simple annotation error
\citep{zhang2025diverging}. We therefore adopt a deliberately nonparametric
latent-type model whose purpose is to isolate what is identified by the
\emph{feedback channel} itself.

\begin{table}[t]
\centering
\small
\caption{Core notation for the formal setup.}
\label{tab:formal-setup-notation}
\begin{tabular}{lp{0.73\linewidth}}
\toprule
Symbol & Meaning \\
\midrule
$\mathcal{Q}$ & Finite set of ordered pairwise comparison queries. \\
$q \in \mathcal{Q}$ & A fixed query; labels $\pm 1$ are interpreted relative to the ordering encoded by $q$. \\
$(\mathcal{Z},\mathcal{F},\pi)$ & Latent type space and population distribution over stable hidden types. \\
$Z \sim \pi$ & Latent type of a sampled annotator. \\
$\widetilde Y \in \{+1,0,-1\}$ & Latent ternary response: strict preference for the first alternative, non-binary response, or strict preference for the second alternative. \\
$Y \in \{+1,-1\}$ & Observed binary response after forced completion of $\widetilde Y=0$. \\
$a_q(z)$ & Conditional non-binary mass at query $q$ for latent type $z$. \\
$m_q(z)$ & Conditional directional preference mean at query $q$ for latent type $z$. \\
$\mu_q$ & Aggregate directional preference, $\mu_q = \mathbb E[m_q(Z)]$. \\
$\nu_q$ & Second moment of the directional mean, $\nu_q = \mathbb E[m_q(Z)^2]$. \\
$\tau_q$ & Persistent disagreement, $\tau_q = \mathrm{Var}(m_q(Z)) = \nu_q-\mu_q^2$. \\
$\alpha_q$ & Aggregate non-binary mass, $\alpha_q = \mathbb E[a_q(Z)]$. \\
$\mathsf{B}$ & Anonymous binary channel. \\
$\mathsf{R}$ & Same-annotator repeat channel. \\
$\mathsf{A}$ & Abstention-enabled ternary channel. \\
$\operatorname{Law}^{\mathsf C}_{M,q}$ & Observable law induced by model $M$ on query $q$ under channel $\mathsf C$. \\
$M \equiv_{\mathsf C} M'$ & Observational equivalence of models $M$ and $M'$ under channel $\mathsf C$. \\
$\mathcal I^{\mathsf C}_q(\theta;M)$ & Identification region of parameter $\theta_q$ at model $M$ under channel $\mathsf C$. \\
$\mathcal S^{\mathsf C}_q(M)$ & Semantic fiber $\mathcal I^{\mathsf C}_q((\mu_q,\tau_q,\alpha_q);M)$. \\
\bottomrule
\end{tabular}
\end{table}

\subsection{Model class}
\label{app:formal-setup:model-class}

We work querywise over a finite set $\mathcal{Q}$ of \emph{ordered} pairwise
comparison queries. In the intended RLHF/alignment interpretation, each
$q \in \mathcal{Q}$ is a prompt paired with two candidate responses, as in
standard pairwise-feedback pipelines
\citep{christiano2017deep,ouyang2022training,rafailov2023direct}. The order of
the two alternatives is part of the query representation itself, so the sign of a
preference label is always interpreted relative to a fixed orientation.

The latent type space is modeled as a standard Borel probability space
$(\mathcal{Z},\mathcal{F},\pi)$. A draw $Z \sim \pi$ should be interpreted
broadly as a stable hidden source of semantic variation: enduring values,
evaluation rubric, task interpretation, stylistic disposition, or any other
persistent hidden context in the sense discussed by
\citet{siththaranjan2023distributional}. The shared type space allows the same
latent type to induce arbitrarily correlated behavior across different queries,
but we impose no structural restrictions on that dependence, and none of the
identification arguments rely on such structure.

\begin{definition}[Latent semantic preference model]
\label{def:appendix-latent-model}
Fix a finite query set $\mathcal{Q}$. A \emph{latent semantic preference model} is
a tuple
\[
M=\bigl((\mathcal{Z},\mathcal{F},\pi),\{(a_q,m_q)\}_{q\in\mathcal{Q}}\bigr),
\]
where, for each $q \in \mathcal{Q}$,
\[
a_q : \mathcal{Z}\to[0,1],
\qquad
m_q : \mathcal{Z}\to[-1,1]
\]
are measurable and satisfy
\[
|m_q(z)| \le 1-a_q(z)
\qquad
\text{for every } z\in\mathcal{Z}.
\]
Equivalently, for each $(q,z)$ the pair $(a_q(z),m_q(z))$ defines a probability
distribution on $\{+1,0,-1\}$ via
\begin{align*}
p_q^0(z) &:= a_q(z), \\
p_q^+(z) &:= \frac{1-a_q(z)+m_q(z)}{2}, \\
p_q^-(z) &:= \frac{1-a_q(z)-m_q(z)}{2}.
\end{align*}
The nonnegativity of $p_q^+,p_q^0,p_q^-$ and the identity
$p_q^+(z)+p_q^0(z)+p_q^-(z)=1$ follow exactly from the constraint
$|m_q(z)| \le 1-a_q(z)$.

Conditional on latent type $Z=z$ and query $q$, the latent response
$\widetilde Y \in \{+1,0,-1\}$ is distributed as
\[
\mathbb P(\widetilde Y=+1 \mid Z=z,q)=p_q^+(z),\quad
\mathbb P(\widetilde Y=0 \mid Z=z,q)=p_q^0(z),\quad
\mathbb P(\widetilde Y=-1 \mid Z=z,q)=p_q^-(z).
\]
Thus
\[
a_q(z)=\mathbb P(\widetilde Y=0\mid Z=z,q),
\qquad
m_q(z)=\mathbb E[\widetilde Y\mid Z=z,q].
\]
\end{definition}

\begin{remark}[Why \((a_q,m_q)\) are the primitive objects]
\label{rem:appendix-primitive-objects}
The pair $(a_q,m_q)$ is the minimal parameterization of the ternary kernel needed
for the theorems in the paper. The quantity $a_q(z)$ records how often type $z$
does not express a strict comparison at query $q$, while $m_q(z)$ records the
centered directional tendency among the two strict outcomes. This parameterization
is intentionally agnostic to any reward-based, Bradley--Terry, Plackett--Luce, or
random-utility representation. That nonparametric choice is deliberate: the object
of study is channel-level identifiability, not the recovery properties of a
particular parametric score model. This viewpoint is aligned with the CPRD-based,
data-centric perspective of \citet{pukdee2026what}.
\end{remark}

\begin{remark}[Interpretation of the zero state]
\label{rem:appendix-zero-state}
The symbol $0$ is intentionally protocol-agnostic. Depending on the elicitation
interface, it may represent abstention, tie, refusal, indecision, or genuine
incomparability. Recent empirical work shows that disagreement in preference data
often arises from underspecification, response style, or refusals rather than
simple annotation error \citep{zhang2025diverging}. The main body of the paper
therefore treats $0$ as \emph{non-binary response mass}. This is the strongest
coarse-graining for the binary channel: any finer semantic subdivision of the zero
state can only make anonymous binary observations less informative. Appendix~\ref{app:extensions}
returns to this issue and explicitly separates tie mass from incomparability mass.
\end{remark}

\begin{definition}[Semantic functionals]
\label{def:appendix-semantic-functionals}
For a model $M$ and query $q\in\mathcal{Q}$, define
\[
\mu_q(M):=\mathbb E_{\pi}[m_q(Z)],
\qquad
\nu_q(M):=\mathbb E_{\pi}[m_q(Z)^2],
\]
\[
\tau_q(M):=\mathrm{Var}_{\pi}(m_q(Z))=\nu_q(M)-\mu_q(M)^2,
\qquad
\alpha_q(M):=\mathbb E_{\pi}[a_q(Z)].
\]
When the ambient model $M$ is clear from context, we write simply
$\mu_q,\nu_q,\tau_q,\alpha_q$.

We interpret
\begin{itemize}
    \item $\mu_q$ as the \emph{aggregate directional preference},
    \item $\tau_q$ as the \emph{persistent disagreement},
    \item $\alpha_q$ as the \emph{non-binary mass}.
\end{itemize}
We also use the vector notation
\[
\mu(M):=(\mu_q(M))_{q\in\mathcal{Q}},\quad
\nu(M):=(\nu_q(M))_{q\in\mathcal{Q}},\quad
\tau(M):=(\tau_q(M))_{q\in\mathcal{Q}},\quad
\alpha(M):=(\alpha_q(M))_{q\in\mathcal{Q}}.
\]
\end{definition}

\begin{remark}[Meaning of persistent disagreement]
\label{rem:appendix-persistent-disagreement}
The quantity $\tau_q=\mathrm{Var}(m_q(Z))$ is the between-type variance of the
conditional directional mean. It is \emph{not} the variance of the observed label.
Hence $\tau_q$ isolates stable latent variation across annotators and excludes
within-type stochasticity. In particular, $\tau_q$ may reflect enduring value
pluralism, but it may also reflect stable nuisance factors such as different
evaluation rubrics or hidden task interpretations. The theorems identify
persistent latent variation as a statistical object; they do not by themselves
resolve its normative interpretation.
\end{remark}

\subsection{Primitive observation channels}
\label{app:formal-setup:channels}

We now formalize the observation channels. The essential distinction is between
the latent semantic response $\widetilde Y$ and the actually observed feedback.
Anonymous forced-choice interfaces convert the ternary latent response into a
binary observation; repeated elicitation reuses the same latent type; abstention
interfaces expose the zero state directly.

\begin{definition}[Primitive channels]
\label{def:appendix-primitive-channels}
Fix a model $M$ and a query $q\in\mathcal{Q}$.

\paragraph{Anonymous binary channel \(\mathsf B\).}
Draw $Z\sim\pi$, then draw $\widetilde Y\in\{+1,0,-1\}$ according to
Definition~\ref{def:appendix-latent-model}. If $\widetilde Y\in\{+1,-1\}$,
output $Y=\widetilde Y$. If $\widetilde Y=0$, draw an independent completion coin
$C\in\{+1,-1\}$ with
\[
\mathbb P(C=+1)=\mathbb P(C=-1)=\frac12
\]
and output $Y=C$.

\paragraph{Same-annotator repeat channel \(\mathsf R\).}
Draw a single latent type $Z\sim\pi$. Conditional on $(Z,q)$, draw two latent
responses
\[
\widetilde Y^{(1)},\widetilde Y^{(2)} \in \{+1,0,-1\}
\]
independently from the same query-specific kernel. Independently complete each
zero response using an independent fair completion coin exactly as in channel
$\mathsf B$. The observable is the pair
\[
\bigl(Y^{(1)},Y^{(2)}\bigr)\in\{+1,-1\}^2.
\]

\paragraph{Abstention channel \(\mathsf A\).}
Draw $Z\sim\pi$, then draw $\widetilde Y\in\{+1,0,-1\}$ from the latent kernel.
The observable is $\widetilde Y$ itself.

\paragraph{Combined channels.}
When we write a combined channel such as $\mathsf{BR}$, $\mathsf{BA}$, or
$\mathsf{BRA}$, we mean that the analyst has access to independent samples from
each constituent primitive channel.
\end{definition}

\begin{assumption}[Conditional independence of repeated elicitation]
\label{ass:appendix-repeat-independence}
Under channel $\mathsf R$, conditional on $(Z,q)$ the latent responses
$\widetilde Y^{(1)}$ and $\widetilde Y^{(2)}$ are independent draws from the
same kernel. All completion coins are mutually independent and independent of all
latent responses.
\end{assumption}

\begin{remark}[Why the completion rule is symmetric]
\label{rem:appendix-symmetric-completion}
The fair-coin completion rule is a deliberate modeling choice. It makes the binary
channel informationally neutral on latent non-binary responses and therefore
isolates the identifiability phenomenon in its cleanest form. This gives the
binary interface the strongest possible chance. Appendix~\ref{app:extensions}
shows that if the completion rule is asymmetric or unknown, identifiability can
only deteriorate.
\end{remark}

\begin{remark}[Population interpretation of the repeat channel]
\label{rem:appendix-repeat-population}
The repeat channel should be read as a minimal \emph{annotator-linking}
intervention: the analyst learns that two labels came from the same stable latent
source. Related ``same-user'' information has already appeared as a useful signal
in heterogeneous-alignment analyses \citep{shirali2025direct}. In our framework,
however, repeats are used only to identify the between-type component of
disagreement and are otherwise kept agnostic to any downstream alignment
objective.
\end{remark}

\subsection{Statistical experiment}
\label{app:formal-setup:statistical-experiment}

The population-level identification results in the main text are statements about
the observable laws induced by the primitive channels. For the finite-sample
results, we adopt a product experiment with independent samples from each channel.

Let
\[
\mathcal Y_{\mathsf B}:=\{+1,-1\},
\qquad
\mathcal Y_{\mathsf R}:=\{+1,-1\}^2,
\qquad
\mathcal Y_{\mathsf A}:=\{+1,0,-1\}.
\]
For a sample-allocation vector
\[
\mathbf n := \bigl((n_{B,q},n_{R,q},n_{A,q})\bigr)_{q\in\mathcal Q}
\in (\mathbb N_0^3)^{|\mathcal Q|},
\]
define the global sample space
\[
\Omega_{\mathbf n}
:=
\prod_{q\in\mathcal Q}
\left(
\mathcal Y_{\mathsf B}^{\,n_{B,q}}
\times
\mathcal Y_{\mathsf R}^{\,n_{R,q}}
\times
\mathcal Y_{\mathsf A}^{\,n_{A,q}}
\right),
\]
with the convention $S^0=\{\varnothing\}$ for any set $S$.

For each fixed model $M$ and query $q$, let
\[
P^{\mathsf B}_{M,q}:=\operatorname{Law}^{\mathsf B}_{M,q},
\qquad
P^{\mathsf R}_{M,q}:=\operatorname{Law}^{\mathsf R}_{M,q},
\qquad
P^{\mathsf A}_{M,q}:=\operatorname{Law}^{\mathsf A}_{M,q}
\]
be the corresponding observable laws. The channel-separated product experiment is
then
\begin{equation}
\label{eq:appendix-product-experiment}
\mathbb P^{\mathbf n}_M
:=
\bigotimes_{q\in\mathcal Q}
\left(
(P^{\mathsf B}_{M,q})^{\otimes n_{B,q}}
\otimes
(P^{\mathsf R}_{M,q})^{\otimes n_{R,q}}
\otimes
(P^{\mathsf A}_{M,q})^{\otimes n_{A,q}}
\right).
\end{equation}
All concentration and minimax statements in the appendices are taken with respect
to the experiment in \eqref{eq:appendix-product-experiment}, unless explicitly
stated otherwise.

\begin{remark}[Querywise reduction]
\label{rem:appendix-querywise}
Every population identification statement proved in this paper is querywise. Since
$\mathcal Q$ is finite, vector-valued statements over all queries follow
coordinatewise. Likewise, uniform finite-sample guarantees are obtained from the
corresponding pointwise bounds by a union bound over $q\in\mathcal Q$.
\end{remark}

\begin{definition}[Channel-measurable estimators and decision rules]
\label{def:appendix-measurable-rules}
Fix a parameter map $\theta_q:\mathcal M(\mathcal Q)\to\Theta$ and an action space
$\mathcal A$.

A \emph{querywise estimator} of $\theta_q$ under sample allocation $\mathbf n$ is a
measurable map
\[
\widehat \theta_q : \Omega_{\mathbf n}\to\Theta.
\]
A \emph{querywise decision rule} is a measurable map
\[
\delta_q : \Omega_{\mathbf n}\to\mathcal A.
\]
When we speak of a rule that uses only a specified channel, such as a
\(\mathsf B\)-measurable or \(\mathsf A\)-measurable rule, we mean that the rule
is measurable with respect to the sigma-field generated by the corresponding
coordinate block of $\Omega_{\mathbf n}$.
\end{definition}

\subsection{Explicit observable laws}
\label{app:formal-setup:observable-laws}

The next proposition makes the dependence of each primitive channel on the latent
model completely explicit. This proposition is the algebraic backbone of the later
identification results.

\begin{proposition}[Observable laws under the primitive channels]
\label{prop:appendix-observable-laws}
Fix a model $M$ and a query $q\in\mathcal Q$.

\textbf{(i) Binary channel.} Under $\mathsf B$, the observable law on
$\mathcal Y_{\mathsf B}=\{+1,-1\}$ is
\[
P^{\mathsf B}_{M,q}(Y=y)
=
\frac{1+y\,\mu_q(M)}{2},
\qquad y\in\{+1,-1\}.
\]

\textbf{(ii) Repeat channel.} Under $\mathsf R$, the observable law on
$\mathcal Y_{\mathsf R}=\{+1,-1\}^2$ is
\[
P^{\mathsf R}_{M,q}(Y^{(1)}=y_1,Y^{(2)}=y_2)
=
\frac{1+(y_1+y_2)\mu_q(M)+y_1y_2\nu_q(M)}{4},
\qquad y_1,y_2\in\{+1,-1\}.
\]

\textbf{(iii) Abstention channel.} Under $\mathsf A$, the observable law on
$\mathcal Y_{\mathsf A}=\{+1,0,-1\}$ is
\[
P^{\mathsf A}_{M,q}(\widetilde Y=0)=\alpha_q(M),
\]
and for $y\in\{+1,-1\}$,
\[
P^{\mathsf A}_{M,q}(\widetilde Y=y)
=
\frac{1-\alpha_q(M)+y\,\mu_q(M)}{2}.
\]
\end{proposition}

\begin{proof}
For channel $\mathsf B$, conditional on $(Z=z,q)$, symmetric completion contributes
zero mean on the event $\{\widetilde Y=0\}$. Hence
\[
\mathbb E[Y\mid Z=z,q]
=
\mathbb P(\widetilde Y=+1\mid Z=z,q)
-
\mathbb P(\widetilde Y=-1\mid Z=z,q)
=
m_q(z).
\]
Taking expectations over $Z\sim\pi$ yields
\[
\mathbb E[Y\mid q]=\mathbb E_{\pi}[m_q(Z)]=\mu_q(M).
\]
Since $Y\in\{+1,-1\}$, the binary law is uniquely determined by its mean:
\[
P^{\mathsf B}_{M,q}(Y=y)=\frac{1+y\,\mathbb E[Y\mid q]}{2}
=
\frac{1+y\,\mu_q(M)}{2}.
\]

For channel $\mathsf R$, conditional on $(Z=z,q)$ the repeated observations are
independent and each has conditional mean $m_q(z)$. Therefore
\[
\mathbb E[Y^{(1)}\mid Z=z,q]
=
\mathbb E[Y^{(2)}\mid Z=z,q]
=
m_q(z),
\qquad
\mathbb E[Y^{(1)}Y^{(2)}\mid Z=z,q]
=
m_q(z)^2.
\]
Taking expectations gives
\[
\mathbb E[Y^{(1)}\mid q]=\mathbb E[Y^{(2)}\mid q]=\mu_q(M),
\qquad
\mathbb E[Y^{(1)}Y^{(2)}\mid q]=\nu_q(M).
\]
Now use the identity
\[
\mathbf 1\{Y^{(1)}=y_1\}\mathbf 1\{Y^{(2)}=y_2\}
=
\frac{1+y_1Y^{(1)}}{2}\cdot\frac{1+y_2Y^{(2)}}{2},
\qquad y_1,y_2\in\{+1,-1\},
\]
and take expectations to obtain
\[
P^{\mathsf R}_{M,q}(y_1,y_2)
=
\frac{1+y_1\mu_q(M)+y_2\mu_q(M)+y_1y_2\nu_q(M)}{4}.
\]

For channel $\mathsf A$, we have directly
\[
P^{\mathsf A}_{M,q}(\widetilde Y=0)
=
\mathbb E_{\pi}[a_q(Z)]
=
\alpha_q(M).
\]
For $y=+1$,
\[
P^{\mathsf A}_{M,q}(\widetilde Y=+1)
=
\mathbb E_{\pi}\!\left[\frac{1-a_q(Z)+m_q(Z)}{2}\right]
=
\frac{1-\alpha_q(M)+\mu_q(M)}{2},
\]
and the formula for $y=-1$ follows analogously.
\end{proof}

\begin{corollary}[Primitive-channel sufficient statistics]
\label{cor:appendix-primitive-summaries}
For each fixed query $q$, the observable law under
\begin{itemize}
    \item channel $\mathsf B$ depends on the model $M$ only through $\mu_q(M)$;
    \item channel $\mathsf R$ depends on the model $M$ only through
    $(\mu_q(M),\nu_q(M))$;
    \item channel $\mathsf A$ depends on the model $M$ only through
    $(\mu_q(M),\alpha_q(M))$.
\end{itemize}
Consequently, for any two models $M,M' \in \mathcal M(\mathcal Q)$,
\[
M \equiv_{\mathsf B} M'
\iff
\mu(M)=\mu(M'),
\]
\[
M \equiv_{\mathsf R} M'
\iff
(\mu,\nu)(M)=(\mu,\nu)(M'),
\]
\[
M \equiv_{\mathsf A} M'
\iff
(\mu,\alpha)(M)=(\mu,\alpha)(M').
\]
\end{corollary}

\begin{proof}
The first claim follows immediately from
Proposition~\ref{prop:appendix-observable-laws}. The equivalences are obtained by
applying the proposition coordinatewise for each $q\in\mathcal Q$.
\end{proof}

\begin{remark}[Relation to CPRD]
\label{rem:appendix-cprd}
Under the anonymous binary channel, the observable law is fully characterized by
the pairwise choice probability
\[
\mathbb P(Y=+1\mid q)=\frac{1+\mu_q}{2}.
\]
This is the channel-level object analogous to the aggregate comparison law studied
by \citet{pukdee2026what}. The conceptual distinction is important: CPRD captures
\emph{what binary pairwise data say}, whereas the latent semantic parameters
$(\tau_q,\alpha_q)$ describe \emph{how the same aggregate law could have arisen}.
The paper's central question is exactly whether those latent explanations are
identifiable from the observable law induced by a given channel.
\end{remark}

\begin{remark}[Why the main text uses \(\mathsf{BR}\)]
\label{rem:appendix-why-br}
At the level of population identification, channel $\mathsf R$ alone already
determines $(\mu_q,\nu_q)$ and hence $(\mu_q,\tau_q)$. Nevertheless, the main text
uses the notation $\mathsf{BR}$ to emphasize the intended experimental design:
anonymous binary data are the default supervision source, while same-annotator
repeats are collected only on a smaller calibration subset. This distinction
matters for finite-sample allocation, even though it does not matter for the
population characterization of the repeat law itself.
\end{remark}

\subsection{Identification under a channel}
\label{app:formal-setup:identification}

We now formalize the identification language used throughout the paper.

\begin{definition}[Observational equivalence]
\label{def:appendix-observational-equivalence}
Let $\mathsf C$ be any primitive or combined channel. Two models
$M,M' \in \mathcal M(\mathcal Q)$ are \emph{observationally equivalent} under
$\mathsf C$, written
\[
M \equiv_{\mathsf C} M',
\]
if
\[
\operatorname{Law}^{\mathsf C}_{M,q}
=
\operatorname{Law}^{\mathsf C}_{M',q}
\qquad
\text{for every } q\in\mathcal Q.
\]
\end{definition}

\begin{definition}[Identification regions]
\label{def:appendix-identification-regions}
Let $\theta_q:\mathcal M(\mathcal Q)\to\Theta$ be any querywise parameter map and
let $\mathsf C$ be any channel. The \emph{identification region} of $\theta_q$ at
model $M$ under channel $\mathsf C$ is
\[
\mathcal I_q^{\mathsf C}(\theta;M)
:=
\left\{
\theta_q(M')
:
M' \in \mathcal M(\mathcal Q),\;
M' \equiv_{\mathsf C} M
\right\}.
\]
We say that $\theta_q$ is
\begin{itemize}
    \item \emph{point-identified} under channel $\mathsf C$ if
    $\mathcal I_q^{\mathsf C}(\theta;M)$ is a singleton for every
    $M \in \mathcal M(\mathcal Q)$;
    \item \emph{partially identified} under channel $\mathsf C$ if
    $\mathcal I_q^{\mathsf C}(\theta;M)$ may contain more than one point.
\end{itemize}
\end{definition}

\begin{definition}[Semantic fiber]
\label{def:appendix-semantic-fiber}
For each model $M$, query $q$, and channel $\mathsf C$, define the
\emph{semantic fiber}
\[
\mathcal S_q^{\mathsf C}(M)
:=
\mathcal I_q^{\mathsf C}\bigl((\mu_q,\tau_q,\alpha_q);M\bigr)
=
\left\{
(\mu_q(M'),\tau_q(M'),\alpha_q(M'))
:
M' \equiv_{\mathsf C} M
\right\}.
\]
Thus $\mathcal S_q^{\mathsf C}(M)$ is the set of all semantic triples compatible
with the same observable law at query $q$ under channel $\mathsf C$.
\end{definition}

\begin{remark}[Pointwise versus vector identification]
\label{rem:appendix-pointwise-vs-vector}
Because the model class is unrestricted across queries, all nontrivial
identification phenomena are already present at a single fixed query $q$.
Accordingly, the main theorems are stated querywise. The corresponding vector
statements over all $q\in\mathcal Q$ are obtained by applying the same argument
coordinatewise.
\end{remark}

\begin{remark}[Interpretive scope]
\label{rem:appendix-interpretive-scope}
The paper studies what the data-collection channel identifies about disagreement.
It does \emph{not} presuppose that persistent disagreement should always be
interpreted as value pluralism, nor that non-binary mass should always be
interpreted as incomparability. Those are downstream semantic judgments that
require either stronger modeling assumptions or richer feedback protocols. The
identification results show exactly which distinctions are and are not available
from the data alone.
\end{remark}

\section{Proof of Theorem 3.1}
\label{app:proof-theorem-3-1}

In the CPRD perspective of \citet{pukdee2026what}, anonymous binary pairwise data
identify an aggregate comparison law. Theorem~3.1 sharpens that viewpoint by
characterizing the entire \emph{semantic fiber} lying above a fixed binary law.
Throughout this appendix, let $\mathcal M(\mathcal Q)$ denote the class of all
latent semantic preference models on the finite query set $\mathcal Q$ in the
sense of Definition~\ref{def:appendix-latent-model}.

\subsection{Restatement in fiber form}
\label{app:proof-theorem-3-1:restatement}

For $\mu \in [-1,1]$, define the feasible set
\begin{equation}
\label{eq:app-feasible-set}
\mathcal F(\mu)
:=
\left\{
(\alpha,\tau)\in \mathbb R_{\ge 0}^2
:
\alpha \le 1-|\mu|,
\quad
\tau \le 1-\alpha-\mu^2
\right\}.
\end{equation}
Since $\alpha \le 1-|\mu| \le 1$, the constraint $\alpha\in[0,1]$ is already
implicit in \eqref{eq:app-feasible-set}.

We restate Theorem~3.1 in the fiber notation of
Definition~\ref{def:appendix-semantic-fiber}.

\begin{theorem}[Exact semantic fiber under anonymous binary feedback]
\label{thm:app-exact-binary-fiber}
For every model $M \in \mathcal M(\mathcal Q)$ and every query
$q \in \mathcal Q$,
\begin{equation}
\label{eq:app-exact-fiber}
\mathcal S_q^{\mathsf B}(M)
=
\left\{
\bigl(\mu_q(M),\tau,\alpha\bigr)
:
(\alpha,\tau)\in \mathcal F\bigl(\mu_q(M)\bigr)
\right\}.
\end{equation}
Equivalently:
\begin{enumerate}
    \item every model $M$ satisfies
    \[
    \bigl(\alpha_q(M),\tau_q(M)\bigr)\in \mathcal F\bigl(\mu_q(M)\bigr);
    \]
    \item for every $\mu \in [-1,1]$ and every $(\alpha,\tau)\in \mathcal F(\mu)$,
    there exists a model whose querywise semantic triple is $(\mu,\tau,\alpha)$
    and whose anonymous binary law is
    \[
    \mathbb P(Y=+1\mid q)=\frac{1+\mu}{2}.
    \]
\end{enumerate}
Hence under channel $\mathsf B$, the aggregate directional preference $\mu_q$ is
point-identified, while persistent disagreement $\tau_q$ and non-binary mass
$\alpha_q$ are only partially identified through the exact region
$\mathcal F(\mu_q)$.
\end{theorem}

\paragraph{Proof strategy.}
The proof has four components. First, the pointwise constraint
$|m_q(z)| \le 1-a_q(z)$ yields necessary upper bounds on $(\alpha_q,\tau_q)$ at
fixed $\mu_q$. Second, a \emph{homogeneous} construction realizes the lower
endpoint $\tau=0$ at any feasible $(\mu,\alpha)$. Third, an \emph{extremal}
three-type construction saturates the upper bound
$\tau=1-\alpha-\mu^2$. Fourth, a mixture argument interpolates every intermediate
value of $\tau$. A final embedding lemma then inserts the resulting one-query
construction into an arbitrary ambient multi-query model without changing the
anonymous binary law on any other query.

\subsection{Auxiliary lemmas}
\label{app:proof-theorem-3-1:lemmas}

\begin{lemma}[Necessary feasibility constraints]
\label{lem:app-feasibility-constraints}
For every model $M \in \mathcal M(\mathcal Q)$ and every query $q\in\mathcal Q$,
\[
0 \le \alpha_q(M) \le 1-|\mu_q(M)|,
\qquad
0 \le \tau_q(M) \le 1-\alpha_q(M)-\mu_q(M)^2.
\]
In particular,
\[
\bigl(\alpha_q(M),\tau_q(M)\bigr)\in \mathcal F\bigl(\mu_q(M)\bigr).
\]
\end{lemma}

\begin{proof}
Fix $M$ and $q$, and suppress the dependence on $(M,q)$ in the notation.
Since $a_q(Z)\in[0,1]$ almost surely,
\[
0\le \alpha_q=\mathbb E[a_q(Z)]\le 1.
\]
Also,
\[
\tau_q=\Var(m_q(Z))\ge 0.
\]

By Definition~\ref{def:appendix-latent-model},
\[
|m_q(z)| \le 1-a_q(z)
\qquad\text{for every } z\in\mathcal Z.
\]
Hence
\[
|\mu_q|
=
\left|\mathbb E[m_q(Z)]\right|
\le
\mathbb E\!\left[|m_q(Z)|\right]
\le
\mathbb E\!\left[1-a_q(Z)\right]
=
1-\alpha_q.
\]
This proves the upper bound $\alpha_q \le 1-|\mu_q|$.

Next, because $|m_q(Z)|\le 1-a_q(Z)\le 1$ almost surely, we have
\[
m_q(Z)^2 \le |m_q(Z)| \le 1-a_q(Z)
\qquad\text{almost surely.}
\]
Therefore
\[
\nu_q
=
\mathbb E[m_q(Z)^2]
\le
\mathbb E[1-a_q(Z)]
=
1-\alpha_q.
\]
Recalling that $\tau_q=\nu_q-\mu_q^2$, we obtain
\[
\tau_q \le 1-\alpha_q-\mu_q^2.
\]
This proves the claim.
\end{proof}

\begin{lemma}[Homogeneous construction]
\label{lem:app-homogeneous-construction}
Fix a query $q$ and numbers $\mu\in[-1,1]$ and $\alpha\in[0,1-|\mu|]$. Then there
exists a one-query model
\[
\overline M^{\mathrm{hom}}_{q;\mu,\alpha}
\]
on the singleton query set $\{q\}$ such that
\[
\mu_q\!\left(\overline M^{\mathrm{hom}}_{q;\mu,\alpha}\right)=\mu,
\qquad
\alpha_q\!\left(\overline M^{\mathrm{hom}}_{q;\mu,\alpha}\right)=\alpha,
\qquad
\tau_q\!\left(\overline M^{\mathrm{hom}}_{q;\mu,\alpha}\right)=0.
\]
\end{lemma}

\begin{proof}
Let the latent space be the one-point standard Borel space
\[
\overline{\mathcal Z}^{\mathrm{hom}}:=\{z_\star\},
\qquad
\overline\pi^{\mathrm{hom}}:=\delta_{z_\star}.
\]
Define
\[
\overline a_q^{\mathrm{hom}}(z_\star):=\alpha,
\qquad
\overline m_q^{\mathrm{hom}}(z_\star):=\mu.
\]
The constraint
\[
\left|\overline m_q^{\mathrm{hom}}(z_\star)\right|
=
|\mu|
\le
1-\alpha
=
1-\overline a_q^{\mathrm{hom}}(z_\star)
\]
holds by assumption, so this indeed defines a valid latent semantic preference
model on the singleton query set $\{q\}$.

Since the latent type is deterministic,
\[
\mu_q
=
\mathbb E[\overline m_q^{\mathrm{hom}}(Z)]
=
\mu,
\qquad
\alpha_q
=
\mathbb E[\overline a_q^{\mathrm{hom}}(Z)]
=
\alpha,
\]
and
\[
\tau_q
=
\Var\!\left(\overline m_q^{\mathrm{hom}}(Z)\right)
=
0.
\]
\end{proof}

\begin{lemma}[Extremal construction]
\label{lem:app-extremal-construction}
Fix a query $q$ and numbers $\mu\in[-1,1]$ and $\alpha\in[0,1-|\mu|]$. Then there
exists a one-query model
\[
\overline M^{\mathrm{ext}}_{q;\mu,\alpha}
\]
on the singleton query set $\{q\}$ such that
\[
\mu_q\!\left(\overline M^{\mathrm{ext}}_{q;\mu,\alpha}\right)=\mu,
\qquad
\alpha_q\!\left(\overline M^{\mathrm{ext}}_{q;\mu,\alpha}\right)=\alpha,
\qquad
\tau_q\!\left(\overline M^{\mathrm{ext}}_{q;\mu,\alpha}\right)=1-\alpha-\mu^2.
\]
In particular, the upper bound from
Lemma~\ref{lem:app-feasibility-constraints} is sharp.
\end{lemma}

\begin{proof}
Consider the three-point latent space
\[
\overline{\mathcal Z}^{\mathrm{ext}}:=\{z_{+},z_{0},z_{-}\}.
\]
Define a probability distribution on this space by
\begin{equation}
\label{eq:app-extremal-probabilities}
\overline\pi^{\mathrm{ext}}(z_+):=\frac{1-\alpha+\mu}{2},
\qquad
\overline\pi^{\mathrm{ext}}(z_0):=\alpha,
\qquad
\overline\pi^{\mathrm{ext}}(z_-):=\frac{1-\alpha-\mu}{2}.
\end{equation}
These three numbers are nonnegative because $\alpha\le 1-|\mu|$, and they sum to
one. Define the querywise kernel by
\[
\bigl(\overline a_q^{\mathrm{ext}},\overline m_q^{\mathrm{ext}}\bigr)(z_+)
:=
(0,1),
\qquad
\bigl(\overline a_q^{\mathrm{ext}},\overline m_q^{\mathrm{ext}}\bigr)(z_0)
:=
(1,0),
\qquad
\bigl(\overline a_q^{\mathrm{ext}},\overline m_q^{\mathrm{ext}}\bigr)(z_-)
:=
(0,-1).
\]
The pointwise validity constraint
\[
\left|\overline m_q^{\mathrm{ext}}(z)\right|
\le
1-\overline a_q^{\mathrm{ext}}(z)
\]
holds with equality at every latent type.

We now compute the semantic functionals. First,
\[
\alpha_q
=
\mathbb E_{\overline\pi^{\mathrm{ext}}}
\left[\overline a_q^{\mathrm{ext}}(Z)\right]
=
\overline\pi^{\mathrm{ext}}(z_0)
=
\alpha.
\]
Second,
\begin{align*}
\mu_q
&=
\mathbb E_{\overline\pi^{\mathrm{ext}}}
\left[\overline m_q^{\mathrm{ext}}(Z)\right] \\
&=
\overline\pi^{\mathrm{ext}}(z_+)
-
\overline\pi^{\mathrm{ext}}(z_-) \\
&=
\frac{1-\alpha+\mu}{2}
-
\frac{1-\alpha-\mu}{2}
=
\mu.
\end{align*}
Third,
\begin{align*}
\nu_q
&=
\mathbb E_{\overline\pi^{\mathrm{ext}}}
\left[\bigl(\overline m_q^{\mathrm{ext}}(Z)\bigr)^2\right] \\
&=
\overline\pi^{\mathrm{ext}}(z_+)+\overline\pi^{\mathrm{ext}}(z_-) \\
&=
\frac{1-\alpha+\mu}{2}
+
\frac{1-\alpha-\mu}{2}
=
1-\alpha.
\end{align*}
Therefore
\[
\tau_q
=
\nu_q-\mu_q^2
=
1-\alpha-\mu^2.
\]
This proves the claim.
\end{proof}

\begin{lemma}[Mixtures preserve \(\mu\) and \(\alpha\) and interpolate \(\tau\)]
\label{lem:app-mixture-interpolation}
Fix a query $q$. Let
\[
\overline M_0
=
\bigl((\overline{\mathcal Z}_0,\overline{\mathcal F}_0,\overline\pi_0),
(\overline a_{0,q},\overline m_{0,q})\bigr)
\]
and
\[
\overline M_1
=
\bigl((\overline{\mathcal Z}_1,\overline{\mathcal F}_1,\overline\pi_1),
(\overline a_{1,q},\overline m_{1,q})\bigr)
\]
be one-query models on the singleton query set $\{q\}$ satisfying
\[
\mu_q(\overline M_0)=\mu_q(\overline M_1)=\mu,
\qquad
\alpha_q(\overline M_0)=\alpha_q(\overline M_1)=\alpha.
\]
Then for every $\lambda\in[0,1]$ there exists a one-query model
\[
\overline M_\lambda
\]
on $\{q\}$ such that
\[
\mu_q(\overline M_\lambda)=\mu,
\qquad
\alpha_q(\overline M_\lambda)=\alpha,
\qquad
\tau_q(\overline M_\lambda)
=
(1-\lambda)\tau_q(\overline M_0)
+
\lambda \tau_q(\overline M_1).
\]
\end{lemma}

\begin{proof}
Define the disjoint-union latent space
\[
\overline{\mathcal Z}_\lambda
:=
(\{0\}\times \overline{\mathcal Z}_0)
\sqcup
(\{1\}\times \overline{\mathcal Z}_1),
\]
with sigma-field inherited from the disjoint union and probability measure
\[
\overline\pi_\lambda
:=
(1-\lambda)\,(\delta_0\otimes \overline\pi_0)
+
\lambda\,(\delta_1\otimes \overline\pi_1).
\]
Define the querywise functions by
\[
\overline a_{\lambda,q}(j,z):=
\overline a_{j,q}(z),
\qquad
\overline m_{\lambda,q}(j,z):=
\overline m_{j,q}(z),
\qquad
j\in\{0,1\}.
\]
Pointwise validity follows immediately because each component model is valid.

Now compute:
\begin{align*}
\mu_q(\overline M_\lambda)
&=
\mathbb E_{\overline\pi_\lambda}
[\overline m_{\lambda,q}(Z)] \\
&=
(1-\lambda)\,
\mathbb E_{\overline\pi_0}[\overline m_{0,q}(Z)]
+
\lambda\,
\mathbb E_{\overline\pi_1}[\overline m_{1,q}(Z)] \\
&=
(1-\lambda)\mu+\lambda\mu
=
\mu.
\end{align*}
Similarly,
\begin{align*}
\alpha_q(\overline M_\lambda)
&=
\mathbb E_{\overline\pi_\lambda}
[\overline a_{\lambda,q}(Z)] \\
&=
(1-\lambda)\,
\mathbb E_{\overline\pi_0}[\overline a_{0,q}(Z)]
+
\lambda\,
\mathbb E_{\overline\pi_1}[\overline a_{1,q}(Z)] \\
&=
(1-\lambda)\alpha+\lambda\alpha
=
\alpha.
\end{align*}
For the second moment,
\begin{align*}
\nu_q(\overline M_\lambda)
&=
\mathbb E_{\overline\pi_\lambda}
\left[\overline m_{\lambda,q}(Z)^2\right] \\
&=
(1-\lambda)\,
\mathbb E_{\overline\pi_0}
\left[\overline m_{0,q}(Z)^2\right]
+
\lambda\,
\mathbb E_{\overline\pi_1}
\left[\overline m_{1,q}(Z)^2\right] \\
&=
(1-\lambda)\nu_q(\overline M_0)
+
\lambda\nu_q(\overline M_1).
\end{align*}
Since $\nu_q(\overline M_j)=\mu^2+\tau_q(\overline M_j)$ for $j\in\{0,1\}$, we get
\begin{align*}
\tau_q(\overline M_\lambda)
&=
\nu_q(\overline M_\lambda)-\mu^2 \\
&=
(1-\lambda)\tau_q(\overline M_0)
+
\lambda\tau_q(\overline M_1),
\end{align*}
as claimed.
\end{proof}

\begin{lemma}[Embedding a one-query construction into an ambient model]
\label{lem:app-embedding-construction}
Let $M \in \mathcal M(\mathcal Q)$, fix a query $q\in\mathcal Q$, and let
\[
\overline M
=
\bigl((\overline{\mathcal Z},\overline{\mathcal F},\overline\pi),
(\overline a_q,\overline m_q)\bigr)
\]
be a one-query model on the singleton query set $\{q\}$ satisfying
\[
\mu_q(\overline M)=\mu_q(M).
\]
Then there exists a model $M' \in \mathcal M(\mathcal Q)$ such that
\[
\bigl(\mu_q(M'),\tau_q(M'),\alpha_q(M')\bigr)
=
\bigl(\mu_q(\overline M),\tau_q(\overline M),\alpha_q(\overline M)\bigr),
\]
and for every $r\in\mathcal Q\setminus\{q\}$,
\[
\bigl(\mu_r(M'),\tau_r(M'),\alpha_r(M')\bigr)
=
\bigl(\mu_r(M),\tau_r(M),\alpha_r(M)\bigr).
\]
Consequently,
\[
M' \equiv_{\mathsf B} M.
\]
\end{lemma}

\begin{proof}
Write
\[
M
=
\bigl((\mathcal Z,\mathcal F,\pi),\{(a_r,m_r)\}_{r\in\mathcal Q}\bigr).
\]
Define the product latent space
\[
\mathcal Z' := \mathcal Z \times \overline{\mathcal Z},
\qquad
\mathcal F' := \mathcal F \otimes \overline{\mathcal F},
\qquad
\pi' := \pi \otimes \overline\pi.
\]
For the target query $q$, define
\[
a'_q(z,\overline z):=\overline a_q(\overline z),
\qquad
m'_q(z,\overline z):=\overline m_q(\overline z).
\]
For every non-target query $r\neq q$, define
\[
a'_r(z,\overline z):=a_r(z),
\qquad
m'_r(z,\overline z):=m_r(z).
\]
Because both $M$ and $\overline M$ satisfy the pointwise validity constraint
$|m|\le 1-a$, the resulting family
$\{(a'_r,m'_r)\}_{r\in\mathcal Q}$ also defines a valid latent semantic
preference model on $\mathcal Q$.

For the target query,
\[
\mu_q(M')
=
\mathbb E_{\pi'}[m'_q(Z,\overline Z)]
=
\mathbb E_{\overline\pi}[\overline m_q(\overline Z)]
=
\mu_q(\overline M),
\]
and similarly
\[
\alpha_q(M')=\alpha_q(\overline M),
\qquad
\tau_q(M')=\tau_q(\overline M).
\]
For any $r\neq q$,
\[
\mu_r(M')
=
\mathbb E_{\pi'}[m'_r(Z,\overline Z)]
=
\mathbb E_{\pi}[m_r(Z)]
=
\mu_r(M),
\]
and likewise
\[
\alpha_r(M')=\alpha_r(M),
\qquad
\tau_r(M')=\tau_r(M).
\]
Thus the two models agree on the full vector $\mu(M')=\mu(M)$. By
Corollary~\ref{cor:appendix-primitive-summaries}, anonymous binary equivalence is
equivalent to equality of $\mu$, hence
\[
M' \equiv_{\mathsf B} M.
\]
\end{proof}

\subsection{Proof of the theorem}
\label{app:proof-theorem-3-1:main-proof}

\begin{proof}[Proof of Theorem~\ref{thm:app-exact-binary-fiber}]
Fix a model $M \in \mathcal M(\mathcal Q)$ and a query $q\in\mathcal Q$. Write
\[
\mu^\star := \mu_q(M).
\]

We first prove the inclusion
\begin{equation}
\label{eq:app-fiber-inclusion-one}
\mathcal S_q^{\mathsf B}(M)
\subseteq
\left\{
(\mu^\star,\tau,\alpha)
:
(\alpha,\tau)\in \mathcal F(\mu^\star)
\right\}.
\end{equation}
Take any
\[
(\mu',\tau',\alpha') \in \mathcal S_q^{\mathsf B}(M).
\]
By Definition~\ref{def:appendix-semantic-fiber}, there exists a model
$M' \in \mathcal M(\mathcal Q)$ such that
\[
M' \equiv_{\mathsf B} M
\]
and
\[
(\mu_q(M'),\tau_q(M'),\alpha_q(M'))
=
(\mu',\tau',\alpha').
\]
By Corollary~\ref{cor:appendix-primitive-summaries},
\[
M' \equiv_{\mathsf B} M
\quad\Longrightarrow\quad
\mu(M')=\mu(M).
\]
In particular,
\[
\mu'=\mu_q(M')=\mu_q(M)=\mu^\star.
\]
Applying Lemma~\ref{lem:app-feasibility-constraints} to the model $M'$ at query
$q$ yields
\[
(\alpha',\tau')\in \mathcal F(\mu').
\]
Since $\mu'=\mu^\star$, this proves \eqref{eq:app-fiber-inclusion-one}.

We now prove the reverse inclusion
\begin{equation}
\label{eq:app-fiber-inclusion-two}
\left\{
(\mu^\star,\tau,\alpha)
:
(\alpha,\tau)\in \mathcal F(\mu^\star)
\right\}
\subseteq
\mathcal S_q^{\mathsf B}(M).
\end{equation}
Fix any pair $(\alpha,\tau)\in \mathcal F(\mu^\star)$. We must construct a model
$M' \in \mathcal M(\mathcal Q)$ such that
\[
M' \equiv_{\mathsf B} M
\]
and
\[
\bigl(\mu_q(M'),\tau_q(M'),\alpha_q(M')\bigr)
=
(\mu^\star,\tau,\alpha).
\]

Set
\[
D := 1-\alpha-(\mu^\star)^2.
\]
Because $(\alpha,\tau)\in \mathcal F(\mu^\star)$, we have $D\ge 0$ and
$0\le \tau \le D$.

\paragraph{Case 1: \(D=0\).}
Then necessarily $\tau=0$. By
Lemma~\ref{lem:app-homogeneous-construction}, there exists a one-query model
\[
\overline M
=
\overline M^{\mathrm{hom}}_{q;\mu^\star,\alpha}
\]
on the singleton query set $\{q\}$ such that
\[
\mu_q(\overline M)=\mu^\star,
\qquad
\alpha_q(\overline M)=\alpha,
\qquad
\tau_q(\overline M)=0=\tau.
\]
Applying Lemma~\ref{lem:app-embedding-construction}, we obtain a model
$M' \in \mathcal M(\mathcal Q)$ satisfying
\[
M' \equiv_{\mathsf B} M
\]
and
\[
\bigl(\mu_q(M'),\tau_q(M'),\alpha_q(M')\bigr)
=
(\mu^\star,\tau,\alpha).
\]
Hence $(\mu^\star,\tau,\alpha)\in \mathcal S_q^{\mathsf B}(M)$.

\paragraph{Case 2: \(D>0\).}
Let
\[
\lambda := \frac{\tau}{D}\in[0,1].
\]
By Lemma~\ref{lem:app-homogeneous-construction}, there exists a one-query model
\[
\overline M_0
=
\overline M^{\mathrm{hom}}_{q;\mu^\star,\alpha}
\]
with
\[
\mu_q(\overline M_0)=\mu^\star,
\qquad
\alpha_q(\overline M_0)=\alpha,
\qquad
\tau_q(\overline M_0)=0.
\]
By Lemma~\ref{lem:app-extremal-construction}, there exists a one-query model
\[
\overline M_1
=
\overline M^{\mathrm{ext}}_{q;\mu^\star,\alpha}
\]
with
\[
\mu_q(\overline M_1)=\mu^\star,
\qquad
\alpha_q(\overline M_1)=\alpha,
\qquad
\tau_q(\overline M_1)=D.
\]
Applying Lemma~\ref{lem:app-mixture-interpolation} with this value of $\lambda$,
we obtain a one-query model $\overline M_\lambda$ such that
\[
\mu_q(\overline M_\lambda)=\mu^\star,
\qquad
\alpha_q(\overline M_\lambda)=\alpha,
\]
and
\begin{align*}
\tau_q(\overline M_\lambda)
&=
(1-\lambda)\tau_q(\overline M_0)
+
\lambda\tau_q(\overline M_1) \\
&=
(1-\lambda)\cdot 0 + \lambda D \\
&=
\tau.
\end{align*}
Now apply Lemma~\ref{lem:app-embedding-construction} to embed
$\overline M_\lambda$ into an ambient model $M' \in \mathcal M(\mathcal Q)$.
Then
\[
M' \equiv_{\mathsf B} M
\]
and
\[
\bigl(\mu_q(M'),\tau_q(M'),\alpha_q(M')\bigr)
=
(\mu^\star,\tau,\alpha).
\]
Thus $(\mu^\star,\tau,\alpha)\in \mathcal S_q^{\mathsf B}(M)$.

The two inclusions \eqref{eq:app-fiber-inclusion-one} and
\eqref{eq:app-fiber-inclusion-two} prove
\eqref{eq:app-exact-fiber}. The equivalent formulation follows immediately from
the definition of $\mathcal S_q^{\mathsf B}(M)$.
\end{proof}

\begin{corollary}[Identification consequences under the anonymous binary channel]
\label{cor:app-binary-identification}
For every model $M \in \mathcal M(\mathcal Q)$ and every query $q\in\mathcal Q$,
\[
\mathcal I_q^{\mathsf B}(\mu_q;M)=\{\mu_q(M)\},
\]
and
\[
\mathcal I_q^{\mathsf B}\bigl((\alpha_q,\tau_q);M\bigr)
=
\mathcal F\bigl(\mu_q(M)\bigr).
\]
In particular, $\mu_q$ is point-identified under $\mathsf B$.

Moreover, if $|\mu_q(M)|<1$, then neither $\alpha_q$ nor $\tau_q$ is
point-identified under $\mathsf B$.
\end{corollary}

\begin{proof}
The first two identities are immediate from
Theorem~\ref{thm:app-exact-binary-fiber}.

To prove the final statement, assume $|\mu_q(M)|<1$ and write
\[
\mu^\star:=\mu_q(M).
\]
Then the two pairs
\[
(\alpha,\tau)=(0,0)
\qquad\text{and}\qquad
(\alpha,\tau)=(1-|\mu^\star|,0)
\]
both belong to $\mathcal F(\mu^\star)$, so $\alpha_q$ is not point-identified.
Likewise, the two pairs
\[
(\alpha,\tau)=(0,0)
\qquad\text{and}\qquad
(\alpha,\tau)=(0,1-(\mu^\star)^2)
\]
both belong to $\mathcal F(\mu^\star)$, and the second has strictly positive
$\tau$ because $|\mu^\star|<1$. Hence $\tau_q$ is not point-identified either.
\end{proof}

\subsection{Coordinatewise realization over finite query sets}
\label{app:proof-theorem-3-1:coordinatewise}

Theorem~\ref{thm:app-exact-binary-fiber} is querywise. Because the model class is
unrestricted across queries, there is no hidden cross-query compatibility
constraint. The next proposition makes that fact explicit and is useful when
constructing matched worlds for the synthetic experiments.

\begin{proposition}[Simultaneous coordinatewise realization]
\label{prop:app-coordinatewise-realization}
Fix a finite query set $\mathcal Q$. Suppose that for each $q\in\mathcal Q$ we are
given a triple $(\mu_q,\tau_q,\alpha_q)$ satisfying
\[
(\alpha_q,\tau_q)\in \mathcal F(\mu_q).
\]
Then there exists a model $M \in \mathcal M(\mathcal Q)$ such that for every
$q\in\mathcal Q$,
\[
\bigl(\mu_q(M),\tau_q(M),\alpha_q(M)\bigr)
=
(\mu_q,\tau_q,\alpha_q).
\]
\end{proposition}

\begin{proof}
For each fixed $q\in\mathcal Q$, Theorem~\ref{thm:app-exact-binary-fiber} yields a
one-query model
\[
\overline M_q
=
\bigl((\overline{\mathcal Z}_q,\overline{\mathcal F}_q,\overline\pi_q),
(\overline a_q,\overline m_q)\bigr)
\]
on the singleton query set $\{q\}$ such that
\[
\mu_q(\overline M_q)=\mu_q,
\qquad
\tau_q(\overline M_q)=\tau_q,
\qquad
\alpha_q(\overline M_q)=\alpha_q.
\]

Define the finite product latent space
\[
\mathcal Z := \prod_{q\in\mathcal Q} \overline{\mathcal Z}_q,
\qquad
\mathcal F := \bigotimes_{q\in\mathcal Q}\overline{\mathcal F}_q,
\qquad
\pi := \bigotimes_{q\in\mathcal Q}\overline\pi_q.
\]
For a latent profile $z=(z_r)_{r\in\mathcal Q}\in\mathcal Z$, define
\[
a_q(z):=\overline a_q(z_q),
\qquad
m_q(z):=\overline m_q(z_q).
\]
Because each coordinate model is valid, the pointwise constraint
$|m_q(z)|\le 1-a_q(z)$ holds for every query $q$ and every profile $z$.

Now fix any query $q\in\mathcal Q$. By construction and the product structure,
\[
\mu_q(M)
=
\mathbb E_{\pi}[m_q(Z)]
=
\mathbb E_{\overline\pi_q}[\overline m_q(Z_q)]
=
\mu_q,
\]
and similarly
\[
\alpha_q(M)=\alpha_q,
\qquad
\nu_q(M)=\nu_q(\overline M_q).
\]
Therefore
\[
\tau_q(M)
=
\nu_q(M)-\mu_q(M)^2
=
\nu_q(\overline M_q)-\mu_q^2
=
\tau_q.
\]
Since $q$ was arbitrary, the claim follows.
\end{proof}

\begin{remark}[Interpretation of the theorem]
\label{rem:app-binary-fiber-interpretation}
Theorem~\ref{thm:app-exact-binary-fiber} says more than that anonymous binary data
fail to identify semantic sources of disagreement. It characterizes the failure
\emph{exactly}. For a fixed binary mean $\mu$, the entire ambiguity about the
latent semantics is summarized by the two-dimensional region
$\mathcal F(\mu)$. This exact fiber characterization is what later makes it
possible to quantify precisely how much same-annotator repeats and abstention
shrink the observational equivalence classes.
\end{remark}

\section{Proof of the Binary-Only Testing Impossibility Result}
\label{app:proof-binary-impossibility}

This appendix proves the binary-only impossibility statement from
Section~\ref{sec:binary-collapse}. The core point is simple but foundational:
if two latent semantic models induce the same anonymous binary law, then no amount
of anonymous binary data can distinguish them, and therefore no semantics-aware
decision rule based only on such data can reliably choose between disjoint actions
that are optimal under the two models. The argument is completely
nonparametric and depends only on observational equivalence under the binary
channel.

\subsection{Binary-only experiment}
\label{app:proof-binary-impossibility:setup}

Fix a query $q\in\mathcal Q$ and a sample size $n\in\mathbb N$. For a model
$M\in\mathcal M(\mathcal Q)$, let
\[
P^{\mathsf B}_{M,q}:=\operatorname{Law}^{\mathsf B}_{M,q}
\]
denote the single-sample anonymous binary law at query $q$, and let
\begin{equation}
\label{eq:app-binary-n-sample-law}
P^{\mathsf B,n}_{M,q}
:=
\bigl(P^{\mathsf B}_{M,q}\bigr)^{\otimes n}
\end{equation}
denote the law of $n$ i.i.d.\ anonymous binary observations
\[
X_q^{(n)}:=(Y_1,\dots,Y_n)\in\{-1,+1\}^n.
\]

\begin{definition}[Binary-only tests and action rules]
\label{def:app-binary-only-tests-actions}
Fix two models $M_0,M_1\in\mathcal M(\mathcal Q)$ and a query $q\in\mathcal Q$.

\paragraph{Tests.}
A \emph{randomized binary-only test} is a measurable map
\[
\varphi:\{-1,+1\}^n\to[0,1],
\]
where $\varphi(x)$ is interpreted as the conditional probability of declaring
model $M_1$ upon observing $x$. Its risks are
\[
R_0(\varphi)
:=
\mathbb E_{P^{\mathsf B,n}_{M_0,q}}[\varphi(X_q^{(n)})],
\qquad
R_1(\varphi)
:=
\mathbb E_{P^{\mathsf B,n}_{M_1,q}}[1-\varphi(X_q^{(n)})].
\]
Thus $R_0(\varphi)$ is the probability of incorrectly declaring $M_1$ under
$M_0$, while $R_1(\varphi)$ is the probability of incorrectly declaring $M_0$
under $M_1$.

\paragraph{Semantics-aware action rules.}
Let $\mathcal A$ be an arbitrary action space, and let
\[
A^\star:\mathcal M(\mathcal Q)\times\mathcal Q \to 2^{\mathcal A}\setminus\{\varnothing\}
\]
be any rule assigning to each model--query pair $(M,q)$ a nonempty set
$A^\star(M,q)\subseteq\mathcal A$ of \emph{model-optimal actions}. A
\emph{binary-only decision rule} is a measurable map
\[
\delta:\{-1,+1\}^n\to\mathcal A.
\]
Its optimal-action loss under model $M_j$ is
\[
L_j(\delta)
:=
P^{\mathsf B,n}_{M_j,q}\!\left(
\delta(X_q^{(n)})\notin A^\star(M_j,q)
\right),
\qquad j\in\{0,1\}.
\]
\end{definition}

The next lemma records the only probabilistic input needed in the proof.

\begin{lemma}[Equality of binary-only $n$-sample laws]
\label{lem:app-equality-binary-n-sample-laws}
Fix a query $q\in\mathcal Q$ and two models
$M_0,M_1\in\mathcal M(\mathcal Q)$. If
\[
M_0 \equiv_{\mathsf B} M_1,
\]
then for every $n\in\mathbb N$,
\[
P^{\mathsf B,n}_{M_0,q}
=
P^{\mathsf B,n}_{M_1,q}.
\]
\end{lemma}

\begin{proof}
By Definition~\ref{def:appendix-observational-equivalence},
\[
M_0 \equiv_{\mathsf B} M_1
\quad\Longrightarrow\quad
P^{\mathsf B}_{M_0,q}=P^{\mathsf B}_{M_1,q}.
\]
Taking $n$-fold product measures yields
\[
P^{\mathsf B,n}_{M_0,q}
=
\bigl(P^{\mathsf B}_{M_0,q}\bigr)^{\otimes n}
=
\bigl(P^{\mathsf B}_{M_1,q}\bigr)^{\otimes n}
=
P^{\mathsf B,n}_{M_1,q}.
\]
\end{proof}

\subsection{Two-point impossibility under identical binary laws}
\label{app:proof-binary-impossibility:main-theorem}

We now state and prove the full binary-only impossibility result. The theorem is
slightly stronger than the statement in the main text: it treats both hypothesis
testing and general semantics-aware action rules in one theorem.

\begin{theorem}[Binary-only impossibility under observational equivalence]
\label{thm:app-binary-only-impossibility}
Fix a query $q\in\mathcal Q$, a sample size $n\in\mathbb N$, and two models
$M_0,M_1\in\mathcal M(\mathcal Q)$ such that
\[
M_0 \equiv_{\mathsf B} M_1.
\]
Then the following hold.

\textbf{(i) Testing lower bound.}
For every randomized binary-only test $\varphi$,
\begin{equation}
\label{eq:app-testing-errors-sum-to-one}
R_0(\varphi)+R_1(\varphi)=1.
\end{equation}
Consequently,
\begin{equation}
\label{eq:app-testing-max-error-lower-bound}
\max\{R_0(\varphi),R_1(\varphi)\}\ge \frac12.
\end{equation}

\textbf{(ii) Action lower bound.}
Let $\mathcal A$ be any action space, let
$A^\star:\mathcal M(\mathcal Q)\times\mathcal Q \to 2^{\mathcal A}\setminus\{\varnothing\}$
be any model-optimal action rule, and assume that the optimal action sets for
$M_0$ and $M_1$ at query $q$ are disjoint:
\begin{equation}
\label{eq:app-disjoint-optimal-action-sets}
A^\star(M_0,q)\cap A^\star(M_1,q)=\varnothing.
\end{equation}
Then every binary-only decision rule $\delta$ satisfies
\begin{equation}
\label{eq:app-action-errors-sum-lower-bound}
L_0(\delta)+L_1(\delta)\ge 1.
\end{equation}
In particular,
\begin{equation}
\label{eq:app-action-max-error-lower-bound}
\max\{L_0(\delta),L_1(\delta)\}\ge \frac12.
\end{equation}

\textbf{(iii) No uniform consistency.}
There does not exist any sequence of binary-only tests
$\{\varphi_n\}_{n\ge 1}$ such that
\[
\max\{R_0(\varphi_n),R_1(\varphi_n)\}\to 0,
\]
and, under \eqref{eq:app-disjoint-optimal-action-sets}, there does not exist any
sequence of binary-only decision rules $\{\delta_n\}_{n\ge 1}$ such that
\[
\max\{L_0(\delta_n),L_1(\delta_n)\}\to 0.
\]
\end{theorem}

\begin{proof}
Fix $q$, $n$, $M_0$, and $M_1$ as in the theorem. By
Lemma~\ref{lem:app-equality-binary-n-sample-laws}, the two models induce the same
law on the binary-only sample space:
\[
P^{\mathsf B,n}_{M_0,q}
=
P^{\mathsf B,n}_{M_1,q}
=:P.
\]

\paragraph{Proof of (i).}
Let $\varphi:\{-1,+1\}^n\to[0,1]$ be any randomized test. Then
\begin{align*}
R_0(\varphi)+R_1(\varphi)
&=
\mathbb E_{P^{\mathsf B,n}_{M_0,q}}[\varphi(X_q^{(n)})]
+
\mathbb E_{P^{\mathsf B,n}_{M_1,q}}[1-\varphi(X_q^{(n)})] \\
&=
\mathbb E_{P}[\varphi(X_q^{(n)})]
+
\mathbb E_{P}[1-\varphi(X_q^{(n)})] \\
&=
1.
\end{align*}
This proves \eqref{eq:app-testing-errors-sum-to-one}. Since the maximum of two
nonnegative numbers is at least half their sum,
\[
\max\{R_0(\varphi),R_1(\varphi)\}
\ge
\frac{R_0(\varphi)+R_1(\varphi)}{2}
=
\frac12,
\]
which proves \eqref{eq:app-testing-max-error-lower-bound}.

\paragraph{Proof of (ii).}
Let $\delta:\{-1,+1\}^n\to\mathcal A$ be any binary-only decision rule. For
brevity, define
\[
A_0^\star:=A^\star(M_0,q),
\qquad
A_1^\star:=A^\star(M_1,q).
\]
By assumption, $A_0^\star\cap A_1^\star=\varnothing$.

We claim that for every sample point $x\in\{-1,+1\}^n$,
\begin{equation}
\label{eq:app-pointwise-indicator-inequality}
\mathbf 1\{\delta(x)\notin A_0^\star\}
+
\mathbf 1\{\delta(x)\notin A_1^\star\}
\ge 1.
\end{equation}
Indeed, if \eqref{eq:app-pointwise-indicator-inequality} were false at some
$x$, then both indicators would equal $0$, which would imply simultaneously that
\[
\delta(x)\in A_0^\star
\qquad\text{and}\qquad
\delta(x)\in A_1^\star.
\]
This contradicts the disjointness assumption
$A_0^\star\cap A_1^\star=\varnothing$.

Taking expectations of \eqref{eq:app-pointwise-indicator-inequality} under the
common law $P$ yields
\begin{align*}
1
&\le
\mathbb E_P\!\left[
\mathbf 1\{\delta(X_q^{(n)})\notin A_0^\star\}
+
\mathbf 1\{\delta(X_q^{(n)})\notin A_1^\star\}
\right] \\
&=
P\!\left(\delta(X_q^{(n)})\notin A_0^\star\right)
+
P\!\left(\delta(X_q^{(n)})\notin A_1^\star\right).
\end{align*}
Since $P=P^{\mathsf B,n}_{M_0,q}=P^{\mathsf B,n}_{M_1,q}$, the last display is
exactly
\[
L_0(\delta)+L_1(\delta)\ge 1,
\]
which proves \eqref{eq:app-action-errors-sum-lower-bound}. Again using the fact
that the maximum of two nonnegative numbers is at least half their sum, we obtain
\[
\max\{L_0(\delta),L_1(\delta)\}
\ge
\frac{L_0(\delta)+L_1(\delta)}{2}
\ge
\frac12,
\]
which proves \eqref{eq:app-action-max-error-lower-bound}.

\paragraph{Proof of (iii).}
Part (i) shows that every test $\varphi_n$ satisfies
\[
\max\{R_0(\varphi_n),R_1(\varphi_n)\}\ge \frac12
\qquad\text{for every } n\ge 1,
\]
so the maximum risk cannot converge to $0$. Likewise, under
\eqref{eq:app-disjoint-optimal-action-sets}, part (ii) shows that every decision
rule $\delta_n$ satisfies
\[
\max\{L_0(\delta_n),L_1(\delta_n)\}\ge \frac12
\qquad\text{for every } n\ge 1,
\]
so the maximum action error cannot converge to $0$ either.
\end{proof}

\begin{corollary}[Main-text binary-only testing impossibility]
\label{cor:app-main-text-binary-impossibility}
Corollary~\ref{cor:binary-testing-impossibility-main} follows immediately from
part~(i) of Theorem~\ref{thm:app-binary-only-impossibility}. If the main text
instead states the binary-only testing impossibility as a theorem, the same proof
applies verbatim.
\end{corollary}

\begin{proof}
Part~(i) shows that every binary-only randomized test has worst-case error at
least $1/2$. Deterministic tests are the special case $\varphi\in\{0,1\}$ almost
surely, so the same lower bound holds for them as well.
\end{proof}

\subsection{Extension to multi-query binary-only experiments}
\label{app:proof-binary-impossibility:multi-query}

The main text states the impossibility result querywise because that is the
natural unit of semantics-aware decision-making in preference learning. For
completeness, we record the finite multi-query extension under the product
experiment of Appendix~\ref{app:formal-setup}.

\begin{proposition}[Multi-query binary-only impossibility]
\label{prop:app-multi-query-binary-impossibility}
Fix two models $M_0,M_1\in\mathcal M(\mathcal Q)$ satisfying
\[
M_0 \equiv_{\mathsf B} M_1,
\]
and let $\mathbf n_B=(n_{B,q})_{q\in\mathcal Q}$ be any finite binary-only sample
allocation. Let
\[
\mathbb P^{\mathbf n_B,\mathsf B}_{M_j}
:=
\bigotimes_{q\in\mathcal Q}
\bigl(P^{\mathsf B}_{M_j,q}\bigr)^{\otimes n_{B,q}},
\qquad j\in\{0,1\},
\]
be the induced product laws on the full binary-only sample space. Then
\[
\mathbb P^{\mathbf n_B,\mathsf B}_{M_0}
=
\mathbb P^{\mathbf n_B,\mathsf B}_{M_1}.
\]
Consequently, every testing and action lower bound from
Theorem~\ref{thm:app-binary-only-impossibility} remains valid with
$X_q^{(n)}$ replaced by the full binary-only dataset over $\mathcal Q$.
\end{proposition}

\begin{proof}
By $M_0 \equiv_{\mathsf B} M_1$, we have
\[
P^{\mathsf B}_{M_0,q}=P^{\mathsf B}_{M_1,q}
\qquad\text{for every } q\in\mathcal Q.
\]
Taking tensor products over the query set and over the corresponding within-query
sample sizes yields
\[
\mathbb P^{\mathbf n_B,\mathsf B}_{M_0}
=
\bigotimes_{q\in\mathcal Q}
\bigl(P^{\mathsf B}_{M_0,q}\bigr)^{\otimes n_{B,q}}
=
\bigotimes_{q\in\mathcal Q}
\bigl(P^{\mathsf B}_{M_1,q}\bigr)^{\otimes n_{B,q}}
=
\mathbb P^{\mathbf n_B,\mathsf B}_{M_1}.
\]
The proof of Theorem~\ref{thm:app-binary-only-impossibility} now applies word for
word with the common multi-query product law in place of the common one-query
law.
\end{proof}

\begin{remark}[Interpretation]
\label{rem:app-binary-impossibility-interpretation}
The theorem does not say that anonymous binary comparisons are uninformative.
Rather, it pinpoints \emph{what they do and do not identify}. Under the anonymous
binary channel, the aggregate directional preference $\mu_q$ is identified, but
semantics-aware distinctions requiring access to $\tau_q$ or $\alpha_q$ remain
provably unavailable. The lower bounds in this appendix therefore express an
epistemic limitation of the channel, not a deficiency of a particular estimator or
optimization procedure.
\end{remark}
\section{Proofs for Section 4}
\label{app:proofs-section-4}

Recent theory on hidden context and heterogeneous feedback argues that richer
feedback protocols and annotator-linked information can substantially change what
preference learning recovers
\citep{siththaranjan2023distributional,shirali2025direct,chidambaram2025direct}.
The purpose of this appendix is more specific: to prove exact
\emph{channel-identification} statements for same-annotator repeats and
abstention-enabled feedback, and to characterize how the induced observational
equivalence classes refine those of the anonymous binary channel.

Throughout, let $\mathcal M(\mathcal Q)$ denote the model class of
Appendix~\ref{app:formal-setup}, and fix a finite query set $\mathcal Q$.
All statements are querywise; vector-valued versions over $\mathcal Q$ follow
coordinatewise.

\subsection{A factorization lemma for combined channels}
\label{app:proofs-section-4:factorization}

We first record a formal consequence of the definition of combined channels from
Definition~\ref{def:appendix-primitive-channels}. Since combined channels provide
independent samples from each constituent primitive channel, their observable laws
factor as product measures.

\begin{lemma}[Factorization of combined-channel laws]
\label{lem:app-combined-channel-factorization}
Let $\mathsf C_1,\mathsf C_2$ be primitive or combined channels, and let
$\mathsf C_1\mathsf C_2$ denote the channel that returns independent samples from
$\mathsf C_1$ and $\mathsf C_2$. Then for every model
$M \in \mathcal M(\mathcal Q)$ and every query $q\in\mathcal Q$,
\[
\operatorname{Law}^{\mathsf C_1\mathsf C_2}_{M,q}
=
\operatorname{Law}^{\mathsf C_1}_{M,q}
\otimes
\operatorname{Law}^{\mathsf C_2}_{M,q}.
\]
Consequently, for any two models $M,M'\in\mathcal M(\mathcal Q)$,
\[
M \equiv_{\mathsf C_1\mathsf C_2} M'
\iff
\bigl(M \equiv_{\mathsf C_1} M' \text{ and } M \equiv_{\mathsf C_2} M'\bigr).
\]
\end{lemma}

\begin{proof}
By Definition~\ref{def:appendix-primitive-channels}, the combined channel
$\mathsf C_1\mathsf C_2$ returns independent outputs from the two constituent
channels. Therefore the corresponding observable law is the product of the two
marginal laws:
\[
\operatorname{Law}^{\mathsf C_1\mathsf C_2}_{M,q}
=
\operatorname{Law}^{\mathsf C_1}_{M,q}
\otimes
\operatorname{Law}^{\mathsf C_2}_{M,q}.
\]

For the equivalence statement, if
$M \equiv_{\mathsf C_1\mathsf C_2} M'$, then the two product measures agree for
every $q$. Equality of product measures implies equality of each marginal law, so
$M \equiv_{\mathsf C_1} M'$ and $M \equiv_{\mathsf C_2} M'$.
Conversely, if the two marginal laws agree under both channels for every $q$,
their product laws also agree for every $q$, so
$M \equiv_{\mathsf C_1\mathsf C_2} M'$.
\end{proof}

\begin{remark}[Population redundancy of anonymous binary data]
\label{rem:app-population-redundancy}
At the population level, the repeat channel already contains enough information to
recover the binary mean through its one-dimensional marginals, and the abstention
channel already contains enough information to recover the binary mean through its
strict-outcome marginals. Thus, strictly as an identification statement,
$\mathsf{BR}$ is equivalent to $\mathsf R$, and $\mathsf{BA}$ is equivalent to
$\mathsf A$. We nevertheless retain the notation $\mathsf{BR}$ and
$\mathsf{BRA}$ to mirror the intended experimental design: anonymous binary labels
are the default supervision source, while repeats and abstention-enabled labels
are treated as channel refinements or calibration augmentations.
\end{remark}

\subsection{Proof of Theorem 4.1: repeat covariance identifies persistent disagreement}
\label{app:proofs-section-4:repeat-covariance}

We restate the theorem in a form that makes both its covariance interpretation and
its identification consequence explicit.

\begin{theorem}[Repeat covariance identifies persistent disagreement]
\label{thm:app-repeat-covariance}
Fix a model $M \in \mathcal M(\mathcal Q)$ and a query $q\in\mathcal Q$.
Under the combined channel $\mathsf{BR}$, let
\[
Y \sim \operatorname{Law}^{\mathsf B}_{M,q},
\qquad
\bigl(Y^{(1)},Y^{(2)}\bigr)\sim \operatorname{Law}^{\mathsf R}_{M,q},
\]
with the two channel outputs independent. Then
\begin{equation}
\label{eq:app-repeat-covariance-cov}
\Cov\!\bigl(Y^{(1)},Y^{(2)} \mid q\bigr) = \tau_q(M).
\end{equation}
Equivalently,
\begin{equation}
\label{eq:app-repeat-covariance-moment}
\tau_q(M)
=
\mathbb E\!\left[Y^{(1)}Y^{(2)} \mid q\right]
-
\Bigl(\mathbb E[Y\mid q]\Bigr)^2.
\end{equation}
In particular, $\tau_q$ is point-identified under channel $\mathsf{BR}$.
\end{theorem}

\begin{proof}
By Proposition~\ref{prop:appendix-observable-laws},
\[
\mathbb E[Y\mid q] = \mu_q(M),
\qquad
\mathbb E\!\left[Y^{(1)}Y^{(2)}\mid q\right] = \nu_q(M).
\]
Also, the one-dimensional marginals of the repeat channel agree with the binary
channel mean, so
\[
\mathbb E\!\left[Y^{(1)}\mid q\right]
=
\mathbb E\!\left[Y^{(2)}\mid q\right]
=
\mu_q(M).
\]
Therefore
\begin{align*}
\Cov\!\bigl(Y^{(1)},Y^{(2)}\mid q\bigr)
&=
\mathbb E\!\left[Y^{(1)}Y^{(2)}\mid q\right]
-
\mathbb E\!\left[Y^{(1)}\mid q\right]
\mathbb E\!\left[Y^{(2)}\mid q\right] \\
&=
\nu_q(M)-\mu_q(M)^2 \\
&=
\tau_q(M),
\end{align*}
which proves \eqref{eq:app-repeat-covariance-cov}. Substituting
$\mathbb E[Y\mid q]=\mu_q(M)$ yields
\eqref{eq:app-repeat-covariance-moment}. Since the right-hand side depends only on
the observable law under $\mathsf{BR}$, the parameter $\tau_q$ is point-identified
under that channel.
\end{proof}

\begin{remark}[Law-of-total-covariance interpretation]
\label{rem:app-total-covariance}
Equation~\eqref{eq:app-repeat-covariance-cov} is the precise channel-level sense
in which same-annotator repeats expose the \emph{persistent} component of
disagreement. Conditional on the latent type $Z$, repeated responses are
independent by Assumption~\ref{ass:appendix-repeat-independence}, so the
conditional covariance vanishes. All remaining covariance therefore comes from the
between-type variation in the conditional mean $m_q(Z)$, namely
$\Var(m_q(Z))=\tau_q$.
\end{remark}

\subsection{Proof of Corollary 4.2: disagreement-gap identity}
\label{app:proofs-section-4:disagreement-gap}

We now prove the disagreement-gap representation of persistent disagreement.

\begin{corollary}[Same-annotator versus different-annotator disagreement gap]
\label{cor:app-disagreement-gap}
Fix a model $M \in \mathcal M(\mathcal Q)$ and a query $q\in\mathcal Q$.
Let $Y$ and $Y'$ be two independent anonymous binary observations on query $q$
from two independently sampled annotators, and let
$(Y^{(1)},Y^{(2)})$ be a same-annotator repeat pair from channel $\mathsf R$.
Define
\[
D^{\mathrm{diff}}_q(M)
:=
\mathbb P(Y\neq Y' \mid q),
\qquad
D^{\mathrm{same}}_q(M)
:=
\mathbb P\!\bigl(Y^{(1)}\neq Y^{(2)} \mid q\bigr).
\]
Then
\begin{equation}
\label{eq:app-disagreement-gap-formula}
D^{\mathrm{diff}}_q(M)-D^{\mathrm{same}}_q(M)=\frac{\tau_q(M)}{2}.
\end{equation}
More explicitly,
\[
D^{\mathrm{diff}}_q(M)=\frac{1-\mu_q(M)^2}{2},
\qquad
D^{\mathrm{same}}_q(M)=\frac{1-\mu_q(M)^2-\tau_q(M)}{2}.
\]
\end{corollary}

\begin{proof}
For any pair of random variables $U,V\in\{+1,-1\}$,
\[
\mathbf 1\{U\neq V\}=\frac{1-UV}{2}.
\]
Taking expectations yields
\[
\mathbb P(U\neq V)=\frac{1-\mathbb E[UV]}{2}.
\]

Apply this first to the independent different-annotator pair $(Y,Y')$.
Because $Y$ and $Y'$ are independent and each has mean $\mu_q(M)$ under the binary
channel,
\[
\mathbb E[YY'\mid q]
=
\mathbb E[Y\mid q]\mathbb E[Y'\mid q]
=
\mu_q(M)^2.
\]
Hence
\[
D^{\mathrm{diff}}_q(M)
=
\frac{1-\mu_q(M)^2}{2}.
\]

Now apply the same identity to the same-annotator repeat pair
$(Y^{(1)},Y^{(2)})$:
\[
D^{\mathrm{same}}_q(M)
=
\frac{1-\mathbb E[Y^{(1)}Y^{(2)}\mid q]}{2}
=
\frac{1-\nu_q(M)}{2},
\]
where the last equality uses Proposition~\ref{prop:appendix-observable-laws}.
Since $\nu_q(M)=\mu_q(M)^2+\tau_q(M)$,
\[
D^{\mathrm{same}}_q(M)
=
\frac{1-\mu_q(M)^2-\tau_q(M)}{2}.
\]
Subtracting the two expressions gives
\eqref{eq:app-disagreement-gap-formula}.
\end{proof}

\subsection{Proof of Theorem 4.3: abstention identifies non-binary mass}
\label{app:proofs-section-4:abstention}

The abstention channel reveals the zero state directly. At the population level,
this makes the non-binary mass an observable quantity.

\begin{theorem}[Abstention identifies non-binary mass]
\label{thm:app-abstention-identification}
Fix a model $M \in \mathcal M(\mathcal Q)$ and a query $q\in\mathcal Q$.
Under the abstention channel $\mathsf A$,
\begin{equation}
\label{eq:app-abstention-identifies-alpha}
\alpha_q(M)
=
\mathbb P(\widetilde Y=0\mid q),
\end{equation}
and
\begin{equation}
\label{eq:app-abstention-identifies-mu}
\mu_q(M)
=
\mathbb P(\widetilde Y=+1\mid q)
-
\mathbb P(\widetilde Y=-1\mid q).
\end{equation}
Consequently, the pair $(\mu_q,\alpha_q)$ is point-identified under channel
$\mathsf A$.
\end{theorem}

\begin{proof}
The identities are immediate from the trinomial observable law in
Proposition~\ref{prop:appendix-observable-laws}. Specifically,
\[
\mathbb P(\widetilde Y=0\mid q)=\alpha_q(M),
\]
and
\[
\mathbb P(\widetilde Y=+1\mid q)
=
\frac{1-\alpha_q(M)+\mu_q(M)}{2},
\qquad
\mathbb P(\widetilde Y=-1\mid q)
=
\frac{1-\alpha_q(M)-\mu_q(M)}{2}.
\]
Subtracting the last two displays gives
\eqref{eq:app-abstention-identifies-mu}. Since the right-hand sides depend only on
the observable law under $\mathsf A$, the pair $(\mu_q,\alpha_q)$ is
point-identified.
\end{proof}

\subsection{Proof of Theorem 4.4: channel-refinement frontier}
\label{app:proofs-section-4:frontier}

We now prove the exact characterization of observational equivalence classes under
the refined channels and then show that the resulting equivalence relations are
strict refinements of one another.

\begin{theorem}[Channel-refinement frontier]
\label{thm:app-channel-frontier}
For any two models $M,M' \in \mathcal M(\mathcal Q)$,
\begin{align}
M \equiv_{\mathsf B} M'
&\iff
\mu(M)=\mu(M'),
\label{eq:app-frontier-B}
\\
M \equiv_{\mathsf{BR}} M'
&\iff
\bigl(\mu,\tau\bigr)(M)=\bigl(\mu,\tau\bigr)(M'),
\label{eq:app-frontier-BR}
\\
M \equiv_{\mathsf A} M'
&\iff
\bigl(\mu,\alpha\bigr)(M)=\bigl(\mu,\alpha\bigr)(M'),
\label{eq:app-frontier-A}
\\
M \equiv_{\mathsf{BRA}} M'
&\iff
\bigl(\mu,\tau,\alpha\bigr)(M)=\bigl(\mu,\tau,\alpha\bigr)(M').
\label{eq:app-frontier-BRA}
\end{align}
Consequently, the observational equivalence relations satisfy
\begin{equation}
\label{eq:app-frontier-strict-inclusions}
\equiv_{\mathsf{BRA}}
\subsetneq
\equiv_{\mathsf{BR}}
\subsetneq
\equiv_{\mathsf B},
\qquad
\equiv_{\mathsf{BRA}}
\subsetneq
\equiv_{\mathsf A}
\subsetneq
\equiv_{\mathsf B}.
\end{equation}
\end{theorem}

To prove strictness, it is convenient to record three explicit one-query witness
models.

\begin{lemma}[Witness models for strict inclusions]
\label{lem:app-frontier-witness-models}
Fix a singleton query set $\mathcal Q=\{q\}$. Define the following three models.

\paragraph{(i) The coin model \(M^{\mathrm{coin}}\).}
Let the latent space be a singleton $\{z_\star\}$ with
\[
a_q(z_\star)=0,
\qquad
m_q(z_\star)=0.
\]
Then
\[
\bigl(\mu_q,\tau_q,\alpha_q\bigr)(M^{\mathrm{coin}})=(0,0,0).
\]
Its observable laws are:
\[
\operatorname{Law}^{\mathsf B}_{M^{\mathrm{coin}},q}
=
\text{fair coin on }\{+1,-1\},
\]
\[
\operatorname{Law}^{\mathsf R}_{M^{\mathrm{coin}},q}
=
\text{uniform law on }\{+1,-1\}^2,
\]
\[
\operatorname{Law}^{\mathsf A}_{M^{\mathrm{coin}},q}
=
\frac12\delta_{+1}+\frac12\delta_{-1}.
\]

\paragraph{(ii) The zero model \(M^{\mathrm{zero}}\).}
Let the latent space be a singleton $\{z_\star\}$ with
\[
a_q(z_\star)=1,
\qquad
m_q(z_\star)=0.
\]
Then
\[
\bigl(\mu_q,\tau_q,\alpha_q\bigr)(M^{\mathrm{zero}})=(0,0,1).
\]
Its observable laws are:
\[
\operatorname{Law}^{\mathsf B}_{M^{\mathrm{zero}},q}
=
\text{fair coin on }\{+1,-1\},
\]
\[
\operatorname{Law}^{\mathsf R}_{M^{\mathrm{zero}},q}
=
\text{uniform law on }\{+1,-1\}^2,
\]
\[
\operatorname{Law}^{\mathsf A}_{M^{\mathrm{zero}},q}
=
\delta_{0}.
\]

\paragraph{(iii) The split model \(M^{\mathrm{split}}\).}
Let the latent space be $\{z_+,z_-\}$ with equal weights
$\pi(z_+)=\pi(z_-)=1/2$, and define
\[
(a_q,m_q)(z_+)=(0,1),
\qquad
(a_q,m_q)(z_-)=(0,-1).
\]
Then
\[
\bigl(\mu_q,\tau_q,\alpha_q\bigr)(M^{\mathrm{split}})=(0,1,0).
\]
Its observable laws are:
\[
\operatorname{Law}^{\mathsf B}_{M^{\mathrm{split}},q}
=
\text{fair coin on }\{+1,-1\},
\]
\[
\operatorname{Law}^{\mathsf R}_{M^{\mathrm{split}},q}
=
\frac12\delta_{(+1,+1)}+\frac12\delta_{(-1,-1)},
\]
\[
\operatorname{Law}^{\mathsf A}_{M^{\mathrm{split}},q}
=
\frac12\delta_{+1}+\frac12\delta_{-1}.
\]

Therefore,
\[
M^{\mathrm{coin}} \equiv_{\mathsf B} M^{\mathrm{zero}}
\equiv_{\mathsf B} M^{\mathrm{split}},
\]
\[
M^{\mathrm{coin}} \equiv_{\mathsf{BR}} M^{\mathrm{zero}},
\qquad
M^{\mathrm{coin}} \not\equiv_{\mathsf{BR}} M^{\mathrm{split}},
\]
\[
M^{\mathrm{coin}} \equiv_{\mathsf A} M^{\mathrm{split}},
\qquad
M^{\mathrm{coin}} \not\equiv_{\mathsf A} M^{\mathrm{zero}},
\]
and no two of the three models are equivalent under $\mathsf{BRA}$.
\end{lemma}

\begin{proof}
For $M^{\mathrm{coin}}$, the latent response is strict and symmetric:
\[
p_q^+(z_\star)=p_q^-(z_\star)=\frac12,
\qquad
p_q^0(z_\star)=0.
\]
Hence channel $\mathsf A$ outputs $\pm1$ equiprobably. Since the same conditional
distribution is redrawn independently under channel $\mathsf R$, the repeat law is
uniform on $\{+1,-1\}^2$. The semantic triple follows from the definitions:
\[
\mu_q=0,\qquad \alpha_q=0,\qquad \tau_q=\Var(0)=0.
\]

For $M^{\mathrm{zero}}$, the latent response is always $0$, so channel
$\mathsf A$ is the point mass at $0$. Under channels $\mathsf B$ and $\mathsf R$,
every latent $0$ is completed by an independent fair coin, hence the binary law is
a fair coin and the repeat law is again uniform on $\{+1,-1\}^2$. The semantic
triple is
\[
\mu_q=0,\qquad \alpha_q=1,\qquad \tau_q=\Var(0)=0.
\]

For $M^{\mathrm{split}}$, the abstention law has no zero mass and is symmetric
between the two strict outcomes:
\[
\operatorname{Law}^{\mathsf A}_{M^{\mathrm{split}},q}
=
\frac12\delta_{+1}+\frac12\delta_{-1}.
\]
Because channel $\mathsf B$ resamples the latent type on each draw, its binary law
is also a fair coin. Under channel $\mathsf R$, however, the same latent type is
reused across the two repeated labels. Conditional on $z_+$ the pair is
deterministically $(+1,+1)$, and conditional on $z_-$ it is deterministically
$(-1,-1)$, so
\[
\operatorname{Law}^{\mathsf R}_{M^{\mathrm{split}},q}
=
\frac12\delta_{(+1,+1)}+\frac12\delta_{(-1,-1)}.
\]
The semantic triple is
\[
\mu_q
=
\frac12(1)+\frac12(-1)=0,
\qquad
\alpha_q=0,
\qquad
\tau_q
=
\Var(m_q(Z))
=
\Var(\pm1)
=
1.
\]

The equivalence and non-equivalence statements follow immediately from the computed
observable laws.
\end{proof}

\begin{proof}[Proof of Theorem~\ref{thm:app-channel-frontier}]
We prove the four equivalences first.

\paragraph{Step 1: Anonymous binary channel.}
Equation~\eqref{eq:app-frontier-B} is exactly
Corollary~\ref{cor:appendix-primitive-summaries}:
\[
M \equiv_{\mathsf B} M'
\iff
\mu(M)=\mu(M').
\]

\paragraph{Step 2: Combined channel \(\mathsf{BR}\).}
By Lemma~\ref{lem:app-combined-channel-factorization},
\[
M \equiv_{\mathsf{BR}} M'
\iff
\bigl(M \equiv_{\mathsf B} M' \text{ and } M \equiv_{\mathsf R} M'\bigr).
\]
Using Corollary~\ref{cor:appendix-primitive-summaries},
\[
M \equiv_{\mathsf B} M'
\iff
\mu(M)=\mu(M'),
\]
and
\[
M \equiv_{\mathsf R} M'
\iff
(\mu,\nu)(M)=(\mu,\nu)(M').
\]
Hence
\[
M \equiv_{\mathsf{BR}} M'
\iff
(\mu,\nu)(M)=(\mu,\nu)(M').
\]
Since $\tau=\nu-\mu^2$ coordinatewise, equality of $(\mu,\nu)$ is equivalent to
equality of $(\mu,\tau)$, proving \eqref{eq:app-frontier-BR}.

\paragraph{Step 3: Abstention channel.}
Equation~\eqref{eq:app-frontier-A} is again immediate from
Corollary~\ref{cor:appendix-primitive-summaries}:
\[
M \equiv_{\mathsf A} M'
\iff
(\mu,\alpha)(M)=(\mu,\alpha)(M').
\]

\paragraph{Step 4: Combined channel \(\mathsf{BRA}\).}
Applying Lemma~\ref{lem:app-combined-channel-factorization} twice,
\[
M \equiv_{\mathsf{BRA}} M'
\iff
\bigl(
M \equiv_{\mathsf B} M',
\;
M \equiv_{\mathsf R} M',
\;
M \equiv_{\mathsf A} M'
\bigr).
\]
Using the primitive-channel characterizations, this is equivalent to
\[
\mu(M)=\mu(M'),
\quad
(\mu,\nu)(M)=(\mu,\nu)(M'),
\quad
(\mu,\alpha)(M)=(\mu,\alpha)(M').
\]
Equivalently,
\[
(\mu,\nu,\alpha)(M)=(\mu,\nu,\alpha)(M').
\]
Since $\tau=\nu-\mu^2$ coordinatewise, this is equivalent to
\[
(\mu,\tau,\alpha)(M)=(\mu,\tau,\alpha)(M'),
\]
which proves \eqref{eq:app-frontier-BRA}.

It remains to prove the strict inclusions in
\eqref{eq:app-frontier-strict-inclusions}.

\paragraph{Step 5: Strictness of \(\equiv_{\mathsf{BR}} \subsetneq \equiv_{\mathsf B}\).}
By Lemma~\ref{lem:app-frontier-witness-models},
\[
M^{\mathrm{coin}} \equiv_{\mathsf B} M^{\mathrm{split}}
\]
but
\[
M^{\mathrm{coin}} \not\equiv_{\mathsf{BR}} M^{\mathrm{split}}.
\]
Hence
\[
\equiv_{\mathsf{BR}} \subsetneq \equiv_{\mathsf B}.
\]

\paragraph{Step 6: Strictness of \(\equiv_{\mathsf A} \subsetneq \equiv_{\mathsf B}\).}
Again by Lemma~\ref{lem:app-frontier-witness-models},
\[
M^{\mathrm{coin}} \equiv_{\mathsf B} M^{\mathrm{zero}}
\]
but
\[
M^{\mathrm{coin}} \not\equiv_{\mathsf A} M^{\mathrm{zero}}.
\]
Therefore
\[
\equiv_{\mathsf A} \subsetneq \equiv_{\mathsf B}.
\]

\paragraph{Step 7: Strictness of \(\equiv_{\mathsf{BRA}} \subsetneq \equiv_{\mathsf{BR}}\).}
By Lemma~\ref{lem:app-frontier-witness-models},
\[
M^{\mathrm{coin}} \equiv_{\mathsf{BR}} M^{\mathrm{zero}}
\]
but
\[
M^{\mathrm{coin}} \not\equiv_{\mathsf{BRA}} M^{\mathrm{zero}}.
\]
Hence
\[
\equiv_{\mathsf{BRA}} \subsetneq \equiv_{\mathsf{BR}}.
\]

\paragraph{Step 8: Strictness of \(\equiv_{\mathsf{BRA}} \subsetneq \equiv_{\mathsf A}\).}
Finally, Lemma~\ref{lem:app-frontier-witness-models} gives
\[
M^{\mathrm{coin}} \equiv_{\mathsf A} M^{\mathrm{split}}
\]
but
\[
M^{\mathrm{coin}} \not\equiv_{\mathsf{BRA}} M^{\mathrm{split}}.
\]
Therefore
\[
\equiv_{\mathsf{BRA}} \subsetneq \equiv_{\mathsf A}.
\]

Because these witness models live on a singleton query set, they witness the
strict inclusions already at the querywise level. By the embedding argument of
Lemma~\ref{lem:app-embedding-construction}, the same strictness extends to any
finite query set.
\end{proof}

\begin{corollary}[Residual semantic fibers under refined channels]
\label{cor:app-residual-fibers-refined}
Fix a model $M \in \mathcal M(\mathcal Q)$ and a query $q\in\mathcal Q$.

\textbf{(i) Under \(\mathsf{BR}\):}
\[
\mathcal S_q^{\mathsf{BR}}(M)
=
\left\{
\bigl(\mu_q(M),\tau_q(M),\alpha\bigr)
:
0 \le \alpha \le
\min\!\bigl\{1-|\mu_q(M)|,\,
1-\mu_q(M)^2-\tau_q(M)\bigr\}
\right\}.
\]

\textbf{(ii) Under \(\mathsf A\):}
\[
\mathcal S_q^{\mathsf A}(M)
=
\left\{
\bigl(\mu_q(M),\tau,\alpha_q(M)\bigr)
:
0 \le \tau \le 1-\alpha_q(M)-\mu_q(M)^2
\right\}.
\]

\textbf{(iii) Under \(\mathsf{BRA}\):}
\[
\mathcal S_q^{\mathsf{BRA}}(M)
=
\left\{
\bigl(\mu_q(M),\tau_q(M),\alpha_q(M)\bigr)
\right\}.
\]
\end{corollary}

\begin{proof}
Part (iii) follows immediately from
Theorem~\ref{thm:app-channel-frontier}: under $\mathsf{BRA}$ the full triple
$(\mu_q,\tau_q,\alpha_q)$ is point-identified.

For part (ii), Theorem~\ref{thm:app-channel-frontier} gives that under channel
$\mathsf A$ the quantities $\mu_q$ and $\alpha_q$ are fixed across the
observational equivalence class. The remaining feasible values of $\tau_q$ are
then exactly those allowed by the binary semantic-fiber characterization of
Theorem~\ref{thm:app-exact-binary-fiber}, namely
\[
0\le \tau \le 1-\alpha_q(M)-\mu_q(M)^2.
\]

For part (i), Theorem~\ref{thm:app-channel-frontier} gives that under channel
$\mathsf{BR}$ the quantities $\mu_q$ and $\tau_q$ are fixed across the
observational equivalence class. Theorem~\ref{thm:app-exact-binary-fiber} then
implies that a candidate $\alpha$ is admissible if and only if the triple
\[
\bigl(\mu_q(M),\tau_q(M),\alpha\bigr)
\]
lies in the binary semantic fiber. This is equivalent to the pair of inequalities
\[
\alpha \le 1-|\mu_q(M)|,
\qquad
\tau_q(M) \le 1-\alpha-\mu_q(M)^2,
\]
or equivalently
\[
0 \le \alpha \le
\min\!\bigl\{1-|\mu_q(M)|,\,
1-\mu_q(M)^2-\tau_q(M)\bigr\}.
\]
This proves the claim.
\end{proof}

\begin{remark}[Minimality of the two refinements]
\label{rem:app-minimality}
The two refinements considered in Section~4 recover complementary semantic
dimensions. Same-annotator repeats identify the between-type component of
disagreement $\tau_q$ but remain blind to non-binary mass $\alpha_q$; the
abstention channel identifies $\alpha_q$ but remains blind to $\tau_q$.
The combined channel $\mathsf{BRA}$ is therefore minimal, within this framework,
for point-identifying the full semantic triple $(\mu_q,\tau_q,\alpha_q)$.
\end{remark}

\section{Proofs for Finite-Sample Results}
\label{app:proofs-finite-sample}

All finite-sample statements in this appendix are taken with respect to the
channel-separated product experiment in \eqref{eq:appendix-product-experiment}.
The concentration bounds use only the boundedness of the channel outputs and
Hoeffding's classical inequality for averages of independent bounded random
variables \citep{hoeffding1963probability}.

Throughout, fix a finite query set $\mathcal Q$. For a query $q\in\mathcal Q$,
assume that the corresponding sample sizes satisfy
\[
n_{B,q}\ge 1,\qquad n_{R,q}\ge 1,\qquad n_{A,q}\ge 1
\]
whenever the associated estimator is defined. We write
\[
\mu_q:=\mu_q(M),\qquad \nu_q:=\nu_q(M),\qquad \tau_q:=\tau_q(M),\qquad
\alpha_q:=\alpha_q(M)
\]
for the true semantic parameters of the data-generating model
$M\in\mathcal M(\mathcal Q)$.

\subsection{Querywise estimators and exact finite-sample moments}
\label{app:proofs-finite-sample:estimators}

For each query $q\in\mathcal Q$, let
\[
Y^{B}_{q,1},\dots,Y^{B}_{q,n_{B,q}}
\stackrel{\text{i.i.d.}}{\sim}
\operatorname{Law}^{\mathsf B}_{M,q},
\]
\[
\bigl(Y^{R,(1)}_{q,1},Y^{R,(2)}_{q,1}\bigr),\dots,
\bigl(Y^{R,(1)}_{q,n_{R,q}},Y^{R,(2)}_{q,n_{R,q}}\bigr)
\stackrel{\text{i.i.d.}}{\sim}
\operatorname{Law}^{\mathsf R}_{M,q},
\]
and
\[
\widetilde Y^{A}_{q,1},\dots,\widetilde Y^{A}_{q,n_{A,q}}
\stackrel{\text{i.i.d.}}{\sim}
\operatorname{Law}^{\mathsf A}_{M,q},
\]
with the three channel blocks mutually independent, as specified by
\eqref{eq:appendix-product-experiment}.

Define the querywise estimators
\begin{equation}
\label{eq:app-estimators-mu}
\widehat\mu_q
:=
\frac{1}{n_{B,q}}
\sum_{i=1}^{n_{B,q}} Y^{B}_{q,i},
\end{equation}
\begin{equation}
\label{eq:app-estimators-nu}
\widehat\nu_q
:=
\frac{1}{n_{R,q}}
\sum_{i=1}^{n_{R,q}}
U_{q,i},
\qquad
U_{q,i}:=
Y^{R,(1)}_{q,i}Y^{R,(2)}_{q,i}\in\{-1,+1\},
\end{equation}
\begin{equation}
\label{eq:app-estimators-alpha}
\widehat\alpha_q
:=
\frac{1}{n_{A,q}}
\sum_{i=1}^{n_{A,q}} V_{q,i},
\qquad
V_{q,i}:=
\mathbf 1\{\widetilde Y^{A}_{q,i}=0\}\in\{0,1\},
\end{equation}
and the plug-in estimator
\begin{equation}
\label{eq:app-estimators-tau}
\widehat\tau_q
:=
\widehat\nu_q-\widehat\mu_q^2.
\end{equation}

When $n_{B,q}\ge 2$, it is occasionally useful to define the unbiased estimator of
$\mu_q^2$,
\begin{equation}
\label{eq:app-estimators-mu2-unbiased}
\widehat{\mu_q^2}^{\,\mathrm{unb}}
:=
\frac{n_{B,q}\widehat\mu_q^2-1}{n_{B,q}-1},
\end{equation}
and the bias-corrected estimator
\begin{equation}
\label{eq:app-estimators-tau-unbiased}
\widetilde\tau_q
:=
\widehat\nu_q-\widehat{\mu_q^2}^{\,\mathrm{unb}}.
\end{equation}

\begin{proposition}[Exact moments of the querywise estimators]
\label{prop:app-estimator-moments}
Fix a model $M\in\mathcal M(\mathcal Q)$ and a query $q\in\mathcal Q$.

\textbf{(i) First-order parameters.}
\[
\mathbb E[\widehat\mu_q]=\mu_q,
\qquad
\Var(\widehat\mu_q)=\frac{1-\mu_q^2}{n_{B,q}},
\]
\[
\mathbb E[\widehat\nu_q]=\nu_q,
\qquad
\Var(\widehat\nu_q)=\frac{1-\nu_q^2}{n_{R,q}},
\]
\[
\mathbb E[\widehat\alpha_q]=\alpha_q,
\qquad
\Var(\widehat\alpha_q)=\frac{\alpha_q(1-\alpha_q)}{n_{A,q}}.
\]

\textbf{(ii) Plug-in estimator of persistent disagreement.}
\[
\mathbb E[\widehat\tau_q]
=
\tau_q-\frac{1-\mu_q^2}{n_{B,q}}.
\]
Thus $\widehat\tau_q$ has an exact finite-sample downward bias equal to
$(1-\mu_q^2)/n_{B,q}$.

\textbf{(iii) Bias-corrected estimator.}
If $n_{B,q}\ge 2$, then
\[
\mathbb E\!\left[\widehat{\mu_q^2}^{\,\mathrm{unb}}\right]=\mu_q^2,
\qquad
\mathbb E[\widetilde\tau_q]=\tau_q.
\]
\end{proposition}

\begin{proof}
By Proposition~\ref{prop:appendix-observable-laws},
\[
Y^{B}_{q,i}\in\{-1,+1\},
\qquad
\mathbb E[Y^{B}_{q,i}]=\mu_q,
\qquad
\mathbb E[(Y^{B}_{q,i})^2]=1.
\]
Hence
\[
\Var(Y^{B}_{q,i})=1-\mu_q^2,
\]
and, since the binary samples are i.i.d.,
\[
\mathbb E[\widehat\mu_q]=\mu_q,
\qquad
\Var(\widehat\mu_q)=\frac{1-\mu_q^2}{n_{B,q}}.
\]

Similarly, each
\[
U_{q,i}=Y^{R,(1)}_{q,i}Y^{R,(2)}_{q,i}\in\{-1,+1\}
\]
satisfies
\[
\mathbb E[U_{q,i}]
=
\mathbb E[Y^{R,(1)}_{q,i}Y^{R,(2)}_{q,i}]
=
\nu_q
\]
by Proposition~\ref{prop:appendix-observable-laws}, and
\[
U_{q,i}^2=1
\quad\text{almost surely.}
\]
Therefore
\[
\Var(U_{q,i})=1-\nu_q^2,
\]
and
\[
\mathbb E[\widehat\nu_q]=\nu_q,
\qquad
\Var(\widehat\nu_q)=\frac{1-\nu_q^2}{n_{R,q}}.
\]

For the abstention channel,
\[
V_{q,i}=\mathbf 1\{\widetilde Y^{A}_{q,i}=0\}\in\{0,1\},
\qquad
\mathbb E[V_{q,i}]
=
\mathbb P(\widetilde Y^{A}_{q,i}=0)
=
\alpha_q
\]
by Theorem~\ref{thm:app-abstention-identification}. Hence
\[
\Var(V_{q,i})=\alpha_q(1-\alpha_q),
\]
and
\[
\mathbb E[\widehat\alpha_q]=\alpha_q,
\qquad
\Var(\widehat\alpha_q)=\frac{\alpha_q(1-\alpha_q)}{n_{A,q}}.
\]

We now prove part (ii). Since
\[
\widehat\tau_q=\widehat\nu_q-\widehat\mu_q^2,
\]
we have
\[
\mathbb E[\widehat\tau_q]
=
\mathbb E[\widehat\nu_q]-\mathbb E[\widehat\mu_q^2].
\]
Using
\[
\mathbb E[\widehat\mu_q^2]
=
\Var(\widehat\mu_q)+\bigl(\mathbb E[\widehat\mu_q]\bigr)^2
=
\frac{1-\mu_q^2}{n_{B,q}}+\mu_q^2,
\]
we obtain
\begin{align*}
\mathbb E[\widehat\tau_q]
&=
\nu_q-\left(\mu_q^2+\frac{1-\mu_q^2}{n_{B,q}}\right) \\
&=
(\nu_q-\mu_q^2)-\frac{1-\mu_q^2}{n_{B,q}} \\
&=
\tau_q-\frac{1-\mu_q^2}{n_{B,q}}.
\end{align*}

Finally, for part (iii), rearranging the identity
\[
\mathbb E[\widehat\mu_q^2]
=
\frac{1}{n_{B,q}}+\frac{n_{B,q}-1}{n_{B,q}}\mu_q^2
\]
gives
\[
\mathbb E\!\left[
\frac{n_{B,q}\widehat\mu_q^2-1}{n_{B,q}-1}
\right]
=
\mu_q^2.
\]
Thus
\[
\mathbb E\!\left[\widehat{\mu_q^2}^{\,\mathrm{unb}}\right]=\mu_q^2,
\]
and therefore
\[
\mathbb E[\widetilde\tau_q]
=
\mathbb E[\widehat\nu_q]
-
\mathbb E\!\left[\widehat{\mu_q^2}^{\,\mathrm{unb}}\right]
=
\nu_q-\mu_q^2
=
\tau_q.
\]
\end{proof}

\begin{remark}[Why the main text uses the plug-in estimator]
\label{rem:app-why-plugin}
The plug-in estimator \eqref{eq:app-estimators-tau} is algebraically simpler and
admits especially transparent concentration bounds. The bias-corrected estimator
\eqref{eq:app-estimators-tau-unbiased} is useful when exact unbiasedness matters,
but it is not needed for the main identification or sample-complexity statements.
\end{remark}

\subsection{Querywise concentration bounds}
\label{app:proofs-finite-sample:querywise-concentration}

We now prove finite-sample deviation bounds for the estimators above. The only
input is Hoeffding's inequality for averages of independent bounded random
variables \citep{hoeffding1963probability}.

\begin{theorem}[Querywise concentration]
\label{thm:app-querywise-concentration}
Fix a model $M\in\mathcal M(\mathcal Q)$ and a query $q\in\mathcal Q$.

\textbf{(i) Mean, second moment, and non-binary mass.}
For every $t_\mu,t_\nu,t_\alpha>0$,
\begin{equation}
\label{eq:app-concentration-mu}
\mathbb P\!\left(
|\widehat\mu_q-\mu_q|>t_\mu
\right)
\le
2\exp\!\left(-\frac{n_{B,q}t_\mu^2}{2}\right),
\end{equation}
\begin{equation}
\label{eq:app-concentration-nu}
\mathbb P\!\left(
|\widehat\nu_q-\nu_q|>t_\nu
\right)
\le
2\exp\!\left(-\frac{n_{R,q}t_\nu^2}{2}\right),
\end{equation}
\begin{equation}
\label{eq:app-concentration-alpha}
\mathbb P\!\left(
|\widehat\alpha_q-\alpha_q|>t_\alpha
\right)
\le
2\exp\!\left(-2n_{A,q}t_\alpha^2\right).
\end{equation}

\textbf{(ii) Persistent disagreement.}
For every $t_\mu,t_\nu>0$,
\begin{equation}
\label{eq:app-concentration-tau-general}
\mathbb P\!\left(
|\widehat\tau_q-\tau_q|>t_\nu+2t_\mu
\right)
\le
2\exp\!\left(-\frac{n_{R,q}t_\nu^2}{2}\right)
+
2\exp\!\left(-\frac{n_{B,q}t_\mu^2}{2}\right).
\end{equation}
In particular, for every $\varepsilon>0$,
\begin{equation}
\label{eq:app-concentration-tau-oneparam}
\mathbb P\!\left(
|\widehat\tau_q-\tau_q|>\varepsilon
\right)
\le
2\exp\!\left(-\frac{n_{R,q}\varepsilon^2}{8}\right)
+
2\exp\!\left(-\frac{n_{B,q}\varepsilon^2}{32}\right).
\end{equation}
\end{theorem}

\begin{proof}
Since each $Y^{B}_{q,i}\in[-1,1]$, Hoeffding's inequality gives
\[
\mathbb P\!\left(
|\widehat\mu_q-\mu_q|>t_\mu
\right)
\le
2\exp\!\left(
-\frac{2n_{B,q}t_\mu^2}{(1-(-1))^2}
\right)
=
2\exp\!\left(-\frac{n_{B,q}t_\mu^2}{2}\right),
\]
which proves \eqref{eq:app-concentration-mu}.

Exactly the same argument applies to
\[
U_{q,i}=Y^{R,(1)}_{q,i}Y^{R,(2)}_{q,i}\in[-1,1],
\]
whose mean is $\nu_q$, giving \eqref{eq:app-concentration-nu}.

For the abstention channel, the variables
\[
V_{q,i}=\mathbf 1\{\widetilde Y^{A}_{q,i}=0\}\in[0,1]
\]
have mean $\alpha_q$, so Hoeffding yields
\[
\mathbb P\!\left(
|\widehat\alpha_q-\alpha_q|>t_\alpha
\right)
\le
2\exp\!\left(-2n_{A,q}t_\alpha^2\right),
\]
proving \eqref{eq:app-concentration-alpha}.

We now prove \eqref{eq:app-concentration-tau-general}. Since
\[
\widehat\tau_q-\tau_q
=
(\widehat\nu_q-\nu_q)-(\widehat\mu_q^2-\mu_q^2),
\]
the triangle inequality gives
\[
|\widehat\tau_q-\tau_q|
\le
|\widehat\nu_q-\nu_q|
+
|\widehat\mu_q^2-\mu_q^2|.
\]
Moreover, since both $\widehat\mu_q$ and $\mu_q$ lie in $[-1,1]$,
\[
|\widehat\mu_q^2-\mu_q^2|
=
|\widehat\mu_q-\mu_q|\,|\widehat\mu_q+\mu_q|
\le
2|\widehat\mu_q-\mu_q|.
\]
Therefore, on the event
\[
\bigl\{|\widehat\nu_q-\nu_q|\le t_\nu\bigr\}
\cap
\bigl\{|\widehat\mu_q-\mu_q|\le t_\mu\bigr\},
\]
we have
\[
|\widehat\tau_q-\tau_q|
\le
t_\nu+2t_\mu.
\]
Taking complements and using a union bound,
\begin{align*}
\mathbb P\!\left(
|\widehat\tau_q-\tau_q|>t_\nu+2t_\mu
\right)
&\le
\mathbb P\!\left(
|\widehat\nu_q-\nu_q|>t_\nu
\right)
+
\mathbb P\!\left(
|\widehat\mu_q-\mu_q|>t_\mu
\right) \\
&\le
2\exp\!\left(-\frac{n_{R,q}t_\nu^2}{2}\right)
+
2\exp\!\left(-\frac{n_{B,q}t_\mu^2}{2}\right),
\end{align*}
which proves \eqref{eq:app-concentration-tau-general}.

Finally, setting
\[
t_\nu=\frac{\varepsilon}{2},
\qquad
t_\mu=\frac{\varepsilon}{4}
\]
in \eqref{eq:app-concentration-tau-general} yields
\eqref{eq:app-concentration-tau-oneparam}.
\end{proof}

\begin{corollary}[Querywise confidence intervals]
\label{cor:app-querywise-confidence-intervals}
Fix a model $M\in\mathcal M(\mathcal Q)$ and a query $q\in\mathcal Q$.
For $\delta\in(0,1)$, define
\[
r_{\mu,q}(\delta)
:=
\sqrt{\frac{2\log(2/\delta)}{n_{B,q}}},
\qquad
r_{\nu,q}(\delta)
:=
\sqrt{\frac{2\log(2/\delta)}{n_{R,q}}},
\qquad
r_{\alpha,q}(\delta)
:=
\sqrt{\frac{\log(2/\delta)}{2n_{A,q}}},
\]
and
\[
r_{\tau,q}(\delta)
:=
\sqrt{\frac{2\log(4/\delta)}{n_{R,q}}}
+
2\sqrt{\frac{2\log(4/\delta)}{n_{B,q}}}.
\]
Then
\[
\mathbb P\!\left(
\mu_q\in[\widehat\mu_q\pm r_{\mu,q}(\delta)]
\right)\ge 1-\delta,
\]
\[
\mathbb P\!\left(
\nu_q\in[\widehat\nu_q\pm r_{\nu,q}(\delta)]
\right)\ge 1-\delta,
\]
\[
\mathbb P\!\left(
\alpha_q\in[\widehat\alpha_q\pm r_{\alpha,q}(\delta)]
\right)\ge 1-\delta,
\]
and
\[
\mathbb P\!\left(
\tau_q\in[\widehat\tau_q\pm r_{\tau,q}(\delta)]
\right)\ge 1-\delta.
\]
\end{corollary}

\begin{proof}
The first three claims are obtained by solving the inequalities in
Theorem~\ref{thm:app-querywise-concentration} for the deviation radius.
For the interval for $\tau_q$, apply
\eqref{eq:app-concentration-tau-general} with failure probability split equally
between the $\nu$- and $\mu$-terms:
\[
t_\nu=\sqrt{\frac{2\log(4/\delta)}{n_{R,q}}},
\qquad
t_\mu=\sqrt{\frac{2\log(4/\delta)}{n_{B,q}}}.
\]
Then
\[
\mathbb P\!\left(
|\widehat\tau_q-\tau_q|>t_\nu+2t_\mu
\right)
\le
\frac{\delta}{2}+\frac{\delta}{2}
=
\delta.
\]
\end{proof}

\subsection{Uniform concentration over a finite query set}
\label{app:proofs-finite-sample:uniform}

We now pass from querywise bounds to simultaneous bounds over the finite query set
$\mathcal Q$.

\begin{theorem}[Uniform concentration over \(\mathcal Q\)]
\label{thm:app-uniform-concentration}
Fix a model $M\in\mathcal M(\mathcal Q)$ and a finite query set $\mathcal Q$.
Then for every $t_\mu,t_\nu,t_\alpha>0$,
\begin{equation}
\label{eq:app-uniform-mu}
\mathbb P\!\left(
\max_{q\in\mathcal Q}|\widehat\mu_q-\mu_q|>t_\mu
\right)
\le
\sum_{q\in\mathcal Q}
2\exp\!\left(-\frac{n_{B,q}t_\mu^2}{2}\right),
\end{equation}
\begin{equation}
\label{eq:app-uniform-nu}
\mathbb P\!\left(
\max_{q\in\mathcal Q}|\widehat\nu_q-\nu_q|>t_\nu
\right)
\le
\sum_{q\in\mathcal Q}
2\exp\!\left(-\frac{n_{R,q}t_\nu^2}{2}\right),
\end{equation}
\begin{equation}
\label{eq:app-uniform-alpha}
\mathbb P\!\left(
\max_{q\in\mathcal Q}|\widehat\alpha_q-\alpha_q|>t_\alpha
\right)
\le
\sum_{q\in\mathcal Q}
2\exp\!\left(-2n_{A,q}t_\alpha^2\right),
\end{equation}
and
\begin{equation}
\label{eq:app-uniform-tau-general}
\mathbb P\!\left(
\max_{q\in\mathcal Q}|\widehat\tau_q-\tau_q|>t_\nu+2t_\mu
\right)
\le
\sum_{q\in\mathcal Q}
\left[
2\exp\!\left(-\frac{n_{R,q}t_\nu^2}{2}\right)
+
2\exp\!\left(-\frac{n_{B,q}t_\mu^2}{2}\right)
\right].
\end{equation}
In particular, for every $\varepsilon>0$,
\begin{equation}
\label{eq:app-uniform-tau-oneparam}
\mathbb P\!\left(
\max_{q\in\mathcal Q}|\widehat\tau_q-\tau_q|>\varepsilon
\right)
\le
\sum_{q\in\mathcal Q}
\left[
2\exp\!\left(-\frac{n_{R,q}\varepsilon^2}{8}\right)
+
2\exp\!\left(-\frac{n_{B,q}\varepsilon^2}{32}\right)
\right].
\end{equation}
\end{theorem}

\begin{proof}
For \eqref{eq:app-uniform-mu}, define the bad event
\[
E_\mu
:=
\left\{
\max_{q\in\mathcal Q}|\widehat\mu_q-\mu_q|>t_\mu
\right\}
=
\bigcup_{q\in\mathcal Q}
\left\{
|\widehat\mu_q-\mu_q|>t_\mu
\right\}.
\]
A union bound and \eqref{eq:app-concentration-mu} yield
\[
\mathbb P(E_\mu)
\le
\sum_{q\in\mathcal Q}
\mathbb P\!\left(
|\widehat\mu_q-\mu_q|>t_\mu
\right)
\le
\sum_{q\in\mathcal Q}
2\exp\!\left(-\frac{n_{B,q}t_\mu^2}{2}\right).
\]
The proofs of \eqref{eq:app-uniform-nu} and \eqref{eq:app-uniform-alpha} are
identical.

For \eqref{eq:app-uniform-tau-general}, define
\[
E_\tau
:=
\left\{
\max_{q\in\mathcal Q}|\widehat\tau_q-\tau_q|>t_\nu+2t_\mu
\right\}
=
\bigcup_{q\in\mathcal Q}
\left\{
|\widehat\tau_q-\tau_q|>t_\nu+2t_\mu
\right\}.
\]
Applying a union bound and then \eqref{eq:app-concentration-tau-general} querywise
gives
\begin{align*}
\mathbb P(E_\tau)
&\le
\sum_{q\in\mathcal Q}
\mathbb P\!\left(
|\widehat\tau_q-\tau_q|>t_\nu+2t_\mu
\right) \\
&\le
\sum_{q\in\mathcal Q}
\left[
2\exp\!\left(-\frac{n_{R,q}t_\nu^2}{2}\right)
+
2\exp\!\left(-\frac{n_{B,q}t_\mu^2}{2}\right)
\right].
\end{align*}
Setting $t_\nu=\varepsilon/2$ and $t_\mu=\varepsilon/4$ yields
\eqref{eq:app-uniform-tau-oneparam}.
\end{proof}

\begin{corollary}[Balanced-allocation specialization]
\label{cor:app-balanced-allocation}
Assume that the sample sizes do not depend on the query:
\[
n_{B,q}\equiv n_B,\qquad
n_{R,q}\equiv n_R,\qquad
n_{A,q}\equiv n_A,
\qquad
Q:=|\mathcal Q|.
\]
Then for every $\varepsilon>0$,
\[
\mathbb P\!\left(
\max_{q\in\mathcal Q}|\widehat\mu_q-\mu_q|>\varepsilon
\right)
\le
2Q\exp\!\left(-\frac{n_B\varepsilon^2}{2}\right),
\]
\[
\mathbb P\!\left(
\max_{q\in\mathcal Q}|\widehat\alpha_q-\alpha_q|>\varepsilon
\right)
\le
2Q\exp\!\left(-2n_A\varepsilon^2\right),
\]
and
\[
\mathbb P\!\left(
\max_{q\in\mathcal Q}|\widehat\tau_q-\tau_q|>\varepsilon
\right)
\le
2Q\exp\!\left(-\frac{n_R\varepsilon^2}{8}\right)
+
2Q\exp\!\left(-\frac{n_B\varepsilon^2}{32}\right).
\]
\end{corollary}

\begin{proof}
Substitute constant sample sizes into
Theorem~\ref{thm:app-uniform-concentration}.
\end{proof}

\subsection{Finite-sample recovery of semantic regimes}
\label{app:proofs-finite-sample:regime-selection}

We now formalize the semantic regime-selection rule used in the experiments and in
the decision-theoretic discussion of Section~5.

\begin{definition}[Threshold-based semantic action rule]
\label{def:app-semantic-action-rule}
Fix thresholds
\[
t_\tau\in[0,1],
\qquad
t_\alpha\in[0,1].
\]
Define the population action map
\[
g_{t_\tau,t_\alpha}:[0,1]\times[0,1]\to
\{\mathsf{denoise},\mathsf{personalize},\mathsf{query}\}
\]
by
\[
g_{t_\tau,t_\alpha}(\tau,\alpha)
=
\begin{cases}
\mathsf{query}, & \alpha>t_\alpha,\\[3pt]
\mathsf{personalize}, & \alpha\le t_\alpha \text{ and } \tau>t_\tau,\\[3pt]
\mathsf{denoise}, & \alpha\le t_\alpha \text{ and } \tau\le t_\tau.
\end{cases}
\]
For each query $q\in\mathcal Q$, the oracle action is
\[
a_q^\star:=g_{t_\tau,t_\alpha}(\tau_q,\alpha_q),
\]
and the plug-in action is
\[
\widehat a_q:=g_{t_\tau,t_\alpha}(\widehat\tau_q,\widehat\alpha_q).
\]
\end{definition}

\begin{remark}[Why \(\alpha\) takes precedence]
\label{rem:app-alpha-precedence}
The rule in Definition~\ref{def:app-semantic-action-rule} assigns precedence to
large non-binary mass. This is the conservative choice: if the feedback channel
suggests that a substantial fraction of responses are non-binary, then the
appropriate semantics-aware action is to collect richer feedback or abstain before
committing to denoising or personalization.
\end{remark}

\begin{lemma}[Deterministic stability under a margin condition]
\label{lem:app-deterministic-stability}
Let
\begin{equation}
\label{eq:app-margin-definition}
\Gamma
:=
\min_{q\in\mathcal Q}
\min\!\left\{
|\tau_q-t_\tau|,
|\alpha_q-t_\alpha|
\right\}.
\end{equation}
Assume $\Gamma>0$. Then on the event
\begin{equation}
\label{eq:app-stability-event}
\left\{
\max_{q\in\mathcal Q}|\widehat\tau_q-\tau_q|\le \frac{\Gamma}{2}
\right\}
\cap
\left\{
\max_{q\in\mathcal Q}|\widehat\alpha_q-\alpha_q|\le \frac{\Gamma}{2}
\right\},
\end{equation}
we have
\[
\widehat a_q=a_q^\star
\qquad\text{for every } q\in\mathcal Q.
\]
\end{lemma}

\begin{proof}
Fix any query $q\in\mathcal Q$. By definition of $\Gamma$,
\[
|\tau_q-t_\tau|\ge \Gamma,
\qquad
|\alpha_q-t_\alpha|\ge \Gamma.
\]
On the event \eqref{eq:app-stability-event},
\[
|\widehat\tau_q-\tau_q|\le \frac{\Gamma}{2},
\qquad
|\widehat\alpha_q-\alpha_q|\le \frac{\Gamma}{2}.
\]
Hence $\widehat\tau_q$ lies on the same side of the threshold $t_\tau$ as $\tau_q$,
and $\widehat\alpha_q$ lies on the same side of the threshold $t_\alpha$ as
$\alpha_q$. Because the map $g_{t_\tau,t_\alpha}$ depends only on these two
threshold comparisons, it follows that
\[
g_{t_\tau,t_\alpha}(\widehat\tau_q,\widehat\alpha_q)
=
g_{t_\tau,t_\alpha}(\tau_q,\alpha_q),
\]
that is,
\[
\widehat a_q=a_q^\star.
\]
Since $q$ was arbitrary, the claim holds simultaneously for every query.
\end{proof}

\begin{theorem}[Finite-sample exact recovery of semantic actions]
\label{thm:app-finite-sample-regime-recovery}
Assume the margin condition \eqref{eq:app-margin-definition} holds with
$\Gamma>0$. Then
\begin{align}
\label{eq:app-regime-recovery-prob}
\mathbb P\!\left(
\exists q\in\mathcal Q:\widehat a_q\neq a_q^\star
\right)
&\le
\mathbb P\!\left(
\max_{q\in\mathcal Q}|\widehat\tau_q-\tau_q|>\frac{\Gamma}{2}
\right)
+
\mathbb P\!\left(
\max_{q\in\mathcal Q}|\widehat\alpha_q-\alpha_q|>\frac{\Gamma}{2}
\right) \\
\nonumber
&\le
\sum_{q\in\mathcal Q}
\left[
2\exp\!\left(-\frac{n_{R,q}\Gamma^2}{32}\right)
+
2\exp\!\left(-\frac{n_{B,q}\Gamma^2}{128}\right)
+
2\exp\!\left(-\frac{n_{A,q}\Gamma^2}{2}\right)
\right].
\end{align}
In particular, if the allocation is balanced across queries,
\[
n_{B,q}\equiv n_B,\qquad
n_{R,q}\equiv n_R,\qquad
n_{A,q}\equiv n_A,
\qquad
Q:=|\mathcal Q|,
\]
then
\begin{equation}
\label{eq:app-regime-recovery-balanced}
\mathbb P\!\left(
\exists q\in\mathcal Q:\widehat a_q\neq a_q^\star
\right)
\le
2Q\exp\!\left(-\frac{n_R\Gamma^2}{32}\right)
+
2Q\exp\!\left(-\frac{n_B\Gamma^2}{128}\right)
+
2Q\exp\!\left(-\frac{n_A\Gamma^2}{2}\right).
\end{equation}
\end{theorem}

\begin{proof}
By Lemma~\ref{lem:app-deterministic-stability},
\[
\left\{
\exists q\in\mathcal Q:\widehat a_q\neq a_q^\star
\right\}
\subseteq
\left\{
\max_{q\in\mathcal Q}|\widehat\tau_q-\tau_q|>\frac{\Gamma}{2}
\right\}
\cup
\left\{
\max_{q\in\mathcal Q}|\widehat\alpha_q-\alpha_q|>\frac{\Gamma}{2}
\right\}.
\]
Taking probabilities yields the first inequality in
\eqref{eq:app-regime-recovery-prob}.

For the second inequality, apply
Theorem~\ref{thm:app-uniform-concentration} with $\varepsilon=\Gamma/2$ in the
one-parameter bound \eqref{eq:app-uniform-tau-oneparam}:
\[
\mathbb P\!\left(
\max_{q\in\mathcal Q}|\widehat\tau_q-\tau_q|>\frac{\Gamma}{2}
\right)
\le
\sum_{q\in\mathcal Q}
\left[
2\exp\!\left(-\frac{n_{R,q}\Gamma^2}{32}\right)
+
2\exp\!\left(-\frac{n_{B,q}\Gamma^2}{128}\right)
\right].
\]
Likewise, \eqref{eq:app-uniform-alpha} with $t_\alpha=\Gamma/2$ gives
\[
\mathbb P\!\left(
\max_{q\in\mathcal Q}|\widehat\alpha_q-\alpha_q|>\frac{\Gamma}{2}
\right)
\le
\sum_{q\in\mathcal Q}
2\exp\!\left(-\frac{n_{A,q}\Gamma^2}{2}\right).
\]
Adding the two bounds proves \eqref{eq:app-regime-recovery-prob}. The balanced
specialization \eqref{eq:app-regime-recovery-balanced} follows immediately.
\end{proof}

\begin{corollary}[Sufficient sample sizes for exact recovery]
\label{cor:app-regime-recovery-sample-complexity}
Under the balanced allocation of
Theorem~\ref{thm:app-finite-sample-regime-recovery}, it is sufficient for
\[
\mathbb P\!\left(
\exists q\in\mathcal Q:\widehat a_q\neq a_q^\star
\right)\le \delta
\]
that
\[
n_R
\ge
\frac{32}{\Gamma^2}\log\!\left(\frac{6Q}{\delta}\right),
\qquad
n_B
\ge
\frac{128}{\Gamma^2}\log\!\left(\frac{6Q}{\delta}\right),
\qquad
n_A
\ge
\frac{2}{\Gamma^2}\log\!\left(\frac{6Q}{\delta}\right).
\]
\end{corollary}

\begin{proof}
If each of the three terms on the right-hand side of
\eqref{eq:app-regime-recovery-balanced} is at most $\delta/3$, then their sum is
at most $\delta$. Solving
\[
2Q\exp\!\left(-\frac{n_R\Gamma^2}{32}\right)\le \frac{\delta}{3},
\qquad
2Q\exp\!\left(-\frac{n_B\Gamma^2}{128}\right)\le \frac{\delta}{3},
\qquad
2Q\exp\!\left(-\frac{n_A\Gamma^2}{2}\right)\le \frac{\delta}{3}
\]
for $n_R,n_B,n_A$ yields the claimed sufficient conditions.
\end{proof}

\begin{remark}[Interpretation of the rates]
\label{rem:app-rates-interpretation}
The sample-complexity scaling in
Corollary~\ref{cor:app-regime-recovery-sample-complexity} is exactly what one
would expect from the structure of the problem. The dependence on
$\Gamma^{-2}$ is the usual bounded-observation threshold-testing rate; the
dependence on $\log Q$ is the price of simultaneous control over a finite query
set; and the asymmetry between the constants for $n_R$ and $n_A$ reflects that
$\widehat\tau_q$ is estimated through a composite moment identity involving both
repeat and binary samples, whereas $\widehat\alpha_q$ is directly observable under
the abstention channel.
\end{remark}

\section{Strong Extensions}
\label{app:extensions}

This appendix records three extensions of the main framework. They are not needed
for the core identifiability frontier, but they show that the channel-theoretic
view developed in the paper extends naturally to richer elicitation protocols.
Conceptually, these extensions connect our analysis to two recurring themes in
recent work: first, richer-than-binary feedback can be necessary for identifying
latent heterogeneity \citep{chidambaram2025direct}; second, empirical preference
datasets often conflate ties, refusals, and other forms of underspecification
\citep{zhang2025diverging}. The results below formalize these phenomena inside the
same latent-semantic framework.

\subsection{Asymmetric and unknown completion of non-binary responses}
\label{app:extensions:asymmetric-completion}

The main paper uses the fair completion rule because it isolates the semantic
collapse phenomenon in its cleanest form. We now generalize the anonymous binary
channel by allowing latent zero responses to be completed asymmetrically.

\begin{definition}[Asymmetric completion channel]
\label{def:app-asymmetric-completion}
Fix a query $q\in\mathcal Q$ and a completion probability
$\lambda_q\in[0,1]$. The \emph{asymmetric completion channel}
$\mathsf B_{\lambda}$ is defined exactly as the anonymous binary channel
$\mathsf B$, except that whenever the latent response satisfies
$\widetilde Y=0$, the observed binary output is
\[
Y^{(\lambda)}=
\begin{cases}
+1, & \text{with probability } \lambda_q,\\
-1, & \text{with probability } 1-\lambda_q.
\end{cases}
\]
Write
\[
c_q:=2\lambda_q-1\in[-1,1].
\]
\end{definition}

\begin{proposition}[Observable law under asymmetric completion]
\label{prop:app-asymmetric-completion-law}
Fix a model $M\in\mathcal M(\mathcal Q)$ and a query $q\in\mathcal Q$. Under the
channel $\mathsf B_{\lambda}$,
\begin{equation}
\label{eq:app-asymmetric-beta}
\mathbb E[Y^{(\lambda)}\mid q]
=
\beta_q^{(\lambda)}(M)
:=
\mu_q(M)+c_q\,\alpha_q(M),
\end{equation}
and therefore
\begin{equation}
\label{eq:app-asymmetric-binary-law}
\mathbb P(Y^{(\lambda)}=+1\mid q)
=
\frac{1+\beta_q^{(\lambda)}(M)}{2}.
\end{equation}
\end{proposition}

\begin{proof}
Conditional on $(Z=z,q)$, the strict outcomes contribute the directional mean
$m_q(z)$ exactly as before, while a latent zero contributes conditional mean
\[
(+1)\lambda_q + (-1)(1-\lambda_q)=2\lambda_q-1=c_q.
\]
Hence
\[
\mathbb E[Y^{(\lambda)}\mid Z=z,q]
=
m_q(z)+c_q\,a_q(z).
\]
Taking expectations over $Z\sim\pi$ gives
\[
\mathbb E[Y^{(\lambda)}\mid q]
=
\mathbb E[m_q(Z)]+c_q\,\mathbb E[a_q(Z)]
=
\mu_q(M)+c_q\,\alpha_q(M),
\]
which proves \eqref{eq:app-asymmetric-beta}. Since
$Y^{(\lambda)}\in\{-1,+1\}$, the binary law is determined by its mean, giving
\eqref{eq:app-asymmetric-binary-law}.
\end{proof}

The next theorem is the exact analogue of Theorem~\ref{thm:app-exact-binary-fiber}
for a known asymmetric completion rule.

\begin{theorem}[Exact semantic fiber under known completion bias]
\label{thm:app-asymmetric-exact-fiber}
Fix a query $q\in\mathcal Q$, a known completion probability $\lambda_q\in[0,1]$,
and let $c_q=2\lambda_q-1$. For every model $M\in\mathcal M(\mathcal Q)$, the
set of semantic triples compatible with the same $\mathsf B_{\lambda}$ law at
query $q$ is exactly
\begin{equation}
\label{eq:app-asymmetric-exact-fiber}
\left\{
(\mu,\tau,\alpha)\in\mathbb R^3:
\begin{array}{l}
0\le \alpha \le 1-|\mu|,\\
0\le \tau \le 1-\alpha-\mu^2,\\
\mu + c_q \alpha = \beta_q^{(\lambda)}(M)
\end{array}
\right\}.
\end{equation}
Equivalently, under known completion bias, the channel point-identifies only the
modified binary mean $\beta_q^{(\lambda)}=\mu_q+c_q\alpha_q$.
\end{theorem}

\begin{proof}
Fix $M$ and $q$.

\paragraph{Necessity.}
Let $(\mu,\tau,\alpha)$ be the semantic triple of any model $M'$ satisfying
\[
M' \equiv_{\mathsf B_{\lambda}} M.
\]
By Proposition~\ref{prop:app-asymmetric-completion-law},
observational equivalence under $\mathsf B_{\lambda}$ implies equality of the
modified binary means:
\[
\mu + c_q \alpha
=
\beta_q^{(\lambda)}(M').
=
\beta_q^{(\lambda)}(M).
\]
Moreover, $(\mu,\tau,\alpha)$ must satisfy the feasibility constraints from
Theorem~\ref{thm:app-exact-binary-fiber}, namely
\[
0\le \alpha \le 1-|\mu|,
\qquad
0\le \tau \le 1-\alpha-\mu^2.
\]
Hence every observationally equivalent model lies in the set displayed in
\eqref{eq:app-asymmetric-exact-fiber}.

\paragraph{Sufficiency.}
Conversely, fix any triple $(\mu,\tau,\alpha)$ satisfying the constraints in
\eqref{eq:app-asymmetric-exact-fiber}. By
Theorem~\ref{thm:app-exact-binary-fiber}, there exists a model $M'$ whose
querywise semantic triple is exactly $(\mu,\tau,\alpha)$. Applying
Proposition~\ref{prop:app-asymmetric-completion-law} to $M'$ yields
\[
\beta_q^{(\lambda)}(M')
=
\mu + c_q\alpha
=
\beta_q^{(\lambda)}(M).
\]
Since a binary law is determined by its mean,
\[
M' \equiv_{\mathsf B_{\lambda}} M.
\]
This proves \eqref{eq:app-asymmetric-exact-fiber}.
\end{proof}

\begin{corollary}[Identification deteriorates away from fair completion]
\label{cor:app-asymmetric-identification-loss}
Fix a query $q\in\mathcal Q$.

\textbf{(i) Known asymmetric completion.}
If $c_q\neq 0$ and $|\beta_q^{(\lambda)}(M)|<1$, then neither $\mu_q$ nor
$\alpha_q$ is point-identified under $\mathsf B_{\lambda}$.

\textbf{(ii) Unknown completion bias.}
Suppose now that the completion rule is not known and may vary across
observationally equivalent models. Let $\beta_q$ denote the observed binary mean.
Then the set of compatible semantic triples is exactly
\begin{equation}
\label{eq:app-unknown-completion-fiber}
\left\{
(\mu,\tau,\alpha)\in\mathbb R^3:
\begin{array}{l}
0\le \alpha \le 1-|\mu|,\\
0\le \tau \le 1-\alpha-\mu^2,\\
|\beta_q-\mu|\le \alpha
\end{array}
\right\}.
\end{equation}
In particular, if $|\beta_q|<1$, then even the strict-preference mean $\mu_q$ is
not point-identified from the anonymous binary law.
\end{corollary}

\begin{proof}
For part~(i), write
\[
\beta^\star:=\beta_q^{(\lambda)}(M).
\]
Because $|\beta^\star|<1$, we have $1-|\beta^\star|>0$. Consider first the triple
\[
(\mu,\tau,\alpha)=(\beta^\star,0,0),
\]
which is feasible and satisfies $\mu+c_q\alpha=\beta^\star$. Next choose any
$t>0$ sufficiently small that
\[
t \le 1-|\beta^\star-c_q t|.
\]
Such a $t$ exists by continuity because the inequality is strict at $t=0$.
Then the triple
\[
(\mu,\tau,\alpha)=(\beta^\star-c_q t,0,t)
\]
is also feasible and satisfies
\[
\mu + c_q \alpha
=
(\beta^\star-c_q t)+c_q t
=
\beta^\star.
\]
By Theorem~\ref{thm:app-asymmetric-exact-fiber}, both triples are compatible with
the same $\mathsf B_{\lambda}$ law, but they have different values of $\mu$ and
$\alpha$. Hence neither parameter is point-identified.

For part~(ii), if the completion rule is unknown, then a semantic triple
$(\mu,\tau,\alpha)$ is compatible with the observed binary mean $\beta_q$ if and
only if there exists some $c\in[-1,1]$ such that
\[
\beta_q=\mu+c\alpha,
\]
together with the feasibility constraints
\[
0\le \alpha \le 1-|\mu|,
\qquad
0\le \tau \le 1-\alpha-\mu^2.
\]
If $\alpha=0$, this condition reduces to $\beta_q=\mu$. If $\alpha>0$, it is
equivalent to
\[
\left|\frac{\beta_q-\mu}{\alpha}\right|\le 1,
\]
that is,
\[
|\beta_q-\mu|\le \alpha.
\]
This proves \eqref{eq:app-unknown-completion-fiber}.

Finally, if $|\beta_q|<1$, then the two feasible triples
\[
(\mu,\tau,\alpha)=(\beta_q,0,0)
\qquad\text{and}\qquad
(\mu,\tau,\alpha)=(\beta_q,0,t)
\]
with any sufficiently small $t>0$ satisfying $t\le 1-|\beta_q|$ both obey
\eqref{eq:app-unknown-completion-fiber}, and the second also permits nearby
$\mu\neq \beta_q$ as long as $|\beta_q-\mu|\le t$. Thus $\mu_q$ is not
point-identified.
\end{proof}

\subsection{Separating tie mass from incomparability mass}
\label{app:extensions:tie-vs-incomparability}

The main paper deliberately treats the zero state as protocol-agnostic non-binary
mass. We now refine that state into two distinct categories.

\begin{definition}[Four-valued latent semantic model]
\label{def:app-four-valued-model}
Fix a finite query set $\mathcal Q$. A \emph{four-valued latent semantic model} is
defined exactly as in Definition~\ref{def:appendix-latent-model}, except that for
each query $q\in\mathcal Q$ we specify measurable maps
\[
a_q^{\mathrm{tie}}:\mathcal Z\to[0,1],
\qquad
a_q^{\mathrm{inc}}:\mathcal Z\to[0,1],
\qquad
m_q:\mathcal Z\to[-1,1],
\]
satisfying
\[
|m_q(z)|
\le
1-a_q^{\mathrm{tie}}(z)-a_q^{\mathrm{inc}}(z)
\qquad
\text{for every } z\in\mathcal Z.
\]
Conditional on $(Z=z,q)$, the latent response
\[
\widetilde Y^\sharp \in \{+1,0_{\mathrm{tie}},0_{\mathrm{inc}},-1\}
\]
has probabilities
\[
p_q^{\mathrm{tie}}(z)=a_q^{\mathrm{tie}}(z),
\qquad
p_q^{\mathrm{inc}}(z)=a_q^{\mathrm{inc}}(z),
\]
\[
p_q^{+}(z)
=
\frac{1-a_q^{\mathrm{tie}}(z)-a_q^{\mathrm{inc}}(z)+m_q(z)}{2},
\qquad
p_q^{-}(z)
=
\frac{1-a_q^{\mathrm{tie}}(z)-a_q^{\mathrm{inc}}(z)-m_q(z)}{2}.
\]

Define the aggregate masses
\[
\alpha_q^{\mathrm{tie}}
:=
\mathbb E[a_q^{\mathrm{tie}}(Z)],
\qquad
\alpha_q^{\mathrm{inc}}
:=
\mathbb E[a_q^{\mathrm{inc}}(Z)],
\]
their total
\[
\alpha_q^\Sigma
:=
\alpha_q^{\mathrm{tie}}+\alpha_q^{\mathrm{inc}},
\]
and, as before,
\[
\mu_q:=\mathbb E[m_q(Z)],
\qquad
\tau_q:=\Var(m_q(Z)).
\]
\end{definition}

We study two channels built from this four-valued latent response.

\begin{definition}[Merged and separated four-valued abstention channels]
\label{def:app-four-valued-channels}
Fix a four-valued latent semantic model.

\paragraph{Merged channel \(\mathsf A^\Sigma\).}
Observe
\[
\widetilde Y^\Sigma \in \{+1,0,-1\},
\]
where both $0_{\mathrm{tie}}$ and $0_{\mathrm{inc}}$ are merged into the single
symbol $0$.

\paragraph{Separated channel \(\mathsf A^\sharp\).}
Observe the full four-valued response
\[
\widetilde Y^\sharp \in \{+1,0_{\mathrm{tie}},0_{\mathrm{inc}},-1\}.
\]
\end{definition}

\begin{proposition}[Merged four-valued abstention identifies only total non-binary mass]
\label{prop:app-merged-four-valued-channel}
Fix a four-valued latent semantic model $M^\sharp$ and a query $q\in\mathcal Q$.
Under the merged channel $\mathsf A^\Sigma$,
\[
\mathbb P(\widetilde Y^\Sigma=0\mid q)=\alpha_q^\Sigma,
\qquad
\mathbb P(\widetilde Y^\Sigma=\pm1\mid q)=\frac{1-\alpha_q^\Sigma\pm\mu_q}{2}.
\]
Hence the merged channel point-identifies $(\mu_q,\alpha_q^\Sigma)$ but does not
separate $\alpha_q^{\mathrm{tie}}$ from $\alpha_q^{\mathrm{inc}}$.

More precisely, if we define the merged semantic fiber by
\[
\mathcal S_{q,\sharp}^{\mathsf A^\Sigma}(M^\sharp)
:=
\left\{
\bigl(\mu_q(M'),\tau_q(M'),\alpha_q^{\mathrm{tie}}(M'),\alpha_q^{\mathrm{inc}}(M')\bigr)
:
M' \equiv_{\mathsf A^\Sigma} M^\sharp
\right\},
\]
then
\begin{equation}
\label{eq:app-merged-four-valued-fiber}
\mathcal S_{q,\sharp}^{\mathsf A^\Sigma}(M^\sharp)
=
\left\{
(\mu^\star,\tau,\alpha^{\mathrm{tie}},\alpha^{\mathrm{inc}}):
\begin{array}{l}
\alpha^{\mathrm{tie}},\alpha^{\mathrm{inc}}\ge 0,\\
\alpha^{\mathrm{tie}}+\alpha^{\mathrm{inc}}=\alpha_\Sigma^\star,\\
0\le \tau \le 1-\alpha_\Sigma^\star-(\mu^\star)^2
\end{array}
\right\},
\end{equation}
where
\[
\mu^\star:=\mu_q(M^\sharp),
\qquad
\alpha_\Sigma^\star:=\alpha_q^\Sigma(M^\sharp).
\]
\end{proposition}

\begin{proof}
The merged observable law is immediate from Definition~\ref{def:app-four-valued-model}:
\[
\mathbb P(\widetilde Y^\Sigma=0\mid q)
=
\mathbb E[a_q^{\mathrm{tie}}(Z)+a_q^{\mathrm{inc}}(Z)]
=
\alpha_q^\Sigma,
\]
and
\[
\mathbb P(\widetilde Y^\Sigma=\pm1\mid q)
=
\frac{1-\alpha_q^\Sigma\pm\mu_q}{2}.
\]
Thus the merged channel law depends only on $\mu_q$ and the total non-binary mass
$\alpha_q^\Sigma$.

To prove \eqref{eq:app-merged-four-valued-fiber}, fix any observationally
equivalent model $M'$. Equality of the merged trinomial law forces
\[
\mu_q(M')=\mu^\star,
\qquad
\alpha_q^{\mathrm{tie}}(M')+\alpha_q^{\mathrm{inc}}(M')=\alpha_\Sigma^\star.
\]
Since the total non-binary mass in $M'$ is $\alpha_\Sigma^\star$, the feasibility
bound for the corresponding ternary coarse-grained model gives
\[
0\le \tau_q(M')\le 1-\alpha_\Sigma^\star-(\mu^\star)^2.
\]
This proves the inclusion ``$\subseteq$''.

Conversely, fix any quadruple
\[
(\mu^\star,\tau,\alpha^{\mathrm{tie}},\alpha^{\mathrm{inc}})
\]
satisfying the constraints in \eqref{eq:app-merged-four-valued-fiber}. Set
\[
\alpha^\Sigma:=\alpha^{\mathrm{tie}}+\alpha^{\mathrm{inc}}=\alpha_\Sigma^\star.
\]
By Theorem~\ref{thm:app-exact-binary-fiber}, there exists a ternary one-query
model with semantic triple $(\mu^\star,\tau,\alpha^\Sigma)$. To refine it into a
four-valued model with the desired split of zero mass, it suffices to replace each
latent zero state by either $0_{\mathrm{tie}}$ or $0_{\mathrm{inc}}$ in the
required proportions.

More explicitly, the homogeneous construction is obtained by a one-point latent
space with
\[
a_q^{\mathrm{tie}}=\alpha^{\mathrm{tie}},
\qquad
a_q^{\mathrm{inc}}=\alpha^{\mathrm{inc}},
\qquad
m_q=\mu^\star,
\]
and the extremal construction is obtained by four deterministic latent states
corresponding to $+1$, $0_{\mathrm{tie}}$, $0_{\mathrm{inc}}$, and $-1$ with
probabilities
\[
\frac{1-\alpha^\Sigma+\mu^\star}{2},
\qquad
\alpha^{\mathrm{tie}},
\qquad
\alpha^{\mathrm{inc}},
\qquad
\frac{1-\alpha^\Sigma-\mu^\star}{2}.
\]
Mixing these two constructions exactly as in
Lemma~\ref{lem:app-mixture-interpolation} realizes every
\[
0\le \tau \le 1-\alpha^\Sigma-(\mu^\star)^2.
\]
The resulting merged channel law has the same $(\mu^\star,\alpha_\Sigma^\star)$ as
$M^\sharp$, so the new model is $\mathsf A^\Sigma$-equivalent to $M^\sharp$.
This proves the inclusion ``$\supseteq$''.
\end{proof}

\begin{proposition}[Separated four-valued abstention identifies the split of non-binary mass]
\label{prop:app-separated-four-valued-channel}
Fix a four-valued latent semantic model $M^\sharp$ and a query $q\in\mathcal Q$.
Under the fully separated channel $\mathsf A^\sharp$,
\[
\mathbb P(\widetilde Y^\sharp=0_{\mathrm{tie}}\mid q)=\alpha_q^{\mathrm{tie}},
\qquad
\mathbb P(\widetilde Y^\sharp=0_{\mathrm{inc}}\mid q)=\alpha_q^{\mathrm{inc}},
\]
and
\[
\mathbb P(\widetilde Y^\sharp=\pm1\mid q)
=
\frac{1-\alpha_q^{\mathrm{tie}}-\alpha_q^{\mathrm{inc}}\pm\mu_q}{2}.
\]
Hence the channel point-identifies
$(\mu_q,\alpha_q^{\mathrm{tie}},\alpha_q^{\mathrm{inc}})$.

Moreover, if we define the separated semantic fiber by
\[
\mathcal S_{q,\sharp}^{\mathsf A^\sharp}(M^\sharp)
:=
\left\{
\bigl(\mu_q(M'),\tau_q(M'),\alpha_q^{\mathrm{tie}}(M'),\alpha_q^{\mathrm{inc}}(M')\bigr)
:
M' \equiv_{\mathsf A^\sharp} M^\sharp
\right\},
\]
then
\begin{equation}
\label{eq:app-separated-four-valued-fiber}
\mathcal S_{q,\sharp}^{\mathsf A^\sharp}(M^\sharp)
=
\left\{
(\mu^\star,\tau,\alpha_{\mathrm{tie}}^\star,\alpha_{\mathrm{inc}}^\star):
0\le \tau \le
1-\alpha_{\mathrm{tie}}^\star-\alpha_{\mathrm{inc}}^\star-(\mu^\star)^2
\right\},
\end{equation}
where
\[
\mu^\star:=\mu_q(M^\sharp),
\qquad
\alpha_{\mathrm{tie}}^\star:=\alpha_q^{\mathrm{tie}}(M^\sharp),
\qquad
\alpha_{\mathrm{inc}}^\star:=\alpha_q^{\mathrm{inc}}(M^\sharp).
\]
\end{proposition}

\begin{proof}
The displayed observable laws are immediate from
Definition~\ref{def:app-four-valued-model}. Therefore equality of the
$\mathsf A^\sharp$ law forces
\[
\mu_q(M')=\mu^\star,
\qquad
\alpha_q^{\mathrm{tie}}(M')=\alpha_{\mathrm{tie}}^\star,
\qquad
\alpha_q^{\mathrm{inc}}(M')=\alpha_{\mathrm{inc}}^\star
\]
for every observationally equivalent model $M'$. Since the total non-binary mass
is fixed at
\[
\alpha_{\mathrm{tie}}^\star+\alpha_{\mathrm{inc}}^\star,
\]
the feasibility bound gives
\[
0\le \tau_q(M')
\le
1-\alpha_{\mathrm{tie}}^\star-\alpha_{\mathrm{inc}}^\star-(\mu^\star)^2.
\]
This proves the inclusion ``$\subseteq$'' in
\eqref{eq:app-separated-four-valued-fiber}.

Conversely, fix any
\[
0\le \tau \le
1-\alpha_{\mathrm{tie}}^\star-\alpha_{\mathrm{inc}}^\star-(\mu^\star)^2.
\]
The same homogeneous/extremal mixture construction used in the proof of
Proposition~\ref{prop:app-merged-four-valued-channel} yields a four-valued model
whose semantic tuple is
\[
(\mu^\star,\tau,\alpha_{\mathrm{tie}}^\star,\alpha_{\mathrm{inc}}^\star).
\]
Because the separated channel law is determined exactly by these three
point-identified first-order quantities, the constructed model is
$\mathsf A^\sharp$-equivalent to $M^\sharp$. This proves
\eqref{eq:app-separated-four-valued-fiber}.
\end{proof}

Together, Propositions~\ref{prop:app-merged-four-valued-channel} and
\ref{prop:app-separated-four-valued-channel} make precise what is lost by merging
distinct non-binary outcomes into a single abstention symbol: the merged channel
retains only the total mass, while the separated channel identifies the split but
still remains blind to the persistent component $\tau_q$.

\subsection{Repeated abstention and higher-order semantic moments}
\label{app:extensions:repeated-abstention}

The main paper uses repeated binary labels to identify the variance of the latent
directional mean. If abstention-enabled labels are repeated from the same latent
type and observed directly, then the same argument identifies higher-order moments
of the latent semantic vector.

\begin{definition}[Repeated abstention channel]
\label{def:app-repeated-abstention}
Fix a model $M\in\mathcal M(\mathcal Q)$ and a query $q\in\mathcal Q$. The
\emph{repeated abstention channel} $\mathsf A_2$ draws a single latent type
$Z\sim\pi$ and then draws two conditionally independent latent responses
\[
\widetilde Y^{(1)},\widetilde Y^{(2)}\in\{+1,0,-1\}
\]
from the same ternary kernel, both of which are observed directly.
Define the corresponding zero indicators
\[
I^{(1)}:=\mathbf 1\{\widetilde Y^{(1)}=0\},
\qquad
I^{(2)}:=\mathbf 1\{\widetilde Y^{(2)}=0\}.
\]
\end{definition}

\begin{definition}[Higher-order semantic moments]
\label{def:app-higher-order-semantic-moments}
For each query $q\in\mathcal Q$, define
\[
\kappa_q:=\Var(a_q(Z)),
\qquad
\rho_q:=\Cov(a_q(Z),m_q(Z)).
\]
Thus $\kappa_q$ is the between-type variance of non-binary propensity and
$\rho_q$ is the between-type covariance between non-binary propensity and
directional preference.
\end{definition}

\begin{proposition}[Repeated abstention identifies the covariance structure of \((m_q(Z),a_q(Z))\)]
\label{prop:app-repeated-abstention-covariance}
Fix a model $M\in\mathcal M(\mathcal Q)$ and a query $q\in\mathcal Q$. Under the
repeated abstention channel $\mathsf A_2$,
\begin{equation}
\label{eq:app-repeated-abstention-tau}
\Cov(\widetilde Y^{(1)},\widetilde Y^{(2)}\mid q)=\tau_q,
\end{equation}
\begin{equation}
\label{eq:app-repeated-abstention-kappa}
\Cov(I^{(1)},I^{(2)}\mid q)=\kappa_q,
\end{equation}
and
\begin{equation}
\label{eq:app-repeated-abstention-rho}
\Cov(I^{(1)},\widetilde Y^{(2)}\mid q)
=
\Cov(\widetilde Y^{(1)},I^{(2)}\mid q)
=
\rho_q.
\end{equation}
Consequently, the repeated abstention channel point-identifies the full
covariance matrix
\[
\Sigma_q
:=
\Cov
\begin{pmatrix}
m_q(Z)\\
a_q(Z)
\end{pmatrix}
=
\begin{pmatrix}
\tau_q & \rho_q\\
\rho_q & \kappa_q
\end{pmatrix}.
\]
\end{proposition}

\begin{proof}
Conditional on $(Z,q)$, the two repeated latent responses are independent by
Definition~\ref{def:app-repeated-abstention}. Moreover,
\[
\mathbb E[\widetilde Y^{(k)}\mid Z,q]=m_q(Z),
\qquad
\mathbb E[I^{(k)}\mid Z,q]=a_q(Z),
\qquad
k\in\{1,2\}.
\]

For \eqref{eq:app-repeated-abstention-tau}, we have
\[
\mathbb E[\widetilde Y^{(1)}\widetilde Y^{(2)}\mid Z,q]
=
\mathbb E[\widetilde Y^{(1)}\mid Z,q]\,
\mathbb E[\widetilde Y^{(2)}\mid Z,q]
=
m_q(Z)^2.
\]
Taking expectations gives
\[
\mathbb E[\widetilde Y^{(1)}\widetilde Y^{(2)}\mid q]
=
\mathbb E[m_q(Z)^2].
\]
Since
\[
\mathbb E[\widetilde Y^{(1)}\mid q]
=
\mathbb E[\widetilde Y^{(2)}\mid q]
=
\mu_q,
\]
we obtain
\[
\Cov(\widetilde Y^{(1)},\widetilde Y^{(2)}\mid q)
=
\mathbb E[m_q(Z)^2]-\mu_q^2
=
\Var(m_q(Z))
=
\tau_q.
\]

For \eqref{eq:app-repeated-abstention-kappa}, the same argument gives
\[
\mathbb E[I^{(1)}I^{(2)}\mid Z,q]
=
a_q(Z)^2,
\]
hence
\[
\Cov(I^{(1)},I^{(2)}\mid q)
=
\mathbb E[a_q(Z)^2]-\alpha_q^2
=
\Var(a_q(Z))
=
\kappa_q.
\]

For \eqref{eq:app-repeated-abstention-rho},
\[
\mathbb E[I^{(1)}\widetilde Y^{(2)}\mid Z,q]
=
\mathbb E[I^{(1)}\mid Z,q]\,
\mathbb E[\widetilde Y^{(2)}\mid Z,q]
=
a_q(Z)m_q(Z).
\]
Taking expectations yields
\[
\mathbb E[I^{(1)}\widetilde Y^{(2)}\mid q]
=
\mathbb E[a_q(Z)m_q(Z)].
\]
Subtracting the product of marginals gives
\[
\Cov(I^{(1)},\widetilde Y^{(2)}\mid q)
=
\mathbb E[a_q(Z)m_q(Z)]-\alpha_q\mu_q
=
\Cov(a_q(Z),m_q(Z))
=
\rho_q.
\]
The same calculation proves
\[
\Cov(\widetilde Y^{(1)},I^{(2)}\mid q)=\rho_q.
\]
Collecting the three identities yields the covariance matrix claim.
\end{proof}

\begin{corollary}[First and second moments under repeated abstention]
\label{cor:app-repeated-abstention-first-second-moments}
The combined channels $\mathsf A$ and $\mathsf A_2$ point-identify the five
querywise quantities
\[
\mu_q,\qquad \alpha_q,\qquad \tau_q,\qquad \kappa_q,\qquad \rho_q.
\]
Equivalently, they identify the full first and second moment structure of the
latent semantic vector $(m_q(Z),a_q(Z))$.
\end{corollary}

\begin{proof}
The single-sample abstention channel $\mathsf A$ identifies
$(\mu_q,\alpha_q)$ by Theorem~\ref{thm:app-abstention-identification}, while
Proposition~\ref{prop:app-repeated-abstention-covariance} identifies
$(\tau_q,\kappa_q,\rho_q)$ from repeated abstention covariances.
\end{proof}

\begin{remark}[Interpretation]
\label{rem:app-repeated-abstention-interpretation}
The main text identifies whether disagreement is persistent or whether
non-binary mass is present. The repeated abstention extension goes one step
further: it identifies whether the \emph{propensity to abstain} is itself stable
across latent types, and whether latent types that abstain more also exhibit
systematically different directional preferences. In that sense, repeated
abstention recovers a second-order semantics of uncertainty that is invisible
under both anonymous binary feedback and one-shot abstention.
\end{remark}

\section{Experimental Details}
\label{app:exp-details}

This appendix records the exact synthetic data-generating processes, estimators,
evaluation metrics, and random seeds used in all experiments. The main paper uses
the semantic-fiber and action-recovery figures, while this appendix contains only
the remaining theorem-local diagnostics: the strict channel frontier, the
repeat-based recovery of $\tau_q$, and the abstention-based recovery of
$\alpha_q$. All experiments are \emph{theorem-matched}: the latent worlds are
generated directly from the constructive proofs in
Appendices~\ref{app:proof-theorem-3-1}, \ref{app:proofs-section-4}, and
\ref{app:proofs-finite-sample}. No reward model, classifier, or preference model
is fitted in any experiment.

Throughout, the symbols $\mu_q$, $\tau_q$, and $\alpha_q$ have the same meaning
as in the main text: aggregate directional preference, persistent disagreement,
and non-binary mass, respectively. All reported error bars correspond to
$\pm 2$ standard errors across Monte Carlo replications. Random number generation
uses fixed integer seeds specified separately for each experiment.

\subsection{Common constructive generator}
\label{app:exp-details:common-generator}

Several experiments require generating a latent world with prescribed semantic
triple $(\mu_q,\tau_q,\alpha_q)$ satisfying the feasibility constraints of
Theorem~\ref{thm:binary-fiber-main}. We use the exact constructive proof.

For a fixed query $q$, define the \emph{homogeneous component}
\[
H_q:\qquad (a_q,m_q)=(\alpha_q,\mu_q),
\]
which realizes
\[
\tau_q=0.
\]
Define the \emph{extremal component} by the three latent states
\[
(a_q,m_q)\in\{(0,1),(1,0),(0,-1)\},
\]
with probabilities
\[
p_q^{+}=\frac{1-\alpha_q+\mu_q}{2},
\qquad
p_q^{0}=\alpha_q,
\qquad
p_q^{-}=\frac{1-\alpha_q-\mu_q}{2}.
\]
This realizes
\[
\tau_q^{\max}=1-\alpha_q-\mu_q^2.
\]
Then
\[
\lambda_q:=
\begin{cases}
\displaystyle
\frac{\tau_q}{1-\alpha_q-\mu_q^2}, & \text{if } 1-\alpha_q-\mu_q^2>0,\\[8pt]
0, & \text{otherwise},
\end{cases}
\]
and the final querywise latent world is the mixture
\[
(1-\lambda_q)H_q+\lambda_q E_q.
\]
By construction, this world has semantic triple
\[
(\mu_q,\tau_q,\alpha_q).
\]

When $\alpha_q=0$, the extremal component reduces to two latent states
$m_q\in\{+1,-1\}$ with probabilities
\[
\frac{1+\mu_q}{2},\qquad \frac{1-\mu_q}{2},
\]
and the same interpolation formula becomes
\[
\lambda_q=\frac{\tau_q}{1-\mu_q^2}.
\]
This is the construction used in the repeat-recovery experiment.

\subsection{Evaluation metrics}
\label{app:exp-details:metrics}

For the repeat- and abstention-recovery experiments, the querywise mean squared
errors are
\[
\mathrm{MSE}_\tau
:=
\frac{1}{|\mathcal Q|}
\sum_{q\in\mathcal Q}
(\widehat\tau_q-\tau_q)^2,
\qquad
\mathrm{MSE}_\alpha
:=
\frac{1}{|\mathcal Q|}
\sum_{q\in\mathcal Q}
(\widehat\alpha_q-\alpha_q)^2.
\]

For the action-recovery experiment, if $a_q^\star$ denotes the oracle semantic
action and $\widehat a_q$ the plug-in action rule, the action accuracy is
\[
\mathrm{Acc}
:=
\frac{1}{|\mathcal Q|}
\sum_{q\in\mathcal Q}
\mathbf 1\{\widehat a_q=a_q^\star\}.
\]
Confusion matrices are accumulated over all Monte Carlo replications and then
row-normalized, so each row reports the empirical conditional distribution of the
predicted action given the true action.

For the channel-frontier experiment, the exact and empirical distances are total
variation distances between observable laws:
\[
\mathrm{TV}(P,Q)
:=
\frac12\sum_x |P(x)-Q(x)|.
\]

\subsection{Main-text semantic-fiber experiment}
\label{app:exp-details:fig1}

The semantic-fiber experiment fixes
\[
\mu^\star=0.40.
\]
The non-binary mass is gridded as
\[
\alpha \in \left\{0,\frac{1-|\mu^\star|}{12},\ldots,1-|\mu^\star|\right\}
=
\{0,0.05,\ldots,0.60\},
\]
and the interpolation weight is gridded as
\[
\lambda \in \{0,0.1,\ldots,1\}.
\]
For each pair $(\alpha,\lambda)$, we set
\[
\tau=\lambda\bigl(1-\alpha-(\mu^\star)^2\bigr),
\]
construct the corresponding exact latent world using the common generator above,
and then sample
\[
n_B=n_R=n_A=100{,}000
\]
observations from each of the three channels. The random seed is
\[
\texttt{seed}=7.
\]

The plotted Monte Carlo estimates are
\[
\widehat\mu
=
\frac{1}{n_B}\sum_{i=1}^{n_B}Y_i,
\qquad
\widehat\tau
=
\frac{1}{n_R}\sum_{i=1}^{n_R}Y_i^{(1)}Y_i^{(2)}-\widehat\mu^2,
\qquad
\widehat\alpha
=
\frac{1}{n_A}\sum_{i=1}^{n_A}\mathbf 1\{\widetilde Y_i=0\}.
\]
No averaging across replications is used in this figure because the sample sizes
are already large enough that Monte Carlo noise is visually negligible.

\subsection{Channel-frontier experiment}
\label{app:exp-details:fig2}

The channel-frontier experiment uses the three explicit one-query witness worlds
from Appendix~\ref{app:proofs-section-4}:

\paragraph{Coin world.}
One-point latent space with
\[
(a_q,m_q)=(0,0).
\]

\paragraph{Zero world.}
One-point latent space with
\[
(a_q,m_q)=(1,0).
\]

\paragraph{Split world.}
Two-point latent space with equal weights and
\[
(a_q,m_q)\in\{(0,1),(0,-1)\}.
\]

For each pair of witness worlds and each channel
\[
\mathsf B,\qquad \mathsf{BR},\qquad \mathsf A,\qquad \mathsf{BRA},
\]
we compute the exact observable law using the closed-form formulas from
Appendix~\ref{app:formal-setup}, and then compute the exact total variation
distance. This yields the exact matrix
\[
\begin{array}{c|cccc}
& \mathsf B & \mathsf{BR} & \mathsf A & \mathsf{BRA}\\
\hline
\textsc{Coin--Zero}  & 0 & 0   & 1 & 1 \\
\textsc{Coin--Split} & 0 & 1/2 & 0 & 1/2 \\
\textsc{Zero--Split} & 0 & 1/2 & 1 & 1
\end{array}
\]
shown in panel (a). Panel (b) estimates the same matrix empirically from
\[
n_{\mathrm{MC}}=200{,}000
\]
samples per world-channel pair, using the random seed
\[
\texttt{seed}=17.
\]

\begin{figure}[t]
\centering
\includegraphics[width=\textwidth]{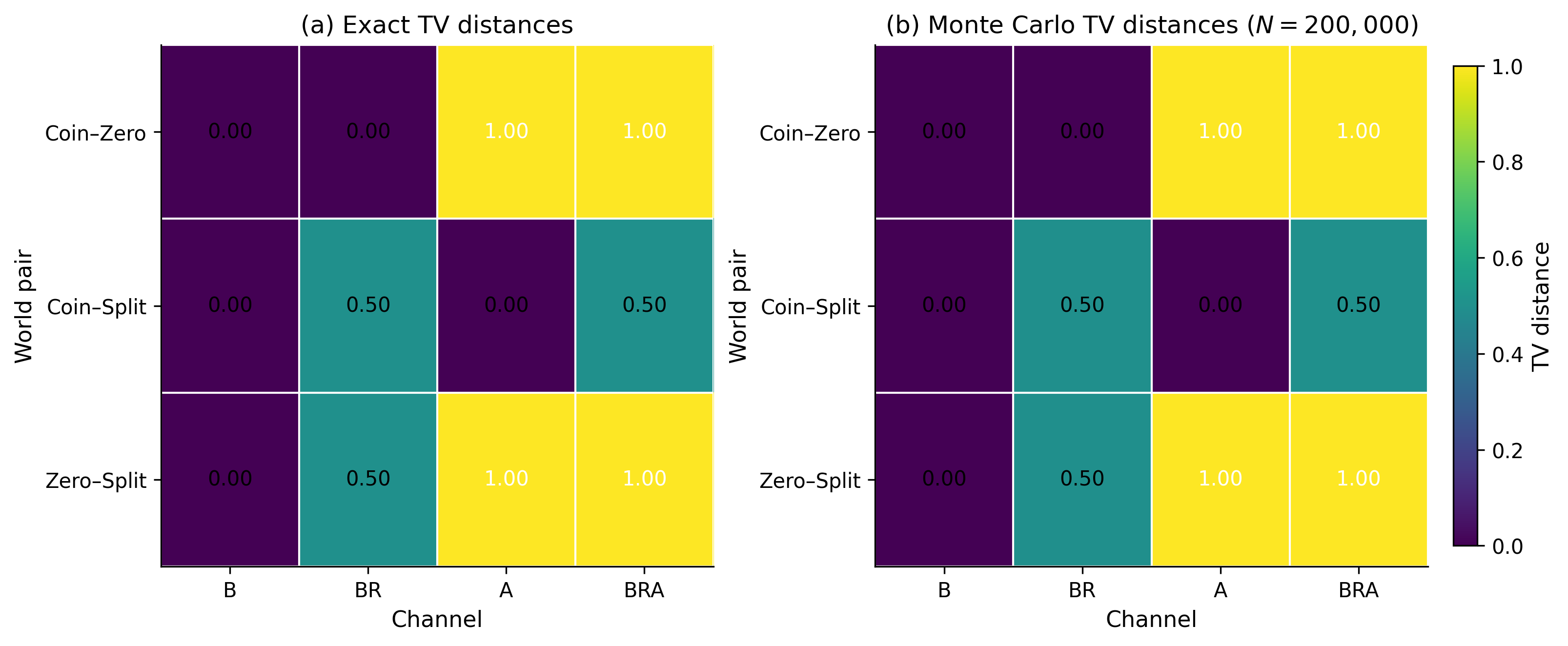}
\caption{\textbf{Strict channel frontier via witness worlds.}
Rows compare the three explicit one-query witness pairs:
\textsc{Coin--Zero}, \textsc{Coin--Split}, and \textsc{Zero--Split}.
Columns correspond to the observable channels
$\mathsf B$, $\mathsf{BR}$, $\mathsf A$, and $\mathsf{BRA}$.
Each cell reports the total variation distance between the two observable laws
under the given channel.
Panel (a) shows the exact distances implied by
Theorem~\ref{thm:frontier-main}, while panel (b) shows Monte Carlo estimates that
recover the same strict-refinement pattern.}
\label{fig:channel-frontier-app}
\end{figure}

\subsection{Repeat-based recovery of persistent disagreement}
\label{app:exp-details:fig3}

For the repeat-recovery experiment, the query set has size
\[
|\mathcal Q|=250.
\]
The latent world is generated by
\[
\mu_q\sim \mathrm{Unif}[-0.8,0.8],
\qquad
\alpha_q=0,
\qquad
\tau_q=c_q(1-\mu_q^2),
\qquad
c_q\sim \mathrm{Unif}[0,0.8].
\]
This ensures
\[
0\le \tau_q\le 1-\mu_q^2,
\]
so the world is feasible querywise. The world-generation seed is
\[
\texttt{world\_seed}=11.
\]

For binary data we use a fixed budget
\[
n_B=512.
\]
For repeat pairs we use the grid
\[
n_R\in\{2,4,8,16,32,64\}.
\]
The representative display budget in panels (a) and (c) is
\[
n_R=32.
\]
The simulation seed is
\[
\texttt{sim\_seed}=123.
\]

The estimators are
\[
\widehat\mu_q
=
\frac{1}{n_B}\sum_{i=1}^{n_B}Y_{q,i}^{B},
\qquad
\widehat\nu_q
=
\frac{1}{n_R}\sum_{i=1}^{n_R}
Y_{q,i}^{R,(1)}Y_{q,i}^{R,(2)},
\]
\[
\widehat\tau_q
=
\widehat\nu_q-\widehat\mu_q^2.
\]
For panel (b) we also report the bias-corrected estimator from
Appendix~\ref{app:proofs-finite-sample}:
\[
\widetilde\tau_q
=
\widehat\nu_q-\frac{n_B\widehat\mu_q^2-1}{n_B-1}.
\]

Panel (c) uses the empirical disagreement-gap estimator
\[
\widehat\tau_q^{\,\mathrm{gap}}
=
2\bigl(\widehat D_q^{\mathrm{diff}}-\widehat D_q^{\mathrm{same}}\bigr),
\]
where
\[
\widehat D_q^{\mathrm{same}}
=
\frac{1}{n_R}\sum_{i=1}^{n_R}
\mathbf 1\{Y_{q,i}^{R,(1)}\neq Y_{q,i}^{R,(2)}\},
\]
and $\widehat D_q^{\mathrm{diff}}$ is formed by pairing consecutive anonymous
binary labels:
\[
\widehat D_q^{\mathrm{diff}}
=
\frac{1}{\lfloor n_B/2\rfloor}
\sum_{i=1}^{\lfloor n_B/2\rfloor}
\mathbf 1\{Y_{q,2i-1}^{B}\neq Y_{q,2i}^{B}\}.
\]

Panel (b) reports the mean and $\pm 2$ standard errors of the querywise MSE over
\[
250
\]
Monte Carlo replications. Panels (a) and (c) report querywise averages of the
corresponding estimators over
\[
250
\]
replications at the representative budget $n_R=32$. The dotted $1/n_R$ line in
panel (b) is a guide-to-eye reference anchored at the first empirical plug-in MSE;
it is not itself a fitted model.

\begin{figure}[t]
\centering
\includegraphics[width=\textwidth]{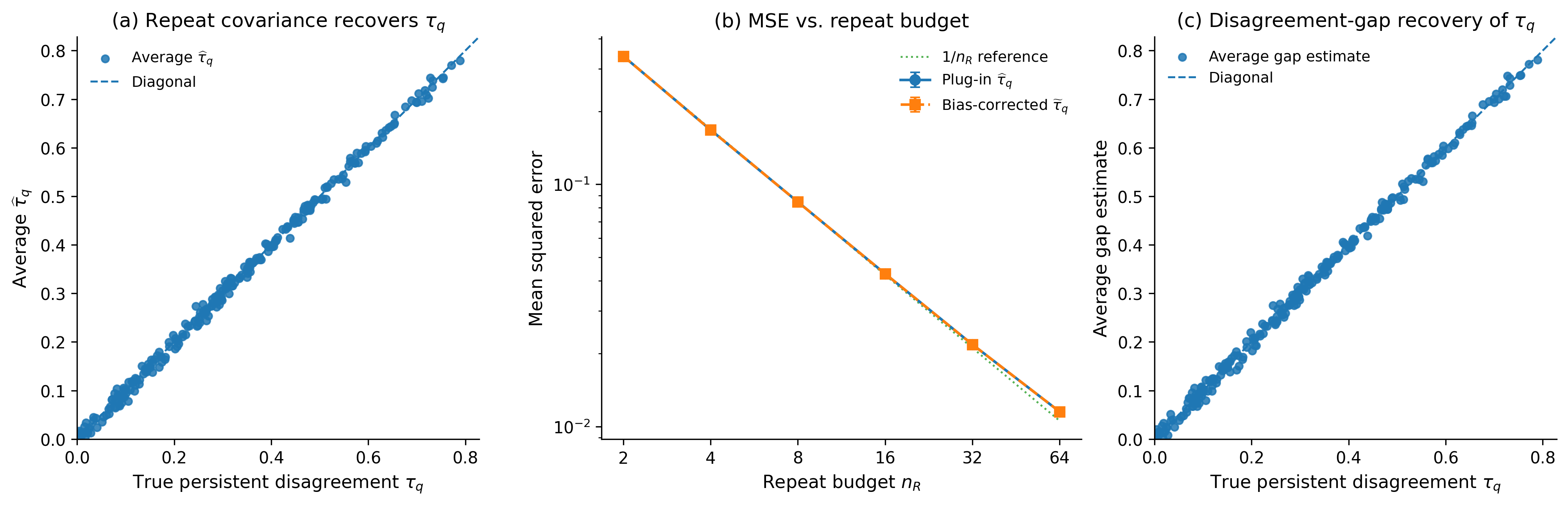}
\caption{\textbf{Recovery of persistent disagreement from repeats.}
Queries are generated with $\alpha_q=0$,
$\mu_q\sim\mathrm{Unif}[-0.8,0.8]$, and
$\tau_q=c_q(1-\mu_q^2)$ with $c_q\sim\mathrm{Unif}[0,0.8]$.
Panel (a) shows querywise recovery of $\tau_q$ from repeat covariance at
representative repeat budget $n_R=32$.
Panel (b) reports mean squared error versus repeat budget $n_R$, together with a
$1/n_R$ reference line and the bias-corrected estimator from
Appendix~\ref{app:proofs-finite-sample}.
Panel (c) validates the disagreement-gap identity
$2(D_q^{\mathrm{diff}}-D_q^{\mathrm{same}})=\tau_q$.
All estimates in panels (a) and (c) are averaged over 250 Monte Carlo
replications.}
\label{fig:tau-recovery-app}
\end{figure}

\subsection{Abstention-based recovery of non-binary mass}
\label{app:exp-details:fig4}

For the abstention-recovery experiment, the query set again has size
\[
|\mathcal Q|=250.
\]
The latent world is generated by
\[
\mu_q\sim \mathrm{Unif}[-0.8,0.8],
\qquad
\alpha_q=b_q(1-|\mu_q|),
\qquad
b_q\sim \mathrm{Unif}[0,0.7],
\qquad
\tau_q=0.
\]
This guarantees
\[
0\le \alpha_q\le 1-|\mu_q|,
\]
so the world is feasible. The world-generation seed is
\[
\texttt{world\_seed}=19.
\]

The abstention budgets are
\[
n_A\in\{2,4,8,16,32,64\},
\]
with representative display budget
\[
n_A=32.
\]
The simulation seed is
\[
\texttt{sim\_seed}=321.
\]

The estimator is
\[
\widehat\alpha_q
=
\frac{1}{n_A}\sum_{i=1}^{n_A}
\mathbf 1\{\widetilde Y_{q,i}^{A}=0\}.
\]
The exact query-averaged theoretical MSE in panel (b) is the Bernoulli variance
formula,
\[
\frac{1}{|\mathcal Q|}
\sum_{q\in\mathcal Q}
\frac{\alpha_q(1-\alpha_q)}{n_A},
\]
which is computed directly from the latent world and plotted without fitting.

For panel (c), we use the Hoeffding confidence interval from
Appendix~\ref{app:proofs-finite-sample} with confidence level
\[
1-\delta=0.95,
\qquad
\delta=0.05,
\]
and radius
\[
r_{A}(n_A,\delta)
=
\sqrt{\frac{\log(2/\delta)}{2n_A}}.
\]
For each replication, we compute the querywise coverage fraction
\[
\frac{1}{|\mathcal Q|}
\sum_{q\in\mathcal Q}
\mathbf 1\{|\widehat\alpha_q-\alpha_q|\le r_A(n_A,\delta)\},
\]
and then average this quantity across replications.

Panel (b) and panel (c) both use
\[
250
\]
Monte Carlo replications. Panel (a) plots the querywise average of
$\widehat\alpha_q$ over
\[
250
\]
replications at the representative budget $n_A=32$.

\begin{figure}[t]
\centering
\includegraphics[width=\textwidth]{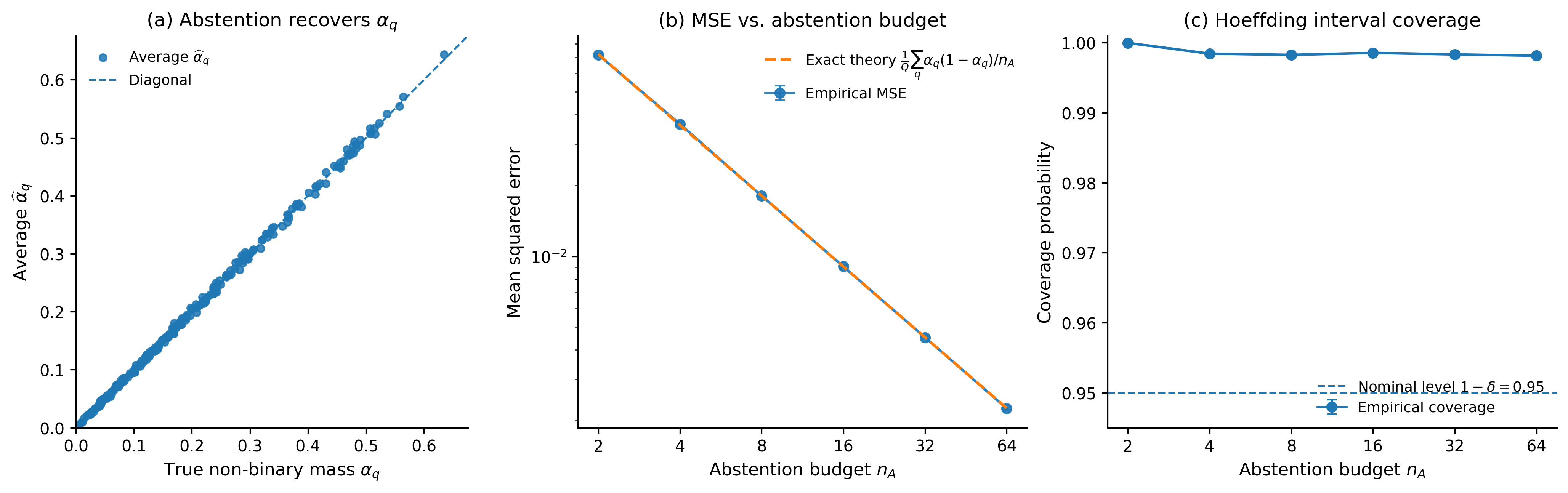}
\caption{\textbf{Recovery of non-binary mass from abstention.}
Queries are generated with
$\mu_q\sim\mathrm{Unif}[-0.8,0.8]$,
$\alpha_q=b_q(1-|\mu_q|)$ with $b_q\sim\mathrm{Unif}[0,0.7]$,
and $\tau_q=0$.
Panel (a) shows querywise recovery of $\alpha_q$ at representative abstention
budget $n_A=32$.
Panel (b) reports mean squared error versus abstention budget $n_A$, together
with the exact theoretical Bernoulli MSE
$\frac{1}{|\mathcal Q|}\sum_q \alpha_q(1-\alpha_q)/n_A$.
Panel (c) shows empirical coverage of the Hoeffding confidence interval for
$\alpha_q$, illustrating valid but conservative finite-sample coverage.}
\label{fig:alpha-recovery-app}
\end{figure}

\subsection{Main-text action-recovery experiment}
\label{app:exp-details:fig5}

The action-recovery experiment evaluates the decision-theoretic consequence of the
channel-refinement results. We define the thresholds
\[
t_\tau=t_\alpha=0.15
\]
and three semantic regimes:
\[
\textsc{Denoise}:\ (\tau,\alpha)=(0.02,0.02),
\]
\[
\textsc{Personalize}:\ (\tau,\alpha)=(0.35,0.02),
\]
\[
\textsc{Query}:\ (\tau,\alpha)=(0.02,0.35).
\]
Each regime contains
\[
Q_{\mathrm{per\,class}}=120
\]
queries, so the total number of queries is
\[
|\mathcal Q|=360.
\]

To make binary underidentification especially sharp, we first draw a base set
\[
\mu_q^{\mathrm{base}}\sim \mathrm{Unif}[-0.6,0.6]
\]
of size $Q_{\mathrm{per\,class}}$ and then reuse exactly the same set of
$\mu$-values across all three regimes. This ensures that the three regimes share
the same empirical binary-law distribution, while differing only in
$(\tau,\alpha)$. After concatenation, the queries are randomly permuted. The
world-generation seed is
\[
\texttt{world\_seed}=23.
\]

The fixed binary budget is
\[
n_B=512,
\]
and the calibration budgets are tied:
\[
n_R=n_A=b,
\qquad
b\in\{2,4,8,16,32,64\}.
\]
The representative confusion-matrix budget is
\[
b=16.
\]
The simulation seed is
\[
\texttt{sim\_seed}=777.
\]

For each replication, we generate a single binary sample of size $n_B$, a repeat
sample of size $\max b$, and an abstention sample of size $\max b$, and then use
the first $b$ samples of each refined channel when evaluating budget $b$. This
coupling reduces Monte Carlo variance across budgets without changing the
distribution at any fixed budget.

The binary-only baseline is the constant rule
\[
\widehat a_q^{\mathsf B}\equiv \textsc{Denoise},
\]
which is the tie-broken majority-class baseline in this balanced three-class
design. The refined-channel action rules are
\[
\widehat a_q^{\mathsf{BR}}
=
\begin{cases}
\textsc{Personalize}, & \widehat\tau_q>t_\tau,\\
\textsc{Denoise}, & \widehat\tau_q\le t_\tau,
\end{cases}
\]
\[
\widehat a_q^{\mathsf A}
=
\begin{cases}
\textsc{Query}, & \widehat\alpha_q>t_\alpha,\\
\textsc{Denoise}, & \widehat\alpha_q\le t_\alpha,
\end{cases}
\]
and
\[
\widehat a_q^{\mathsf{BRA}}
=
\begin{cases}
\textsc{Query}, & \widehat\alpha_q>t_\alpha,\\
\textsc{Personalize}, & \widehat\alpha_q\le t_\alpha \text{ and }\widehat\tau_q>t_\tau,\\
\textsc{Denoise}, & \widehat\alpha_q\le t_\alpha \text{ and }\widehat\tau_q\le t_\tau.
\end{cases}
\]

Panel (a) in the main-text action-recovery figure plots the mean and
$\pm 2$ standard errors of the action accuracy over
\[
250
\]
Monte Carlo replications. Panels (b)--(d) aggregate confusion counts over the same
\[
250
\]
replications at representative budget $b=16$ and then row-normalize them. The
binary-only confusion matrix is omitted because the rule is constant and therefore
uninformative once the accuracy curve is shown.

\subsection{Implementation notes}
\label{app:exp-details:implementation}

All figures were generated directly from deterministic scripts using fixed seeds,
with no hyperparameter search and no external data. The exact filenames are:
\[
\texttt{figure1\_exact\_semantic\_fiber\_final.png},
\]
\[
\texttt{figure2\_channel\_frontier\_final.png},
\]
\[
\texttt{figure3\_recovery\_persistent\_disagreement\_final.png},
\]
\[
\texttt{figure4\_recovery\_nonbinary\_mass\_final.png},
\]
\[
\texttt{figure5\_semantics\_aware\_action\_recovery\_final.png}.
\]
The plotting code computes all displayed quantities directly from the synthetic
samples and the formulas in this appendix; no manual post-processing or selective
smoothing was applied.
\end{document}